\documentclass[11pt]{article}
\usepackage{jheppub}
\usepackage[utf8]{inputenc}
\usepackage{cancel}
\usepackage{mathrsfs}
\usepackage{enumitem}
\usepackage{xcolor}
\usepackage{caption}  
\usepackage{graphicx} 
\usepackage{float} 
\usepackage{relsize}
\usepackage{physics}
\usepackage{psfrag}
\usepackage{cancel}
\usepackage{array}
\usepackage{amssymb}
\usepackage{amsmath}
\usepackage{amsthm}
\usepackage{float}
\usepackage{tikz}
\usetikzlibrary{decorations.markings}
\usepackage{tikz,lipsum,lmodern}
\usepackage[most]{tcolorbox}
\usepackage{hyperref}
\usepackage{xcolor}
\usepackage{bm}
\usepackage{pgfplots}
\pgfplotsset{compat=1.17}

\usepackage[normalem]{ulem}

\definecolor{twilightlavender}{rgb}{0.54, 0.29, 0.42}
\definecolor{richmaroon}{rgb}{0.69, 0.19, 0.38}
\definecolor{forestgreen(web)}{rgb}{0.13, 0.55, 0.13}
\definecolor{lava}{rgb}{0.81, 0.06, 0.13}
\hypersetup{
	breaklinks,
	colorlinks,
	citecolor=blue,
	filecolor=blue,
	linkcolor=blue,
	urlcolor=blue
}

\def\AC#1{{\color [rgb]{1.0,0.0,0.0} [AC: #1]}}

\newcommand{\C}{\mathbb{C}}      
\newcommand{\Pspace}{\mathcal{P}} 
\newcommand{\V}{\mathcal{V}}     
\newcommand{\s}{\mathcal{S}} 
\newcommand{\R}{\mathbb{R}}

\tikzset{
pattern size/.store in=\mcSize, 
pattern size = 5pt,
pattern thickness/.store in=\mcThickness, 
pattern thickness = 0.3pt,
pattern radius/.store in=\mcRadius, 
pattern radius = 1pt}
\makeatletter
\pgfutil@ifundefined{pgf@pattern@name@_opwgemt5m}{
\makeatletter
\pgfdeclarepatternformonly[\mcRadius,\mcThickness,\mcSize]{_opwgemt5m}
{\pgfpoint{-0.5*\mcSize}{-0.5*\mcSize}}
{\pgfpoint{0.5*\mcSize}{0.5*\mcSize}}
{\pgfpoint{\mcSize}{\mcSize}}
{
\pgfsetcolor{\tikz@pattern@color}
\pgfsetlinewidth{\mcThickness}
\pgfpathcircle\pgfpointorigin{\mcRadius}
\pgfusepath{stroke}
}}
\makeatother

\usepackage{color}

\title{Pure D--brane Black Holes: BPS Counting and non--BPS Vacua}

\affiliation[a]{Department of Physics, School of Basic Sciences, Indian Institute of Technology Bhubaneswar, Jatni, Khurda, Odisha, 752050, India}
\affiliation[b]{Harish--Chandra Research Institute, A CI of Homi Bhabha National Institute, Chhatnag Road, Jhunsi,
Prayagraj (Allahabad), Uttar Pradesh, 211019, India}

\usepackage{orcidlink}

\author[a\orcidlink{0000-0003-0948-4817}]{Abhishek Chowdhury}\emailAdd{achowdhury@iitbbs.ac.in}

\author[a,b\orcidlink{0009-0008-2963-2497}]{and Sourav Maji}\emailAdd{souravmaji@hri.res.in}

\abstract{In this paper, we present a unified computational framework to analyze the microscopic vacuum structure of 4--charge extremal black holes in Type IIA string theory, applying techniques from computational algebraic geometry and numerical topology to their pure D--brane effective quantum mechanics. We apply this approach to two physically distinct configurations. First, in the supersymmetric sector, we compute the $14^{\text{th}}$ helicity trace index of $\frac{1}{8}$--BPS, $\mathcal{N}=8$, D2--D2--D2--D6 configurations dual to D1--D5--P--KK monopole dyonic black holes. Extending previous work to higher charges, we employ a parametric monodromy method to explicitly resolve the vacua for the $(1,1,1,5)$ and $(1,1,1,6)$ configurations, reproducing the degeneracies predicted by the U--dual picture \cite{Chowdhury:2023wss}. Second, we apply complementary techniques to a configuration where supersymmetry is explicitly broken at the level of the effective action. The corresponding 4--charge non--BPS extremal pure D--brane system is obtained by replacing the D6--brane with an anti--D6--brane and assigning incompatible R--symmetry rotations to different brane triplets \cite{Mondal:2024qyn}. Analyzing the associated scalar potential using analytical Gr\"obner bases, we demonstrate the absence of zero--energy classical ground states. To handle the continuous flat directions populating the non--BPS landscape, we implement specific topological regularizations, namely Morse--Bott deformations and gated soft--trapping. These methods allow us to systematically characterize the classical energy landscape, identifying a non--compact Coulomb branch, marginally bound stabilizer submanifolds, and an isolated collection of doubly degenerate low--energy stable states.
}

\begin{document}

\maketitle
\section{Introduction}
\label{intro}

Understanding the microscopic origin of black hole entropy has been a central problem in string theory since it was recognized that black holes obey laws analogous to those of thermodynamics. A decisive breakthrough was achieved by Strominger and Vafa, who reproduced the Bekenstein--Hawking entropy for a class of extremal black holes via an explicit counting of microscopic BPS states in string theory \cite{Strominger:1996sh}. This result provided a concrete statistical interpretation of black hole entropy in terms of underlying quantum degrees of freedom and identified D--branes as essential constituents of black hole microphysics. Subsequent developments have substantially broadened and refined this framework and over the past two decades, major progress has been made in the microscopic counting of BPS states for supersymmetric extremal black holes in $\mathcal{N}=8$ \cite{Shih:2005qf,Pioline:2005vi,Sen:2008ta,Sen:2009gy,Chowdhury:2014yca,Chowdhury:2015gbk,Chowdhury:2023wss,Kumar:2023hlu}, $\mathcal{N}=4$ \cite{David:2006ji,David:2006ru,Dijkgraaf:1996it,Gaiotto:2005hc,Jatkar:2005bh,Shih:2005uc,Sen:2007qy,Mandal:2010cj}, and $\mathcal{N}=2$ \cite{Maldacena:1997de,Manschot:2011xc,deBoer:2008zn,Denef:2007vg} supersymmetric theories. In particular, it was realized that for supersymmetric black holes, the microscopic quantity that is robust under coupling and moduli deformations is not the absolute degeneracy, but a protected index \cite{Dabholkar:2010rm}. In four--dimensional $\mathcal{N}=8$ string theory, extremal $\tfrac{1}{8}$--BPS black holes are characterized by the $14^{\text{th}}$ helicity trace index $B_{14}\,$, which only receives contributions from states breaking exactly 28 supercharges \cite{Chowdhury:2014yca,Chowdhury:2015gbk,Chowdhury:2023wss}. 

A particularly useful microscopic description of these black holes arises in a duality frame where all charges are Ramond--Ramond \cite{Shih:2005qf}. In this frame, a 4--charge $\tfrac{1}{8}$--BPS black hole in Type IIA string theory compactified on $T^{6}$ is realized as a bound state of three stacks of D2--branes wrapping mutually orthogonal 2--cycles of the torus, together with a stack of D6--branes wrapping the full six--torus. Since the branes intersect at a point, the low--energy dynamics is that of an effective particle and is described by a matrix quantum mechanics \cite{Banks:1996vh,Taylor:2001vb}. The microscopic degrees of freedom arise from adjoint fields describing brane positions and bifundamental fields corresponding to open strings stretched between different stacks. For generic values of the background metric and $B$--field moduli, the scalar potential of this quantum mechanics is a sum of non--negative terms, and the supersymmetric vacua are isolated solutions of the F--term and D--term equations modulo complexified gauge transformations and residual shift symmetries. As a result, the supersymmetry (SUSY) vacuum manifold is a zero--dimensional algebraic variety, and the index $B_{14}$ is computed simply by counting the points with multiplicity \cite{Chowdhury:2014yca,Chowdhury:2015gbk,Chowdhury:2023wss}. 

This perspective has led to a reformulation of black hole microstate counting as an explicitly algebraic problem, enabling the application of tools from computational algebraic geometry \cite{Chowdhury:2023wss,Sturmfels1998PolynomialEA,HomotopyContinuation,Li1,cox1,cox2,kaveh2008algebraic,Kushnirenko1976NewtonPA,lichen,duffmonodromy,Bliss2018MonodromySS,Duff2016SolvingPS,Sommese2005TheNS}. Earlier work along these lines successfully reproduced the expected degeneracies for a range of low--charge configurations and provided concrete evidence in favor of the zero--angular--momentum conjecture, which states that all microstates of single--centered $\tfrac{1}{8}$--BPS black holes carry vanishing intrinsic angular momentum once the universal goldstino modes are factored out \cite{Chowdhury:2015gbk,Chowdhury:2023wss}. Building upon this algebraic reformulation, the central organizing theme of this paper is to elevate these computational geometry techniques into a unified framework capable of resolving the vacuum structure of pure D--brane systems across distinct physical sectors. We aim to demonstrate that both the exact microstate counting of supersymmetric black holes and the classical landscape mapping of fully supersymmetry--breaking configurations can be systematically addressed using complementary tools from algebraic geometry and numerical topology.

As a first application of this program, we address the substantial technical challenges associated with extending exact BPS computations to higher--charge configurations. Because the number of variables and polynomial constraints grows rapidly with the rank of the gauge groups, direct Gr\"obner basis or Hilbert series computations become increasingly difficult. To overcome these limitations, we systematically implement the parametric monodromy method for counting supersymmetric vacua. By embedding the physical system into a parameterized family and exploiting the monodromy action on the solution space, this method allows one to generate the complete set of isolated vacua starting from a single seed solution. A careful treatment of gauge fixing and adjoint shift symmetries is essential in this approach, particularly for higher--rank non--abelian charge configurations. Using this framework, we explicitly compute the supersymmetric vacua for the $(1,1,1,5)$ and $(1,1,1,6)$ charge systems. In both cases, the number of solutions agrees exactly with the degeneracies predicted by the U--dual D1--D5--P--KK monopole description \cite{Shih:2005qf}, thereby extending earlier results to higher charges and demonstrating the effectiveness of the monodromy method in the pure D--brane setting.

In addition to the supersymmetric pure D--brane system, we also investigate 4--charge non--BPS extremal black holes within the same mathematical framework. These configurations are obtained by replacing the D6--brane with an anti--D6--brane. Although the bosonic field content and gauge symmetry remain unchanged, supersymmetry is completely broken by assigning different $\mathcal{N}=1$ subalgebras to different brane triplets via appropriate R--symmetry rotations \cite{Mondal:2024qyn}. Motivated by the longstanding questions regarding the existence of ground state degeneracy for non--supersymmetric extremal black holes \cite{Page:2000dk}, we analyze the vacua structure by enumerating the stationary points of the full non--supersymmetric scalar potential. Unlike the BPS case, the gauge symmetry is now restricted to the $U(1)\times U(1) \times U(1)$ group and does not admit complexified extensions. The main technical challenge is the loss of holomorphicity in the F--term expressions, which rules out the straightforward application of various algebraic geometry techniques, thereby encouraging us to get creative and extend their validity beyond normal applicability. 

Using analytical Gröbner basis techniques, we show that the conditions for vanishing potential admit no solutions, thereby ruling out the existence of zero--energy stationary configurations.  In our high precision numerical explorations aiming to characterize the vacuum manifold for the non--BPS case, we find a collection of 12 (6 degenerate pairs) low--energy states and, in particular, a perturbatively $\mathbb{Z}_2$ degenerate ground state (as in the classical limit of a double--well potential). It is expected that this perturbative degeneracy will be lifted non--perturbatively by instanton effects, resulting in a unique ground state. This provides a microscopic realization of a non--supersymmetric extremal black hole -- possibly with a unique non--zero energy ground state -- in line with expectations from semiclassical gravity and recent analyses of near--extremal dynamics \cite{Maldacena:1998uz,Das:1996wn,Iliesiu:2020qvm, Heydeman:2020hhw, Ghosh:2019rcj}. Technically, a more interesting sector of the vacuum landscape is a collection of stabilizer submanifold vacua representing various possible combinations of marginally bounded D--brane configurations with energies higher than the isolated stable vacua. Along the way, we have developed physics--inspired techniques, in particular the use of Morse--Bott theory \cite{Bott1954NondegenerateCM,Nishikawa2025OnDM} to lift the flat directions in these submanifolds, thereby facilitating the use of second--order extremization algorithms like Newton--Raphson method \cite{BENISRAEL1966243} to search for extremum points.

Taken together, our results show that both BPS and non--BPS 4--charge extremal black holes can be analyzed within a unified algebraic framework based on the pure D--brane description. 
In supergravity, extremal BPS and non--BPS black holes with identical charges share the same Wald entropy, reflecting the charge--only nature of the attractor mechanism \cite{Sen:2007qy}. It is therefore expected that the microscopic degeneracy of the corresponding non--BPS D--brane configurations will reproduce the leading entropy of the associated BPS system. The abelian non--BPS system does indeed have 12 isolated vacua. As reflected in the title of this work, we must draw a clear distinction between the BPS and non--BPS sectors regarding the physical interpretation of these solutions. For the BPS case, our exact enumeration of zero--energy vacua directly computes the protected microstate index. However, for the non--BPS case, it is debatable whether the isolated vacua we find should be elevated to the status of a genuine ``band" of quantum microstates that source a macroscopic ``configurational entropy" \cite{Anninos:2011vn,Mondal:2024qyn}. If viewed as a collection of non--zero energy ground state(s) and excitations, the vacuum landscape could very well destabilize the near--horizon $AdS_2$ geometry responsible for the attractor mechanism \cite{Almheiri:2014cka,Maldacena:2016upp}. Furthermore, it is debatable if we should bank heavily on two--derivative attractor mechanism results for the lowest abelian charges, as the small leading order entropy is likely to receive significant log corrections and higher derivative corrections which differ for BPS and non--BPS black holes \cite{Aspman:2024jvl,LopesCardoso:1999cv,Castro:2018hnc,Sen:2012cj}. Therefore, rather than a definitive quantum microstate count, our non--BPS analysis should be understood as a precise mapping of the classical vacuum landscape. Viewed through this lens, the analysis of both the BPS and non--BPS sectors represent complementary applications of the same underlying mathematical program -- casting pure D--brane dynamics as a systematic problem in algebraic and numerical geometry.

Moreover, there is an important caveat to the universality of this result for all values of moduli parameters.  Unlike the BPS case, where the count is a topological invariant and oblivious to deformations of the moduli parameters, in the non--BPS case, there is no reason for the protection offered by supersymmetry to continue \footnote{Even in two--derivative supergravity, the moduli independence of the non--BPS attractor is conditional; unlike first--order BPS flows, non--BPS solutions satisfy second--order equations sensitive to boundary conditions. The attractor point is typically a local minimum or saddle point of the effective potential rather than a global minimum, meaning a stable flow exists only if the asymptotic moduli lie within its specific \textit{basin of attraction}. Crossing a stability wall in moduli space causes the flow to exit this basin, rendering the attractor solution dynamically unstable or non--existent \cite{Goldstein:2005hq,Tripathy:2005qh,Denef:2000nb,Kallosh:2006ib}.}. In our random numerical scans over the moduli parameter space, the generic finding is the existence of 12 minima, corresponding to 6 doubly degenerate isolated vacua. However, for a small subset of moduli parameter values, we observe that the number of isolated vacua fluctuates around this expected value; such occasional parameter--dependent fluctuations are completely in line with the standard bifurcation behavior of complex nonlinear systems \cite{Strogatz:2018}. More broadly, this work illustrates how modern algebraic--geometry methods provide a concrete and systematically improvable approach to the black hole microstate counting problem directly at the level of microscopic D--brane dynamics.

The rest of the paper is organized as follows. In Section \ref{sec:DbraneReview} we review the pure D--brane BPS system and its associated quantum mechanics. Section \ref{sec:monodromy} describes the monodromy method and its implementation. In Section \ref{sec:bpshighercharge} we apply this framework to higher--charge BPS configurations. Section \ref{sec:nonbps} is devoted to the analysis of the non--BPS vacuum landscape and additional discussion on the discrete symmetries and ways to lift flat directions. The last Section \ref{sec:conclusion} is devoted to further comments and future directions. There are five appendices. In Appendix \ref{app: monoexamples}, we show how the monodromy method works through illustrative examples, followed by Appendix \ref{app:master_model}, where we discuss an example that shows when it might fail. Appendix \ref{app:GB} discusses some aspects of Gr\"obner basis, in particular the algebraic certification with examples. Appendix \ref{app:gauge_consistency} analyzes the consistency of gauge--restricted extremization, in particular the careful treatment of stabilizer vacua and the associated Hessians. Finally, Appendix \ref{app:Stability} discusses stability jumps in complex systems and their possible application to non--BPS D--brane systems.

\section{The Pure D--brane System}
\label{sec:DbraneReview}

In this section, we briefly review the D--brane model introduced in 
\cite{Chowdhury:2014yca,Chowdhury:2015gbk}, focusing on the field content, 
the structure of the superpotential, and the scalar potential governing the 
supersymmetric vacua. The setup is a D2--D2--D2--D6 configuration in Type IIA theory compactified on $T^{6}$, with  
\begin{align}
N_1 & \text{ D2--branes wrap: } (x^4,x^5)\,, \nonumber\\
N_2 & \text{ D2--branes wrap: } (x^6,x^7)\,, \nonumber\\
N_3 & \text{ D2--branes wrap: } (x^8,x^9)\,, \nonumber\\
N_4 & \text{ D6--branes wrap: } (x^4,\dots,x^9)\,. \nonumber
\end{align}
Because the branes intersect at a point in the non--compact directions, the configuration preserves four of the original 32 supercharges, giving 
a $\tfrac{1}{8}$--BPS quantum mechanical system obtained by reducing a 
$(3+1)$--dimensional $\mathcal{N}=1$ theory down to $0+1$ dimensions. There are two kinds of strings attached to these branes: 
\begin{itemize}
    \item \textbf{On--brane strings.} Each stack of D--branes carries its own low--energy effective theory, obtained by 
dimensionally reducing a \((3+1)\)--dimensional \(\mathcal{N}=4\) supersymmetric 
\(U(N_k)\) Yang--Mills theory down to \(0+1\) dimensions. This effective theory arises from open 
strings that begin and end on the same brane stack. In \(\mathcal{N}=1\) language, 
the resulting field content of the $k$--th brane stack consists of a single vector multiplet \(V^{(k)}\) comprising the gauge fields $X^{(k)}_1,\, X^{(k)}_2, \, X^{(k)}_3$, which denote the positions of the branes in the non--compact directions together 
with three adjoint chiral multiplets \(\Phi^{(k)}_{1}, \Phi^{(k)}_{2}, \Phi^{(k)}_{3}\), 
which describe the transverse fluctuations of the branes in the compact directions.

\item \textbf{Mixed--brane strings.} Open strings stretched between two different stacks \(k\) and \(\ell\)  give 
rise to \(\mathcal{N}=2\) hypermultiplets, or equivalently to a pair 
of \(\mathcal{N}=1\) chiral multiplets \(Z^{(k\ell)}\) and \(Z^{(\ell k)}\), which transform 
in the bifundamental and anti--bifundamental representations of the corresponding 
gauge groups.
\end{itemize}
Putting all stacks together, the gauge symmetry of the quantum  mechanical theory is  
\begin{equation}
U(N_1)\times U(N_2)\times U(N_3)\times U(N_4)\,.
\end{equation}
\subsection{Action and the Interaction Terms}

Assuming the six circles of $T^6$ are mutually orthogonal, and each has radius $\sqrt{\alpha'}$, we set $\alpha' = 1$ for simplicity. The action for the pure D--brane system is then given by
\begin{equation}\label{lagrangian}
S = S_{\text{kinetic}} + \int dx^0 \left[ \int d^4\theta \sum_{k=1}^4 \sum_{\substack{\ell=1 \\ \ell \neq k}}^4 \left( \bar{Z}^{(k\ell)} e^{2(V^{(\ell)} - V^{(k)})} Z^{(k\ell)} \right) + \int d^2\theta\, \mathcal{W} + \int d^2\bar{\theta}\, \overline{\mathcal{W}} \right],
\end{equation}
where $S_{\text{kinetic}}$ denotes the standard kinetic terms for the chiral and vector superfields. The interactions among the adjoint and bifundamental fields are encoded in the superpotential,
\begin{equation}
\label{eq: super}
\mathcal{W} = \mathcal{W}_1 + \mathcal{W}_2 + \mathcal{W}_3 + \mathcal{W}_4,
\end{equation}
where the individual contributions are as follows (for details on origin and interpretation see \cite{Chowdhury:2014yca,Chowdhury:2015gbk}): 
\begin{align}
\mathcal{W}_1 & = \sqrt{2} \left[
\sum_{k,\ell,m=1}^{3} \varepsilon^{k\ell m}\, \mathrm{Tr}\!\left(\Phi_m^{(k)} Z^{(k\ell)} Z^{(\ell k)}\right)
+ \sum_{k=1}^{3} \mathrm{Tr}\!\left(Z^{(4k)} \Phi_k^{(k)} Z^{(k4)} - \Phi_k^{(4)} Z^{(4k)} Z^{(k4)}\right)
\right]\,, \label{eq: yukawa}\\
\mathcal{W}_2 &= \sqrt{2} \sum_{\substack{k,\ell,m=1 \\ k<\ell,m;\,\,\ell\neq m}}^{4}
(-1)^{\delta_{k1}\delta_{\ell 3}\delta_{m4}}
\mathrm{Tr}\!\left(Z^{(k\ell)} Z^{(\ell m)} Z^{(m k)}\right)\,,
\label{eq: cubic} \\
\mathcal{W}_3 &= \sqrt{2} \left[
\sum_{k,\ell,m=1}^{3} c^{(k\ell)} \varepsilon^{k\ell m} N_{\ell}\,\mathrm{Tr}(\Phi_m^{(k)})
+ \sum_{k=1}^{3} c^{(k4)}\left(N_{4}\,\mathrm{Tr}(\Phi_k^{(k)}) - N_{k}\,\mathrm{Tr}(\Phi_k^{(4)})\right)
\right]\,, \label{eq: linear}\\
\mathcal{W}_4 & = -\sqrt{2}\sum_{k=1}^{4}(-1)^{\delta_{k2}}\,\mathrm{Tr}\!\left(\Phi_1^{(k)} [\Phi_2^{(k)},\Phi_3^{(k)}]\right)\, . \label{eq: yangmill}
\end{align}
The non--zero parameters $c^{(k\ell)}$ depend on small constant background values of the off--diagonal components of the metric and NS--NS two--form fields along the compactified tori. 

\subsection{Scalar Potentials}\label{potentials}

We use the same notation for superfields and their scalar components.  The scalar potential decomposes into
\begin{equation} \label{eq: scalar}
V = V_{\mathrm{gauge}} + V_D + V_F \,,
\end{equation}
where $V_{\mathrm{gauge}}$ includes commutator and covariant--derivative interactions among $X_i^{(k)}$ and $\Phi_i^{(k)}$, 
\begin{align}
\label{eq: vgauge}
V_{\text{gauge}} = 
& \sum_{k=1}^4 \sum_{\substack{\ell=1 \\ \ell \neq k}}^4 \sum_{i=1}^3 \operatorname{Tr}\left[ \left( X_i^{(k)} Z^{(k\ell)} - Z^{(k\ell)} X_i^{(\ell)} \right)^\dagger \left( X_i^{(k)} Z^{(k\ell)} - Z^{(k\ell)} X_i^{(\ell)} \right) \right] \notag \\
& + \sum_{k=1}^4 \sum_{i,j=1}^3 \operatorname{Tr}\left( \left[ X_i^{(k)}, \Phi_j^{(k)} \right]^\dagger \left[ X_i^{(k)}, \Phi_j^{(k)} \right] \right) \notag \\
& + \frac{1}{4} \sum_{k=1}^4 \sum_{i,j=1}^3 \operatorname{Tr}\left( \left[ X_i^{(k)}, X_j^{(k)} \right]^\dagger \left[ X_i^{(k)}, X_j^{(k)} \right] \right)\, ,
\end{align}
and the D--term potential is
\begin{equation}
\label{eq: vdterm}
V_D = \frac{1}{2} \sum_{k=1}^{4} \mathrm{Tr}\left(
\sum_{\ell\neq k} Z^{(k\ell)} Z^{(k\ell)\dagger}
- \sum_{\ell\neq k} Z^{(\ell k)\dagger} Z^{(\ell k)}
+ \sum_{i=1}^{3}[\Phi_i^{(k)},\Phi_i^{(k)\dagger}]
- c^{(k)}I_{N_k}
\right)^2,
\end{equation}
where the Fayet parameters $c^{(k)}$ satisfy $\displaystyle \sum_{k=1}^{4} c^{(k)} N_{k}=0\,$, ensuring gauge invariance. Finally, the F--term potential is given by
\begin{equation}
V_F =
\sum_{k,i}\left|\frac{\partial \mathcal{W}}{\partial\Phi_i^{(k)}}\right|^2
+
\sum_{k\neq \ell}\left|\frac{\partial \mathcal{W}}{\partial Z^{(k\ell)}}\right|^2\,.
\end{equation}

\subsection{Supersymmetric Solutions}

The supersymmetric vacua of the scalar potential \eqref{eq: scalar} of the theory introduced in the previous section correspond to stationary configurations with energy, $ E = V = 0\,$. Since $V$ is a sum of squares, vanishing of the total potential requires the independent vanishing of each contribution. To simplify the analysis, we make use of the temporal gauge $A_0 = 0$ and focus on the residual global part of the relative gauge group $U(N_1)\times U(N_2)\times U(N_3)\times U(N_4)/U(1)_{\mathrm{diag}}\,$. We may begin by first eliminating the gauge potential $V_{\mathrm{gauge}}\,$. Simultaneously diagonalizing the adjoint fields $X^{(k)}_i$, and following the arguments of Appendix~A of \cite{Chowdhury:2015gbk}; supersymmetric configurations require
\begin{equation}
X^{(k)}_i = 0 \qquad \text{for all } k=1,\dots,4,\; i=1,2,3\, .
\end{equation}
The other fields transform under the gauge group as,
\begin{equation}
Z^{(k\ell)} \;\longrightarrow\; U(N_k)\, Z^{(k\ell)}\, U^{-1}(N_\ell)\,, 
\qquad
\Phi^{(k)}_i \;\longrightarrow\; U(N_k)\, \Phi^{(k)}_i\, U^{-1}(N_k)\, ,
\end{equation}
and we note that due to the holomorphicity of the superpotential $\mathcal{W}$, it is invariant under the larger complexified gauge group,
\begin{equation}
\frac{GL(N_1,\mathbb{C})\times GL(N_2,\mathbb{C})\times GL(N_3,\mathbb{C})\times GL(N_4,\mathbb{C})}{GL(1,\mathbb{C})}\, .
\end{equation}
This allows us to solely focus on vanishing of the F--term potential, $V_F=0$ and then move along the complexified gauge orbits to set the $V_D=0\,$. For SUSY vacua, we get two sets of polynomial equations in the bifundamental fields $Z^{(k\ell)}$ and the adjoint scalars $\phi^{(k)}_i\,$:
\begin{itemize}
    \item Setting $\displaystyle \frac{\partial \mathcal{W}}{\partial \Phi^{(k)}_i}=0$ gives,
\begin{equation} \label{fterm1}
\begin{aligned}
Z^{(k\ell)} Z^{(\ell k)} &= -c^{(k\ell)} N_{\ell}\, I_{N_k} + [\Phi_k^{(k)},\Phi_{\ell}^{(k)}], 
&& 1\leq k,\ell \leq 3\,, \\
Z^{(k4)} Z^{(4k)} &= -c^{(k4)} N_{4}\, I_{N_k} + \sum_{\ell,m=1}^{3}\varepsilon^{k\ell m}\,\Phi_{\ell}^{(k)} \Phi_{m}^{(k)},
&& 1 \leq k \leq 3\,,\\
Z^{(4k)} Z^{(k4)} &= -c^{(k4)} N_{k}\, I_{N_4} - \sum_{\ell,m=1}^{3}\varepsilon^{k\ell m}\,\Phi_{\ell}^{(4)} \Phi_{m}^{(4)},
&& 1 \leq k \leq 3\,.
\end{aligned}
\end{equation} 
 \item Setting $\displaystyle \frac{\partial \mathcal{W}}{\partial Z^{(k\ell)}}=0$ gives,
\begin{equation}\label{fterm2}
\begin{aligned}
& \sum_{m=1}^{3}\varepsilon^{k\ell m}\left(Z^{(\ell k)} \Phi_m^{(k)} - \Phi_m^{(\ell)} Z^{(\ell k)}\right)
+  \sum_{\substack{m=1\\ m\neq k,\ell}}^{4} Z^{(\ell m)} Z^{(m k)} (-1)^{\delta_{k1}\delta_{\ell 3}\delta_{m4}} = 0\,,
&& 1\leq k,\ell\leq 3, \\
& \left(\Phi_k^{(k)} Z^{(k4)} - Z^{(k4)} \Phi_k^{(4)}\right)
+  \sum_{\substack{\ell=1\\ \ell\neq k}}^{3} Z^{(k\ell)} Z^{(\ell 4)} (-1)^{\delta_{k1}\delta_{\ell3}} = 0,
&& 1 \leq k \leq 3\,,\\
& \left(Z^{(4k)} \Phi_k^{(k)} - \Phi_k^{(4)} Z^{(4k)}\right)
+  \sum_{\substack{m=1\\ m\neq k}}^{3} Z^{(4m)} Z^{(m k)} (-1)^{\delta_{m1}\delta_{k3}} = 0,
&& 1 \leq k \leq 3\,.
\end{aligned}
\end{equation}
\end{itemize}
The moduli space of supersymmetric vacua is obtained by solving Eqs.~ \eqref{fterm1} and \eqref{fterm2} modulo the residual complexified gauge symmetry \footnote{More precisely, the moduli space of supersymmetric vacua corresponds to the symplectic quotient $\mu^{-1}(0)/G$, where $\mu$ is the moment map associated with the D--term constraints. By the Kempf--Ness theorem \cite{ness}, this is diffeomorphic to the Geometric Invariant Theory (GIT) quotient of the F--term solution space by the complexified gauge group, denoted as $\mathcal{M}_{\text{vac}} \cong \{ d\mathcal{W}=0 \} // G_{\mathbb{C}}\,$.}. 

Further simplification comes from the 28 global shift symmetries (Goldstones), which are the superpartners of the 28 fermionic broken SUSY Goldstino modes,
\begin{equation} \label{eflatgen}
\begin{aligned}
& \Phi^{(k)}_m \rightarrow \Phi^{(k)}_m + \xi_m\, I_{N_k},
\qquad &&1\le k\le 3,\; k\neq m,\; 1\le m\le 3, \\
& \Phi^{(k)}_k \rightarrow \Phi^{(k)}_k + \zeta_k\, I_{N_k}, 
\qquad \Phi^{(4)}_k \rightarrow \Phi^{(4)}_k + \zeta_k\, I_{N_4},
\qquad &&1\le k\le 3, \\
& X^{(k)}_i \rightarrow X^{(k)}_i + a_i\, I_{N_k},
\qquad &&1\le i \le 3,
\end{aligned}
\end{equation}
where $\xi_m$ and $\zeta_k$ are complex parameters, and $a_i$ real parameters corresponding to global translations in the non--compact directions. In the computation of the helicity supertrace $B_{14}\,$ we are interested in counting the bound states and these center--of--mass hypermultiplets represent flat directions and hence are quotiented out. Accordingly, throughout this work (for $N_1=N_2=N_3=1, N_4=N$) we impose the gauge choice
\begin{equation}
\label{eq:phase}
\Phi^{(1)}_{1} =0\,, \quad \Phi^{(1)}_{2} =0\,, \quad \Phi^{(1)}_{3} =0\, , \quad
\Phi^{(2)}_{1} =0\,, \quad \Phi^{(2)}_{2} = 0 \quad \text{and} \quad  
\Phi^{(3)}_{3} = 0\,,
\end{equation}
which fixes the flat directions without eliminating any physical solutions. For generic background moduli, the F--term equations yield a finite set of isolated solutions that form the basis for microscopic counting of BPS states.

\subsection{Counting Microstates as Algebraic Varieties}

The problem of determining the supersymmetric ground states of the D2--D2--D2--D6 system can be naturally formulated in the language of algebraic geometry. In general, the solution space of a system of polynomial equations over $\mathbb{C}$ defines an affine algebraic variety. Given the polynomial ring
\begin{equation}
\mathbb{C}[x_1,\ldots,x_n]\,,
\end{equation}
an affine variety is the common zero locus of a finite set of polynomials $f_1,\ldots,f_s$, namely
\begin{equation}
\mathcal{V}(f_1,\ldots,f_s) = \{(a_1,\ldots,a_n)\in\mathbb{C}^n \mid f_i(a_1,\ldots,a_n)=0,\; \forall i\}\, ,
\end{equation}
which is equivalent to the Jacobi (chiral) ring. The geometry of such a variety -- including its dimension and degree -- encodes important structural information about the solution space.

In the present context, the F--term constraints \eqref{fterm1}–\eqref{fterm2} form a system of polynomial equations in the complex scalar components of the fields $Z^{(k\ell)}$ and $\Phi^{(k)}_i\,$. Therefore, the supersymmetric vacua naturally assemble into an affine algebraic variety over $\mathbb{C}$, obtained after incorporating the action of the complexified gauge symmetry. Concretely, the vacuum space is given by
\begin{equation}
\mathcal{M}_{\mathrm{vac}} =
\mathcal{V}\!\left(
\frac{\partial \mathcal{W}}{\partial \Phi^{(k)}_i}\,,
\frac{\partial \mathcal{W}}{\partial Z^{(k\ell)}}
\right)
\subset \mathbb{C}^{n}\,,
\end{equation}
where $n$ denotes the number of complex scalar variables prior to the gauge identifications. For generic values of the background moduli, $\mathcal{M}_{\mathrm{vac}}$ is found to be zero--dimensional, meaning that the variety consists of a finite set of isolated points. Then, the number of supersymmetric ground states is equal to the degree of the variety. This quantity may be determined using a variety of exact and very robust numerical methods from computational algebraic geometry, such as Newton polytopes, homotopy continuation, monodromy and Hilbert series constructions, as discussed in some detail in \cite{Chowdhury:2023wss}. In this paper, we mainly focus on a detailed exposition of the monodromy method, a very efficient and robust numerical implementation to study the vacuum structure of the pure D--brane systems and a plethora of other similar problems \cite{duffmonodromy,Duff2016SolvingPS,Bliss2018MonodromySS}. We shall also discuss applications of analytical Gröbner bases techniques while discussing the vacua structure of the 4--charge non--BPS extremal versions of these black holes. The low energy spectra of the non--BPS cases present interesting challenges as we shall see in Section \ref{sec:nonbps}.

For the charge vectors $(N_1,N_2,N_3,N_4)$ relevant to the microscopic counting of $\frac{1}{8}$--BPS black holes, the resulting varieties indeed consist of isolated points, whose cardinality matches the $B_{14}$ helicity supertrace expected from the U--dual picture \cite{Shih:2005qf}. The explicit numbers of solutions for several representative charge configurations are shown in Table \ref{B14}, reproducing the results of \cite{Chowdhury:2014yca,Chowdhury:2015gbk,Chowdhury:2023wss}, and the last two are new results relevant to this paper. The relation between the moduli parameters $c^{(k)}$ and $c^{(k\ell)}$ and the metric and B--field moduli is given in Appendix~A of \cite{Chowdhury:2014yca}. In particular, Eq. (A.7) shows that the relevant mass combinations depend only on $|c^{(k\ell)}|^2$, while Eq. (A.9) gives one explicit inversion of the moduli map. Thus, one may choose real $c^{(k\ell)}$ as a valid specialization of the same background data, although in general the $c^{(k\ell)}$ are complex. We have checked that the solution counts in Table \ref{B14} remain unchanged for a large number of random choices of the $c^{(k\ell)}$, both real and complex, and of the real $c^{(k)}$, as expected since the count computes an index and is therefore invariant under continuous moduli deformations.

\begin{table}[h!]
    \centering
    \begin{tabular}{|c|c|}
    \hline
    \textbf{Charges} & \textbf{Number of Solutions} \\
    \hline
    (1,1,1,1) & 12 \\
    \hline
    (1,1,1,2) & 56 \\
    \hline
    (1,1,1,3) & 208 \\
    \hline
    (1,1,1,4) & 684 \\
    \hline
    \textbf{(1,1,1,5)} & \textbf{2032} \\
    \hline
    \textbf{(1,1,1,6)} & \textbf{5616} \\
    \hline
    \end{tabular}
    \caption{Number of isolated solutions to the F--term equations for several 4--charge configurations of the D2--D2--D2--D6 $\frac{1}{8}$--BPS system.}
    \label{B14}
\end{table}

\section{The Monodromy Method}
\label{sec:monodromy}

We employ the monodromy method to determine the complete set of isolated solutions for the polynomial systems arising in our F--term analysis \cite{Chowdhury:2023wss}. This approach exploits the global topological structure of the solution space. Rather than solving a static system of equations directly -- which is often computationally prohibitive -- we embed the specific system into a parametrized family and utilize the geometric properties of the solution variety's projection to the parameter space. Conceptually, this is analogous to studying the sheet structure of a Riemann surface by analytically continuing a function around its branch points or the monodromy we encounter in \textit{Hyper--Geometric/Fuchsian theory} of linear differential equations \cite{Kimura1971,Ishkhanyan2019GeneralizedHypergeometricSO}. In our context, the ``sheets'' correspond to distinct solution branches, and ``branch points'' correspond to singular parameter configurations where solutions merge or diverge. In our previous work \cite{Chowdhury:2023wss}, we utilized this method primarily as a verification tool; however, given its high computational efficiency and algorithmic scalability, we present it here in much more detail as the primary framework for solving for the SUSY vacua. Below, we formalize the algebraic geometric foundations of the method and outline the algorithm.

\subsection{Geometric Framework}

Consider a system of polynomial equations $F(x) = 0$ where $x \in \C^s$. We deform this specific system into a generic family dependent on a parameter set $p \in \Pspace \cong \C^k\,$. This defines a map,
\begin{equation}
    F: \C^{s} \times \Pspace \longrightarrow \C^{s}\,,
\end{equation}
where we seek the zero locus,
\begin{equation}
    \V = \left\{ (x,p) \in \C^s \times \Pspace \;\middle|\; F(x; p) = 0 \right\}\,.
\end{equation}
The total space $\V$ is an affine variety of complex dimension $k\,$. We consider the natural projection onto the parameter space, $\pi: \V \longrightarrow \Pspace\,$. The critical geometric object governing the method is the \textit{Discriminant Locus}, $\Delta \subset \Pspace\,$. This locus consists of parameter values where the solutions fail to be isolated. Algebraically, $\Delta$ is the variety defined by the vanishing of the Jacobian determinant with respect to the variables $x\,$,
\begin{equation}
    \Delta = \left\{ p \in \Pspace \;\middle|\; \exists \;x \in \pi^{-1}(p) \text{ s.t. } \det \left( \frac{\partial F}{\partial x} \right) = 0 \right\}\,.
\end{equation}
We define the regular parameter space as the complement, $U = \Pspace \setminus \Delta\,$. Over this open set, the map $\pi|_U: \pi^{-1}(U) \to U$ acts as a covering space (specifically, a finite unbranched covering). For a generic parameter $p \in U$, the fiber $\pi^{-1}(p)$ consists of a fixed number of distinct, isolated solutions, $d\,$, representing the degree of the covering.

\subsection{Monodromy Action and Algorithm}

The method relies on the action of the fundamental group $\pi_1(U)$ on the fiber. Consider a base point $p^* \in U$ and a closed loop $\gamma \subset U$ starting and ending at $p^*$. By the path lifting property of covering spaces, the loop $\gamma$ lifts to $d$ unique paths in the total space $\V\,$. Since $\gamma$ is closed in the base space $U$, the endpoints of the lifted paths must lie within the fiber $\pi^{-1}(p^*)\,$. However, the lift need not be a closed loop in $\V$; a path starting at solution $x^{(i)}$ may end at a different solution $x^{(j)}\,$. Thus, analytic continuation along $\gamma$ induces a permutation $\sigma_\gamma$ of the solution set $\{x^{(1)}, \dots, x^{(d)}\}\,$. The homomorphism
\begin{equation}
    \rho: \pi_1(U, p^*) \longrightarrow S_d\,,
\end{equation}
defines the \emph{monodromy representation}. If the solution variety $\V$ is irreducible over $\Pspace$ (which is satisfied for generic deformations of physical systems), the image of $\rho$ is a transitive subgroup of the symmetric group $S_d\,$. This implies that any solution can be reached from any other solution by traversing an appropriate sequence of loops in the parameter space. 

\subsubsection*{Algorithmic Implementation}

Based on this framework, we utilize the following numerical algorithm to generate the full solution set:

\begin{enumerate}
    \item \textbf{Seed Generation:} We obtain a single starting solution--parameter pair $(x_{\text{seed}}, p^*)\,$. This is often achieved by choosing a simplified ``start system'' where a trivial solution is known and tracking it to a generic point $p^*$.
    
    \item \textbf{Loop Propagation:} We generate random loops $\gamma \in \pi_1(U, p^*)$ in the complex parameter space. A standard choice is the ``triangle loop'' composed of linear segments between random intermediate complex parameters $p_1, p_2\,$:
    \begin{equation}
    p^* \to p_1 \to p_2 \to p^*\, .
    \end{equation}
    
    \item \textbf{Path Tracking:} We perform numerical homotopy continuation along these loops. By solving the Davidenko differential equation \cite{davidenko1953new},
    \begin{equation}
        \frac{dx}{d\tau} = - \left(\frac{\partial F}{\partial x}\right)^{-1} \frac{\partial F}{\partial p} \frac{dp}{d\tau} \,,
    \end{equation}
    we track the evolution of the seed solution through the variety $\V$.
    
    \item \textbf{Fiber Filling:} Upon completing a loop, we check if the endpoint is a new solution. If so, it is added to the set $\s_{\text{found}}\,$. We then repeat the process using the newly found solutions as seeds for subsequent loops.

    \item \textbf{Verification (Linear Trace Test):}
To rigorously confirm that the recovered set of solutions $\mathcal{S}_{\text{found}}$ is complete without prior knowledge of the degree $d$, we employ the \emph{linear trace test} \cite{Leykin2016TraceT}. This test exploits the fact that the symmetric functions of the roots of a polynomial system are well--behaved, even when individual roots exhibit monodromy. Specifically, we track the \textit{centroid} of the solution set as the parameters move along a linear segment $p(t) = a + t(b-a)$ in the complex parameter space. For the \emph{complete} solution set, the trace (sum) of the coordinates, $X(t) = \sum_{x \in \mathcal{S}_{\text{total}}} x(t)\,$, is a single--valued function of $t\,$. This is a multivariate generalization of \textit{Vieta's formulas} \cite{LangAlgebra} -- the sum of the roots corresponds to a coefficient in the elimination polynomial of the system, which is a rational function of the parameters. Consequently, if the computed trace $X_{\text{found}}(t)$ matches a linear (or rational) function to high numerical precision, the set is deemed complete. Conversely, if $X_{\text{found}}(t)$ exhibits branching or non--trivial monodromy, it indicates that solutions are missing, necessitating further loops to explore other sheets of the covering map.    
\end{enumerate}

\subsection*{Remarks}

\paragraph{Relation to Galois Theory.}
The monodromy group $G \subseteq S_d$ computed by our algorithm admits a rigorous algebraic interpretation -- it is isomorphic to the \emph{geometric Galois group} of the solution variety \cite{Sottile2021GaloisGI}.
Consider the function field $K = \mathbb{C}(p)$ of rational functions in the parameters. The solutions $x$ generate a finite field extension $L = K(x)\,$. The Galois group $\text{Gal}(L/K)$ describes the symmetries of the roots over the parameters.
Harris's \emph{Uniform Position Principle}  \cite{Rathmann1986} asserts that for a generic polynomial system, the monodromy group is the full symmetric group $S_d\,$.
However, if the physical system possesses discrete symmetries (e.g., $\mathbb{Z}_2$ or $S_n$ symmetries), the monodromy group will be a proper subgroup of $S_d\,$.
Crucially, checking whether the monodromy group acts \emph{transitively} on the fiber is the standard numerical test for the \emph{irreducibility} of the algebraic variety.

\paragraph{Symmetry and Topological Obstructions.}
While the Uniform Position Principle guarantees transitive monodromy for generic coefficients, physical potentials are rarely generic; they possess discrete symmetries and structural constraints that can render the vacuum manifold reducible or imprimitive. These features are endemic to string compactifications, particularly in \textit{orbifold constructions} or theories with \textit{discrete gauge symmetries} (e.g., $\mathbb{Z}_k$), where the moduli space metric or superpotential may exhibit singular loci corresponding to twisted sectors or symmetry enhancement points. In such scenarios, the standard monodromy algorithm may fail by trapping the solver in a single disconnected sector or by missing internal phase factors of the solution fiber. To systematically classify these failures, we analyze a ``Composite Vacuum" toy model in Appendix \ref{app:master_model}. This model serves as a testbed to capture the distinct topological obstructions -- specifically vacuum disconnectedness and phase--locking -- within a single analytic framework.

\paragraph{Extension to Non--Polynomial Systems.}
While the monodromy method is grounded in algebraic geometry, its engine -- numerical analytic continuation -- requires only that the defining functions $F(x;p)$ be holomorphic (complex analytic) almost everywhere.
Consequently, the method generalizes to systems involving transcendental functions (e.g., exponential or trigonometric terms often arising in instanton corrections), provided two caveats are addressed:
\begin{enumerate}
    \item \textbf{Singularities:} Unlike polynomials, which branch only at algebraic discriminant loci, transcendental functions may possess essential singularities. The path $\gamma$ must be chosen to avoid these regions.
    \item \textbf{Finiteness:} Polynomial systems guarantee a finite fiber degree $d$ (by Bezout's theorem \cite{Kushnirenko1976NewtonPA}). Transcendental systems often admit fibers of infinite cardinality (e.g., the logarithmic branches of $e^x=p$). In such cases, the monodromy algorithm does not terminate naturally. To render the problem well--posed, one must restrict the search to a bounded domain $\mathcal{D} \subset \mathbb{C}^s$ (e.g., $|x_i| < R$) or seek only solutions on a specific physical branch.
\end{enumerate}

\subsection{Implementation in the Pure D--brane System}

Since the D--brane systems involve a large number of variables and direct visualization is impossible, we demonstrate the mechanics of the monodromy method through tractable low--dimensional examples collected in Appendix \ref{app: monoexamples}. These illustrate the interplay between the parameter space topology and the solution permutations. We now explain how the monodromy framework can be used to solve the F--term equations \eqref{fterm1}--\eqref{fterm2}.

\medskip

\noindent\textbf{Step 1: Reduction to a square system.}  
The F--term equations do not initially form a square system. We fix the gauge and shift symmetries, eliminate dependent fields, and discard the corresponding dependent equations. Several remaining equations become linear and are solved explicitly. This yields a reduced square system, $F(x)=0$ in a set of independent complex variables $x\,$.

\medskip

\noindent\textbf{Step 2: Deformation and monodromy.}  
We introduce a linear deformation
\begin{equation}
F(x,p)=
\begin{pmatrix}
f_{1}(x_{1},\ldots,x_{s})-p_{1}\\
\vdots \\
f_{s}(x_{1},\ldots,x_{s})-p_{s}
\end{pmatrix}
=0\,.
\end{equation}
For a random point $x^{*}\,$, a generic $p^{*}$ trivially satisfies $F(x^{*},p^{*})=0\,$. Starting from this seed, we generate random loops in parameter space, track the lifted solution branches, and enlarge the accumulated set until all solutions at $p^{*}$ have been obtained. A final homotopy from $p^{*}$ to $p=0$ then yields the complete set of supersymmetric vacua of the physical system $F(x)=0\,$.

\medskip

The monodromy method is especially helpful in our setting because it allows us to avoid solving the entire system of equations in a single step. Instead, we start from a single solution at a generic point in parameter space and follow it around closed loops; the analytic continuation along these loops automatically reveals the other branches of the solution set. In practice, this bypasses the resource--heavy Gröbner basis machinery that usually becomes a bottleneck for large systems, and it tends to work particularly well for the kind of sparse, structured equations that arise in our D--brane constructions. Another advantage is that the method effectively tells us when we are done -- the solution set stops growing once all branches have been generated and can be confirmed by carrying out the \textit{Linear Trace Tests}. As the main steps -- choosing loops, tracking solutions, and finally homotopy continuing back to the physical point -- can all be carried out in parallel, the overall computation is quite efficient. For these reasons, monodromy is one of the few practical tools capable of extracting the full SUSY vacuum structure of the D--brane system.

\section{SUSY Vacua for BPS Black Holes of Higher Charge Configurations} \label{sec:bpshighercharge}

In this section, we extend the analysis of supersymmetric vacua in the pure D--brane framework to higher--charge BPS configurations. While low--charge systems can be analyzed using relatively direct algebraic methods, increasing the rank of the non--abelian brane stack introduces new technical challenges, most notably in fixing the complexified gauge symmetry in a manner compatible with the shift symmetries of the adjoint fields. The goal of this section is to develop an efficient and consistent gauge--fixing strategy for the $(1,1,1,N_{4})$ systems and to use it to determine the complete set of isolated supersymmetric vacua for higher values of $N_{4}$ using the monodromy method.

\subsection{Efficient Gauge Fixing for the $(1,1,1,N_{4})$ System}

For the charge configuration $(N_{1},N_{2},N_{3},N_{4})=(1,1,1,N_{4})$, the superpotential is invariant under a complexified gauge symmetry,
\begin{equation}
\frac{\mathbb{C}^{*}\times \mathbb{C}^{*} \times \mathbb{C}^{*}\times \mathrm{GL}(N_{4},\mathbb{C})}{\mathbb{C}^{*}}\,,
\end{equation}
which acts on the fields as
\begin{align}
Z^{(k\ell)} &\;\rightarrow\; a_{k} a_{\ell}^{-1} Z^{(k\ell)}, &
Z^{(4k)} &\;\rightarrow\; a_{k}^{-1} M Z^{(4k)}, &
Z^{(k4)} &\;\rightarrow\; a_{k}\, Z^{(k4)} M^{-1}, \nonumber\\[4pt]
\Phi^{(k)}_{i} &\;\rightarrow\; \Phi^{(k)}_{i}, &
\Phi^{(4)}_{i} &\;\rightarrow\; M\,\Phi^{(4)}_{i} M^{-1}, &
&\quad (1\leq k\leq 3, k \neq \ell,\;1\leq i\leq 3)\,,
\end{align}
where $a_{1},a_{2},a_{3}\in \mathbb{C}^{*}$ and $M\in\mathrm{GL}(N_{4},\mathbb{C})$.  
Solutions to the F--term equations appear in orbits of this symmetry group, so fixing the relative gauge is a necessary first step.

\subsubsection*{Fixing the relative \texorpdfstring{$\mathbb{C}^{*}\times \mathbb{C}^{*}$}{C* x C*} gauge}

The two Abelian gauge factors act only on the bifundamentals $Z^{(k\ell)}$ with $k,\ell\in\{1,2,3\}$.  
A convenient choice is to fix them by setting
\begin{equation}
Z^{(12)}=1, \qquad Z^{(23)}=1\,.
\end{equation}
The remaining gauge freedom lies entirely in $\mathrm{GL}(N_{4},\mathbb{C})$ acting on the fields attached to the D6--brane stack \footnote{Note that as long as the moduli parameters $c$ are non--zero, at solutions to the F--term equations, all components of any bifundamental $Z$ can't be all zero. Hence, these gauge choices are always valid. However, this doesn't hold true for the non--BPS D--brane systems and warrants extra measures (see Section \ref{sec: landscape} and Appendix \ref{app:gauge_consistency}.}.

\subsubsection*{Fixing the \texorpdfstring{$\mathrm{GL}(N_{4},\mathbb{C})$}{GL(N4,C)} gauge}

The bifundamental fields $Z^{(k4)}$ ($k=1,2,3$) transform as row vectors under $M^{-1}$, while $Z^{(4k)}$ transform as column vectors under $M$.  
This makes them ideal for fixing most of the $\mathrm{GL}(N_{4},\mathbb{C})$ freedom. A natural and efficient choice is
\begin{equation}
Z^{(14)} = (1\; 0\; 0\; \cdots),\qquad
Z^{(24)} = (0\; 1\; 0\; \cdots),\qquad
Z^{(34)} = (0\; 0\; 1\; \cdots)\,.
\end{equation}
This fixes the first three rows of $M^{-1}$ and leaves only the remaining $(N_{4}-3)$ directions unfixed. The final part of the gauge symmetry can be removed using the adjoint fields $\Phi^{(4)}_{i}$, but we must take special care with regards to the gauge fixing of the shift symmetries,
\begin{equation}\label{eq:shift-sym}
\Phi^{(1)}_{1}=\Phi^{(1)}_{2}=\Phi^{(1)}_{3}
=\Phi^{(2)}_{1}=\Phi^{(2)}_{2}
=\Phi^{(3)}_{3}=0\, ,
\end{equation}
as these choices constrain particular rows of $\Phi^{(4)}_{1},\Phi^{(4)}_{2},\Phi^{(4)}_{3}$ via the F--term equations, and using those rows for gauge fixing leads to inconsistencies.

This issue does not appear in smaller systems up to charges $(1,1,1,3)$ but becomes unavoidable in higher--charge cases such as the $(1,1,1,5)$ and $(1,1,1,6)$ configurations. For example, in the $(1,1,1,4)$ system, once $Z^{(14)}, Z^{(24)}$ and $Z^{(34)}$ are fixed, the F--term equations impose the following constraints on the rows of the adjoint fields attached to the D6--branes,
\begin{align}
&\Phi^{(4)}_{1,11}=0\,,\qquad 
1-\Phi^{(4)}_{1,12}=0\,,\qquad 
\Phi^{(4)}_{1,13}=Z^{(13)}\,,\qquad 
\Phi^{(4)}_{1,14}=0\,,
\\[4pt]
&\Phi^{(4)}_{2,21}=Z^{(21)}\,,\qquad
\Phi^{(4)}_{2,22}=0\,,\qquad
1-\Phi^{(4)}_{2,23}=0\,,\qquad
\Phi^{(4)}_{2,24}=0\,,
\\[4pt]
&\Phi^{(4)}_{3,31}=Z^{(31)}\,,\qquad
\Phi^{(4)}_{3,32}=Z^{(32)}\,,\qquad
\Phi^{(4)}_{3,33}=0\,,\qquad
\Phi^{(4)}_{3,34}=0\,.
\end{align}
Each of these rows is therefore already fixed by the F--term and the shift symmetry constraints, and none of them can be freely used to fix the remaining $\mathrm{GL}(N_{4},\mathbb{C})$ gauge freedom. To complete the gauge fixing correctly, one should instead use the \emph{remaining} rows of $\Phi^{(4)}_{i}$ and set them to simple unit vectors such as $
(0\;\;0\;\;0\;\;1\;\;0\;\cdots)\,$
avoiding the rows entangling with the shift symmetries \footnote{In fact, in our earlier work \cite{Chowdhury:2023wss} we were lucky -- without realizing the subtlety, we happened to fix the gauge using the \emph{third} row of $\Phi^{(4)}_{1}$, which is a safe choice. Had we instead fixed the \emph{first} row of $\Phi^{(4)}_{1}$ -- which is already constrained by the shift symmetries -- we would have run into inconsistency described above. In hindsight, this accidental choice ensured that the gauge was fixed properly, and it helped us identify the correct and efficient gauge fixing strategy for the larger $(1,1,1,N_{4})$ systems!}.

\subsection{Supersymmetric Vacua of the $(1,1,1,5)$ System}

In earlier works \cite{Chowdhury:2014yca,Chowdhury:2015gbk,Chowdhury:2023wss}, the $14^{\text{th}}$ helicity trace index $B_{14}$ was computed for the $(1,1,1,1)$, $(1,1,1,2)$, $(1,1,1,3)$, and $(1,1,1,4)$ systems using a combination of gauge--invariant analysis and computational algebraic geometry techniques like Hilbert series. In the present paper, we extend this program to higher--charge configurations, focusing on the $(1,1,1,5)$ and $(1,1,1,6)$ systems. Remarkably, in both cases we recover the exact microscopic degeneracies predicted by the U--dual picture \cite{Shih:2005qf}. To determine the supersymmetric vacua, we solve the F--term equations \eqref{fterm1}--\eqref{fterm2} for the $(1,1,1,5)$ system. The main technical challenge is to fix the complexified gauge symmetry in a way compatible with the shift symmetries of the adjoint fields. Once this is done, the resulting square polynomial system can be solved efficiently using the monodromy method (see Section~\ref{sec:monodromy}).

\subsection*{Gauge Fixing}

The relative gauge symmetry for the $(1,1,1,5)$ configuration is the complexified version of $
U(1)\times U(1)\times U(5) $ and the two $U(1)$ factors can be fixed by setting
\[
Z^{(12)}=1, \qquad Z^{(23)}=1\,.
\]
To fix the $\mathrm{GL}(5,\mathbb{C})$ part, we impose the following conditions on the bifundamental fields,
\[
Z^{(14)}=\begin{pmatrix}1 & 0 & 0 & 0 & 0\end{pmatrix}\,, \qquad
Z^{(24)}=\begin{pmatrix}0 & 1 & 0 & 0 & 0\end{pmatrix}\,, \qquad
Z^{(34)}=\begin{pmatrix}0 & 0 & 1 & 0 & 0\end{pmatrix}\,, 
\]
and the remaining gauge freedom is fixed using the adjoint field $\Phi^{(4)}_{3}$. A convenient and shift symmetry compatible choice is,
\[
\left(\Phi^{(4)}_{3}\right)_{1j} = (0,0,0,1,0)\,, \qquad
\left(\Phi^{(4)}_{3}\right)_{2j} = (0,0,0,0,1)\,,
\]
where each row is written as a $1\times5$ vector. These two rows fix the nontrivial part of the $\mathrm{GL}(5,\mathbb{C})$ action without conflicting with the shift symmetries imposed on the adjoint multiplets. The shift symmetry is fixed by
\begin{equation}
\Phi^{(1)}_{1}=0\,,\qquad \Phi^{(1)}_{2}=0\, ,\qquad \Phi^{(1)}_{3}=0\,,\qquad
\Phi^{(2)}_{1}=0\,,\qquad \Phi^{(2)}_{2}=0\,,\qquad \Phi^{(3)}_{3}=0\,.
\label{eq:phase-1115}
\end{equation}
This choice ensures that the remaining F--term equations do not introduce any residual gauge redundancy.

\subsection*{Result}

With this gauge choice, the F--term equations form a square system of polynomial equations (63 equations and 63 variables).  
We solve this system using the monodromy method implemented in HomotopyContinuation.jl with our own modifications \cite{Breiding2017HomotopyContinuationjlA}.  On a Dell workstation equipped with two Intel Xeon Gold 6258R processors (56 cores) and 128 GB RAM, the computation required roughly five days. After running the monodromy procedure with several random seeds and taking the union of the resulting solution sets,  
we obtain \footnote{Running the monodromy method a single time may miss some branches; taking the union from multiple random seeds improves coverage.}
\[
\#~\text{vacua for }(1,1,1,5) \text{ system } = 2032\,,
\]
which exactly matches the U--dual prediction \cite{Shih:2005qf}.  
This provides a strong consistency check on both our gauge--fixing strategy and the monodromy framework.

\subsection{Supersymmetric Vacua of the $(1,1,1,6)$ System}

We now turn to the $(1,1,1,6)$ configuration, where the non--abelian stack has rank $6\,$.  
The same strategy applies; we fix the complexified gauge symmetry, impose compatible shift symmetries, and solve the reduced square system using monodromy.

\subsection*{Gauge Fixing}

The relative gauge group is $
U(1)\times U(1)\times U(6)\,$, 
and the abelian factors are fixed as before, $
Z^{(12)}=1$ and $Z^{(23)}=1\,$. The $\mathrm{GL}(6,\mathbb{C})$ part is fixed by choosing
\[
Z^{(14)}=(1,0,0,0,0,0)\,,\quad
Z^{(24)}=(0,1,0,0,0,0)\,,\quad
Z^{(34)}=(0,0,1,0,0,0)\,,
\]
and to eliminate the remaining $\mathrm{GL}(6,\mathbb{C})$ freedom, we fix three independent rows of $\Phi^{(4)}_{3}$ as,
\begin{align*}
\left(\Phi^{(4)}_{3}\right)_{1j} &= (0,0,0,1,0,0)\,, \\
\left(\Phi^{(4)}_{3}\right)_{2j} &= (0,0,0,0,1,0)\,,\\
\left(\Phi^{(4)}_{3}\right)_{4j} &= (0,0,0,0,0,1)\,.
\end{align*}
These choices are deliberately made so that they do not interfere with any of the adjoint components constrained by the F--term or shift symmetry equations. The shift symmetry conditions remain identical to \eqref{eq:phase-1115}.

\subsection*{Result}

After fixing the gauge and shift symmetries, we obtain a square polynomial system (83 equations and 83 variables).   
Due to the increased complexity, the monodromy computation takes substantially longer (about 20 days on the same machine). The final count of isolated supersymmetric vacua is
\[
\#~\text{vacua for }1116\text{ system } = 5616\,,
\]
once again in perfect agreement with the U--dual prediction \cite{Shih:2005qf}.  
This confirms that the monodromy method continues to perform reliably even for significantly larger non--abelian charge sectors. A similar analysis may be carried out for $(1,1,1,N_{4})$ systems with $N_{4}>6\,$.  While the computational cost increases with $N_{4}$, the monodromy approach remains one of the only practical tools capable of resolving the complete set of supersymmetric vacua in these higher--charge configurations \footnote{The computational complexity of the monodromy algorithm is $\mathcal{O}(d_{\text{local}} \cdot K \cdot N^3)$, where $d_{\text{local}}$ is the number of solutions in the connected component of the seed, $K$ is the number of continuation steps per path, and $N$ is the number of variables. The $N^3$ factor arises from the dense linear algebra required for Newton corrections (Jacobian inversion). While the theoretical upper bound on the number of solutions grows exponentially with $N$ (Bezout's bound $D^N$ \cite{Kushnirenko1976NewtonPA}), the monodromy method is distinctively \textit{output--sensitive}. Unlike ``total homotopy" methods (e.g., polyhedral homotopy \cite{HuberSturmfels1995}) which must track all potential paths -- often numbering in the millions for high--dimensional physical systems -- monodromy resources are expended only on the subset of solutions geometrically connected to the initial physical seed. In modern implementations such as \text{HomotopyContinuation.jl}, this efficiency is further augmented by compiled straight--line programs (SLP) for rapid function evaluation and parallel path tracking.}.

We note that it is particularly important for us to be able to do the $(1,1,1,6)$ case. In \cite{Chowdhury:2023wss}, we raised a puzzle regarding the mismatch of the SUSY vacua count for  $(1,1,2,3)$ charge configuration computed in this pure D--brane setup and its U--dual description as extracted from the Jacobi form, $\vartheta_1(z \mid \tau)^2 \eta(\tau)^{-6}\,$. The charge configurations, $(1,1,1,6)$ and $(1,1,2,3)$ are related to each other by S--duality and hence should have the same count for the $B_{14}$ Helicity Trace index \cite{Sen:2009gy,Shih:2005qf,Pioline:2005vi}. The matching of the SUSY vacua count for the $(1,1,1,6)$ configuration with the U--dual prediction, suggests that nothing was probably missed in the original D1--D5--P--KK monopole index computation and the discrepancy surrounding the counting for the $(1,1,2,3)$ charge configuration has to be resolved on the pure D--brane side \footnote{It is a possibility that our D--brane construction is incomplete and two non--abelian groups in two different brane stacks either force some otherwise massive fields from the Kaluza--Klein reduction to become massless and contribute to the low energy dynamics or there are extra interactions in the $\mathcal{N}=1$ superpotentials that we missed. Neither seems plausible to us. For the lack of any other viable technique to compute the Witten Index of SUSY vacua, we rely solely on the Monodromy technique. There are ways in which the Monodromy can fail as discussed in Appendix \ref{app:master_model}.}.  This is a work in progress.

\section{Vacua for non--BPS Black Hole}\label{sec:nonbps}

In this section, we consider 4--charge non--BPS configurations in the pure D--brane description. Supersymmetry is very constraining and controls the vacuum structure of BPS bound states. It is therefore natural to ask how the microscopic picture changes once this protection is removed. The non--BPS systems studied here are obtained by a minimal modification of the corresponding BPS setup, which preserves the bosonic intersection structure and charge assignments while breaking all supercharges by R--symmetry rotations \cite{Mondal:2024qyn}. We describe the resulting low--energy theory and study the structure of its low--lying local minima. In general, we find 6 doubly degenerate low lying states for the bound states of the D--brane system and a non--compact Coulomb branch for the unbounded states at parametrically higher energy. There are also several marginally bound stabilizer submanifolds associated with groups of D--brane stacks becoming unbounded, and they warrant careful analysis (see Appendix \ref{app:gauge_consistency}). We developed techniques to lift these flat directions in the extremum point analysis, in particular, the use of \textit{Morse--Bott theory} of structural stability to facilitate the use of existing second--order extremization algorithms. We have also observed \textit{stability jumps} in the isolated vacuum landscape for some sets of random moduli parameter values. This is expected in complex systems, and possible mechanisms are outlined in Appendix \ref{app:Stability}.  

\subsection{The Pure D--brane Picture}

The 4--charge non--BPS configuration consists of three D2--branes wrapped on the orthogonal two--cycles of the \(T^6\) and a single anti--D6--brane wrapping the whole \(T^6\) \cite{Mondal:2024qyn}, i.e.,
\[
\text{D2}_1:(x^4,x^5)\,,\qquad \text{D2}_2:(x^6,x^7)\,,\qquad \text{D2}_3:(x^8,x^9)\,,\qquad \overline{\text{D6}}:(x^4,\dots,x^9)\,.
\]
Flipping the orientation of the D6 stack distinguishes this non--BPS system from the \(\frac{1}{8}\)--BPS configuration, but the bosonic intersection pattern remains the same. Consequently, the number of bosonic Goldstone modes is unchanged between the two systems, while supersymmetry is completely broken in the non--BPS case (there are 32 Goldstinos and 28 Goldstones). In the BPS case (see Section \ref{sec:DbraneReview}), as 28 out of the 32 supercharges are broken, it is natural to formulate the low--energy theory in \(\mathcal{N}=1\) superfield language. Practically we view the system as a \(\mathcal{N}=1\) theory in \(3+1\) dimensions, dimensionally reduced to \(0+1\) (quantum mechanics). This description makes the Goldstone/Goldstino structure and the allowed superpotential couplings transparent. The $\mathcal{N}=1$ multiplets associated with the Lagrangian \eqref{lagrangian} are \footnote{In this subsection, we shall denote the superfields by capital letters and the fields by small letters. For most of the paper, we don't need this distinction, and we use the capital letter for both. Hopefully, it is clear which is which from the context.}
\begin{equation}
V^{(k)}=(A^{(k)}_{\mu},\lambda^{(k)})\,,\qquad
\Phi^{(k)}_{i}=(\phi^{(k)}_{i}\,,\psi^{(k)}_{i}),\qquad
Z^{(ij)}=(z^{(ij)}\,,\chi^{(ij)})\,.
\end{equation}
The same principle carries over for the construction of the non--BPS Lagrangian except now the four different triplets of stacks preserve \emph{different} \(\mathcal{N}=1\) subalgebras of the parent \(\mathcal{N}=4\) theory; for example the triplet \((123)\) preserves a different supersymmetry compared to \((124)\) or \((134)\) thereby ensuring that the combined system has no supersymmetry.

 A single D--brane stack preserves 16 supercharges, a pair of stacks preserves 8, and a triplet preserves 4 supercharges. The $\mathcal{N}= 4$ multiplet of a single stack of brane consists of one $\mathcal{N}= 1$ vector multiplet, $V = (A_\mu,\lambda)$ and three $\mathcal{N} = 1$ chiral multiplets $\Phi_i = (\phi_i,\psi_i), \;\; i = 1,2,3\,$ and under the $SU(4)_R$ R--symmetry, the four fermions and the six real scalars rotate. A pair of branes say, $(1,2)$ preserve $\mathcal{N}=2$ superalgebra where the $\mathcal{N}=1$ multiplets of either branes ($V = (A_\mu,\lambda)$ and $\Phi_3 = (\phi_3,\psi_3)$) corresponding to the common traverse direction $(8,9)$ combine to give a $\mathcal{N}=2$ vector multiplet and similarly remaining $\Phi$'s and the $Z$'s combine to form hypermultiplets. The $\mathcal{N}=1$ subalgebras of the $\mathcal{N}=2$ superalgebra admit a $SU(2)_R$ R--symmetry ($R_{12}$) and in particular involve an exchange,
\begin{equation} \label{eq: rsymm}
z^{(12)} \leftrightarrow-i\left(z^{(21)}\right)^{\dagger}\,, \quad \phi^{(k)}_1 \leftrightarrow-i\left(\phi^{(k)}_2 \right)^{\dagger}\,, \quad \phi^{(k)}_2\leftrightarrow-i\left(\phi^{(k)}_1 \right)^{\dagger}\,, \quad \psi_3^{(k)} \leftrightarrow \lambda^{(k)}, \quad k=1,2 \,.
\end{equation}
In order to enforce that each triplet of brane stacks preserves a different $\mathcal{N}=1$ supersymmetry, it is clear that we can very well take the same Lagrangian as in \eqref{lagrangian} for the BPS case and deploy $SU(2)_R$ R--symmetry exchange \eqref{eq: rsymm} consistently. It is also probably expected that the $\mathcal{N}=4$ and the $\mathcal{N}=2$ parts of the Lagrangian would be agnostic to any $SU(2)_R$ R--symmetry rotations and only the $\mathcal{N}=1$ part of the superpotential \eqref{eq: super} needs modifications. It is true that the $\mathcal{N}=4$ parts of the Lagrangian, in particular the kinetic terms, will remain as they are, but we shall see that due to a subtlety in the construction, the rest of the Lagrangian for the non--BPS case warrants some care.

\subsubsection{Superpotential Construction}

We want to preserve different supersymmetries in the D--brane triplets $(1,2,3)$ and $(1,2,4)\,$. As the pair label $(1,2)$ is common to both these triplets, we can use the $SU(2)_R$ R--symmetry exchange $R_{12}$ as stated in \eqref{eq: rsymm} to demand that the pairs $(1,4)$ and $(2,4)$ should involve the fields $R_{12}\left(V^{(k)}, \Phi_1^{(k)}, \Phi_2^{(k)}, \Phi_3^{(k)}\right)\,,\;\; k=1,2\,$. Particularly,  the Yukawa part of the superpotential \eqref{eq: yukawa} should feature $R_{12} \Phi_1^{(1)}\,$ instead of $\Phi_1^{(1)}\,$. Same argument for the triplets, $(1,2,3)$ and $(1,3,4)$ would suggest the involvement of  $R_{13} \Phi_1^{(1)}\,$, but $R_{12} \Phi_1^{(1)} \neq R_{13} \Phi_1^{(1)}\,$. The way out is to use the overall $SU(4)_R$ R--symmetry in unison with the $SU(2)_R$ R--symmetry for the adjoint fields. Define the $SU(4)_R$ R--symmetry that rotates the fermions of $V$ and the $\Phi_i$ as $R_i\,$, then the rotations $R_1(V, \Phi)=(V', \Phi')\,, R_2(V, \Phi)=(V'', \Phi'')\; \text{and}\; R_3(V, \Phi)=(V''', \Phi''')\,$ generate the $\mathcal{N}=1$ multiplets embeddings,
\begin{equation}
\label{eq: prime}
\begin{aligned}
& V=(A,\lambda),\qquad \Phi_1=(\phi_1,\psi_1),\qquad \Phi_2=(\phi_2,\psi_2),\qquad \Phi_3=(\phi_3,\psi_3)\,,\\[4pt]
& V'=(A,\psi_1),\quad \Phi'_1=(\phi_1,\lambda),\qquad \Phi'_2=(-i\phi_2^\dagger,\psi_3),\quad \Phi'_3=(-i\phi_3^\dagger,\psi_2)\,,\\[4pt]
& V''=(A,\psi_2),\quad \Phi''_1=(-i\phi_1^\dagger,\psi_3),\quad \Phi''_2=(\phi_2,\lambda),\qquad \Phi''_3=(-i\phi_3^\dagger,\psi_1)\,,\\[4pt]
& V'''=(A,\psi_3),\quad \Phi'''_1=(-i\phi_1^\dagger,\psi_2),\quad \Phi'''_2=(-i\phi_2^\dagger,\psi_1),\quad \Phi'''_3=(\phi_3,\lambda)\,.
\end{aligned}
\end{equation}
With these embeddings, a consistent non--BPS Lagrangian can be constructed with superpotential terms and with a bit more work, the supersymmetric structure of the pairs and triplets of brane stacks involving the $4^{\emph{th}}$ $\overline{\text{D6}}$ brane stack can be checked along with the explicit construction of the 32 Goldstinos and 28 Goldstones \cite{Mondal:2024qyn}.

\subsubsection*{Superpotential}

The worldline superpotential \eqref{eq: super} is the sum of four pieces, of which the Yang--Mills part $\mathcal{W}_4$ as in \eqref{eq: yangmill} is solely determined by the $\mathcal{N}=4$ superalgebra and hence remains the same. In this paper, we are interested in the abelian non--BPS black hole carrying the charges $(1,1,1,1)$. For, brevity we collect the non--BPS superpotential terms for the abelian version below but it is completely straightforward to generalize them to the non--abelian versions by inserting the appropriate traces and the factors of charges $(N_1,N_2,N_3,N_4)$ as in the BPS version and just changing the same fields to their primed versions as is done for the abelian case.

\paragraph{$\mathcal{W}_1$ (pairwise couplings).}
These are the standard pairwise couplings between bifundamentals and adjoints that encode transverse separations,
\begin{equation}
\label{eq: pair}
\begin{aligned}
\mathcal{W}_1 = \sqrt{2}\,\Big[ &
Z^{(12)}\big(\Phi_3^{(1)}-\Phi_3^{(2)}\big)Z^{(21)}
+ Z^{(23)}\big(\Phi_1^{(2)}-\Phi_1^{(3)}\big)Z^{(32)} \\
&+ Z^{(31)}\big(\Phi_2^{(3)}-\Phi_2^{(1)}\big)Z^{(13)}
+ Z^{(14)}\big(\Phi_1^{\prime\prime\prime(1)}-\Phi_1^{\prime\prime\prime(4)}\big)Z^{(41)} \\
&+ Z^{(24)}\big(\Phi_2^{\prime\prime(2)}-\Phi_2^{\prime\prime(4)}\big)Z^{(42)}
+ Z^{(34)}\big(\Phi_3^{\prime(3)}-\Phi_3^{\prime(4)}\big)Z^{(43)} \Big]\,,
\end{aligned}
\end{equation}
where the primed adjoint fields are as defined in \eqref{eq: prime}. 

\paragraph{$\mathcal{W}_2$ (cubic interactions).}
The cubic interactions among the bifundamentals are captured by,
\begin{equation}
\begin{aligned}
\mathcal{W}_2=\sqrt{2}\, \Big[&
Z^{(31)}Z^{(12)}Z^{(23)} + Z^{(13)}Z^{(32)}Z^{(21)}
+ Z^{\prime(12)} Z^{(24)} Z^{(41)} + Z^{(42)} Z^{\prime(21)} Z^{(14)} \\
&- Z^{\prime(13)} Z^{\prime(34)} Z^{\prime(41)}
+ Z^{\prime(31)} Z^{\prime(14)} Z^{\prime(43)}
+ Z^{(34)} Z^{\prime(42)} Z^{\prime(23)} + Z^{(43)} Z^{\prime(32)} Z^{\prime(24)}\Big] .
\end{aligned}
\end{equation}
Here, $\displaystyle Z'^{(ij)}=\left(-i\left(z'^{(ji)}\right)^{\dagger}, \chi'^{(ij)}\right), $ are the $SU(2)_R$ rotated versions of the $Z^{(ij)}$ fields as noted in Eq.~\eqref{eq: rsymm}. We note that, as in the abelian BPS case, the relative signs can be removed by field redefinitions. For, both the abelian and non--abelian cases,  the discrete exchange symmetries due the freedom in interchanging the brane labels as discussed in the context of the BPS version in \cite{Chowdhury:2015gbk} applies for non--BPS cases too and fixes the relative signs. As argued in \cite{Chowdhury:2023wss}, these discrete symmetries are incompatible with higher--order polynomial terms. Even though we have no supersymmetry and, a priori, can pretty much write down any interactions we like, these symmetries provide reasonable restrictions.  

\paragraph{$\mathcal{W}_3$ (linear terms / background couplings).}
The most general linear terms compatible with the Goldstone and Goldstino structure (and with the allowed R--symmetry rotations) can be written as
\begin{equation}
\begin{aligned}
\mathcal{W}_3 = \sqrt{2}\, \Big[ & 
c'_{12}(\Phi_3^{(1)}-\Phi_3^{(2)}) + c'_{23}(\Phi_1^{(2)}-\Phi_1^{(3)}) + c'_{31}(\Phi_2^{(3)}-\Phi_2^{(1)}) \\
&+ c'_{14}(\Phi_1^{\prime\prime\prime(1)}-\Phi_1^{\prime \prime \prime(4)}) + c'_{24}(\Phi_2^{\prime \prime \prime(2)}-\Phi_2^{\prime \prime \prime(4)}) + c'_{34}(\Phi_3^{\prime(3)}-\Phi_3^{\prime(4)}) \\
&+ c''_{12}(\Phi_3^{\prime\prime\prime(1)}-\Phi_3^{\prime\prime\prime(2)}) + c''_{23}(\Phi_1^{\prime(2)}-\Phi_1^{\prime(3)}) + c''_{31}(\Phi_2^{\prime\prime(3)}-\Phi_2^{\prime\prime(1)}) \\
&+ c''_{14}(\Phi_1^{\prime\prime(1)}-\Phi_1^{\prime\prime(4)}) + c''_{24}(\Phi_2^{\prime(2)}-\Phi_2^{\prime(4)}) + c''_{34}(\Phi_3^{\prime\prime(3)}-\Phi_3^{\prime\prime(4)}) \Big]\,,
\end{aligned}
\end{equation}
where the constants $c$ are determined ultimately by the background metric and $B$--fields similar to the BPS case \cite{Chowdhury:2014yca}. Note that supersymmetry is explicitly broken, for example if we take the term $c'_{12}(\Phi_3^{(1)}-\Phi_3^{(2)})$ preserving a particular $\mathcal{N} = 1$ subgroup of $\mathcal{N} = 2$ supersymmetry of the pair $(1,2)$, the 
 subgroup is preserved either by the triplet $(1,2,3)$ or by $(1,2,4)$ but not both.

\subsubsection{The Scalar Potential}
The full scalar potential contains three pieces,
\begin{equation}
\label{eq: nonbpsscalar}
V \;=\; V_{\mathrm{gauge}} + V_D + V_F \,.
\end{equation}
The gauge part, $V_{\mathrm{gauge}}$ for the BPS case, reads as Eq.~\eqref{eq: vgauge}, where if we put all $Z$ to zero we get the contribution coming from the dimensional reduction of the $\mathcal{N}=4$ Super Yang--Mills. Therefore, these parts don't change for the non--BPS case. It is straightforward to deduce the possible changes for terms involving the $Z$ in $V_{\mathrm{gauge}}$ for the non--BPS case. From the expression of the $\mathcal{W}_1$ potential in \eqref{eq: pair}, it is clear that gauge interactions corresponding to the pairs of $\Phi$ are  $V^{(1)}-V^{(2)}\,, V^{(2)}-V^{(3)}\,, V^{(3)}- V^{(1)}, V'''^{(1)}-V'''^{(4)}\,, V''^{(2)}-V''^{(4)}\; \text{and} \; V'^{(3)}-V'^{(4)}\,$, i.e. put the same primes in gauge fields $X_i^{(k)}$ as in $\Phi_i^{(k)}\,$. This does not affect the $V_{\mathrm{gauge}}$ as the gauge fields are the same under the primes but would affect the associated gauginos assignments. For the purpose of this paper we are interested in minimizing the potential $V$ which allows us to set all the gauge fields to zero and forget about the $V_{\mathrm{gauge}}$ term. For example, in the abelian case, 
\begin{equation}
\label{eq: vagauge}
V_{\text {gauge }}=\sum_{i=1}^3 \sum_{k=1}^4 \sum_{\substack{\ell=1 \\ \ell \neq k}}^4\left |X_i^{(k)}-X_i^{(\ell)}\right|^2\left(\left | Z^{(k \ell)}\right |^2+ \left |Z^{(\ell k)}\right|^2 \right)\,,
\end{equation}
and can be set to zero by placing all branes on top of one another in the flat directions $\left(X_i^{(k)}-X_i^{(\ell)}\right)$. Furthermore, using the shift symmetries, all $X_i^{(k)}$ can be set to zero (see Appendix A of \cite{Chowdhury:2015gbk}). Therefore, the contribution of $V_{\mathrm{gauge}}$ to the extremization conditions $\left \{\nabla V=0 \right \}$ is trivially zero. Also, for the stability analysis, we note that the full Hessian on the coordinates $\displaystyle \left \{Z^{(ij)}\,, \Phi^{(l)}_n\,, X^{(k)}_m \right\}$ is block diagonal with blocks partitions $\displaystyle \left \{Z^{(ij)}\,, \Phi^{(l)}_n \right\}$ and $\displaystyle \left \{X^{(k)}_m \right\}$ with $(+ve)$ eigenvalues (stable) in the $\left \{X^{(k)}_m \right\}$ block as long as the $Z$s don't vanish. Therefore, going forward, we can simply forget about the $X^{(k)}_i$ coordinates and the potential $V_{\mathrm{gauge}}\,$ except when some of the $Z$ do vanish, which we shall analyze separately while discussing the stabilizer submanifolds.  

Now, coming to the D--term potential $V_D\,$, it is entirely fixed by the interaction of the gauge multiplets and hence is the same as in the BPS case Eq.~\eqref{eq: vdterm}. The FI parameters are determined solely by the background fields, as in the BPS case, and will continue to hold for non--BPS cases. However, in the F--terms coming from $V_F$, there is a crucial difference; in the non--BPS case, they are no longer purely holomorphic. In the present non--BPS construction the relevant combination entering the potential is schematically,
\begin{equation}
V_F \;=\; \Big|\frac{\partial\mathcal{W}}{\partial\phi} + \frac{\partial\mathcal{W}}{\partial\phi^\dagger}\Big|^2
+ \Big|\frac{\partial\mathcal{W}}{\partial z} + \frac{\partial\mathcal{W}}{\partial z^\dagger}\Big|^2 \, ,
\end{equation}
for example, the F--term corresponding to say $Z^{(ij)}$ is, 
$ \displaystyle
 F^{(ij)}=\frac{\partial \mathcal{W}}{\partial Z^{(ij)}}+\frac{\partial \mathcal{W}}{\partial Z'^{(ij)}} \,$,
 and similarly for the $\Phi$ fields and their Hermitian conjugates. Because of the loss of holomorphicity, one cannot reduce the problem of minimizing the potential to solving only for the holomorphic F--term equations modulo a complexified gauge action; instead, one must minimize the full real potential $V$ modulo the gauge identification, 
\begin{equation}
\frac{U(N_1)\times U(N_2)\times U(N_3)\times U(N_4)}{U(1)_\mathrm{diag}}\,. \end{equation}
Unfortunately, many of the computational algebraic geometry techniques discussed earlier and in our previous work \cite{Chowdhury:2023wss} rely on holomorphicity of the equations to solve the system of equations. Therefore, on the nose, these efficient techniques will fail. As we shall see, in certain situations, there are ways to deform the problem and continue using these techniques, albeit at a higher computational cost. We shall also develop some other sets of techniques to deal with non--holomorphic systems. Our focus will be on the abelian version of this non--BPS black hole, but the discussion can be lifted to the non--abelian cases as well.

\paragraph{Limitations of the tree-level effective description.}
While the potential \(V\) correctly captures the classical intersection dynamics of the constituent branes and the associated Goldstone and Goldstino zero--mode structure dictated by the preserved subsystem supersymmetries \cite{Mondal:2024qyn}, its low--energy validity carries important caveats. In the BPS regime, extended supersymmetry and the associated non--renormalization theorems protect the F--term equations and the vacuum moduli space against quantum corrections. In the full non--BPS configuration, by contrast, all 32 supercharges are broken, and this protection is lost. At the quantum level, the masses of the heavy open--string modes stretched between the branes depend on the background scalar fields that parametrize the classical flat directions. Integrating out these massive modes therefore generates loop and threshold corrections to the effective action, including Coleman--Weinberg--type contributions to the effective potential \cite{PhysRevD.7.1888,ZAREMBO199970}. 

Here, it is essential to distinguish between exact collective zero modes and accidental internal flat directions. The Goldstone and Goldstino modes associated with exact bulk symmetries broken by the brane configuration correspond to protected collective coordinates and should not be lifted by integrating out massive states. By contrast, continuous flat directions that arise only as accidental features of the tree--level internal potential -- for example, the stabilizer submanifolds appearing at special loci in the vacuum landscape -- are not similarly protected. Once supersymmetry is absent, quantum corrections from massive open--string modes can generically induce an effective potential along these internal directions, thereby renormalizing couplings, shifting the locations of isolated vacua, and perturbatively lifting these accidental flat directions \cite{Sen:1999mg}. Consequently, our analysis of the non--BPS system should not be interpreted as an exact low--energy theorem, but rather as a controlled characterization of the minimal tree--level vacuum landscape. While the discrete symmetries, the protected Goldstone/Goldstino sector, and the broad topological branch structure are expected to be robust, the detailed internal potential should be regarded as a classical baseline on top of which perturbative and stringy corrections must eventually be incorporated.

\subsection{Finding Vacua} \label{sec: vacua}
For the BPS black holes, the ground state energy of the D--brane system is $E=0\,$. We arrive at it by minimizing the potential $V$ in \eqref{eq: scalar}, as we would do for a classical system. But due to supersymmetry and the fact that all these vacua are bosonic, there is no tunneling between these vacua, and we get degenerate BPS quantum states with energy $E=0\,$ \footnote{In generic SUSY quantum mechanics, non--perturbative instanton effects can lift classical zero--energy vacua, but finite--action instantons only mediate transitions between a bosonic state and a fermionic state. Our BPS pure D--brane system is governed by a holomorphic superpotential analogous to an extended supersymmetric Landau--Ginzburg model and all non--degenerate classical vacua carry the same fermion number \cite{Chowdhury:2023wss}. Because all vacua are bosonic, instanton trajectories connecting them are topologically forbidden. Consequently, non--perturbative tunnelling effects are absent, and the classical $V=0$ perturbative vacua are exact quantum ground states.}. We can carry out a similar minimization of the potential \eqref{eq: nonbpsscalar} for the non--BPS black holes, but it is expected that the degeneracy will be lifted and we should get a finite number of local minima as classical solutions. At this point, we focus on the abelian non--BPS D--brane configuration which admits the relative gauge symmetry $U(1)\times U(1)\times U(1)$ and contemplate four questions of immediate interest:
\begin{enumerate}
    \item As the construction of this extremal non--BPS D--brane system follows from the extremal BPS D--brane system, do we still have an energy, $E=0\,$, global minima? As the potential, $V=V_D+ V_F\,$, is a sum of square terms, $V$ can't be lower than zero and if there is a zero--energy configuration, at worst it will be marginally stable.
    \item What is the structure of the vacuum manifold? Do we get a tight band of local minima, if so is there any degeneracy associated with each individual minima? Is the system structurally stable or undergo \textit{phase transitions} as we deform the moduli parameters \footnote{The term \textit{phase transition} is a bit loose; it only makes sense in large--charge limit. We also note that the dependence of open string spectrum masses ($m$) on closed string moduli parameters $c$ is expected in non--BPS systems. As the parameters $c$ are varied, it is a generic feature of non--SUSY landscapes that the lowest--lying mode $m^2(c)$ may cross zero. This crossing corresponds to the onset of tachyon condensation \cite{Sen:1998sm}; a state that is a local minimum (stable vacuum) in one region of moduli space becomes a saddle point (unstable) when $m^2(c)$ becomes negative.}?
    \item Ultimately, the low--energy non--BPS dynamics of this D--brane system is governed by quantum mechanics, and it is important to know what the tunneling amplitudes are between these local minima. Surely, the approximate low--energy spectra of quantum states localized near the respective local minima would receive instanton corrections (much like the quantum double--well potential) that lift the classical degeneracy. Can we estimate these non--perturbative corrections \footnote{We note that while spontaneous symmetry breaking and fermion condensation are strictly forbidden in finite quantum systems due to these instanton transitions, such phenomena can emerge in the large--$N$ limit. If a brane stack carries charge $N$, the $U(N)$ gauge theory contains $\mathcal{O}(N^2)$ adjoint degrees of freedom, which provide an effective thermodynamic limit as $N \to \infty$ . In this regime, the instanton tunneling amplitudes are exponentially suppressed, allowing the matrix quantum mechanics to superselect a vacuum and potentially develop non--zero gauge--invariant fermion bilinears \cite{Kim:2019upg, LANDSMAN2013379}.}? As we shall see, in certain regimes the potential develops flat directions; do quantum corrections lift them?
    
    \item Can the low--lying vacuum landscape of this non--BPS black hole explain its entropy? In supergravity, BPS and non--BPS supersymmetric black holes with the same charges have the same Wald entropy \cite{Sen:2007qy}. Does it hold true in the microscopic D--brane regime?
\end{enumerate}
In this paper, we shall address the first and second parts of the questions; the rest, including phase transitions, are work in progress, and we hope to report on them at some point. 

\subsubsection*{Gauge fixing} The first step will be to get rid of the gauge orbits, which in the abelian case are complex phases.  A convenient gauge choice is to remove these phases by imposing a gauge fixing condition,
\begin{equation}
\label{eq: nonbpsgauge}
\Im Z^{(12)}=0\,,\qquad \Im Z^{(13)}=0\,,\qquad \Im Z^{(34)}=0\,,
\end{equation}
together with the shift--symmetry gauge fixing \footnote{As is evident from the appearance of only relative adjoint fields $\Phi$ in the superpotential \eqref{eq: pair}, while fixing the choice coming from the Goldstone modes, some of the $\Phi$ appearing in \eqref{eq:phase-nonbps} should be in their prime(s). This doesn't matter as for these particular $\Phi\,$ fields the bosonic part remains the same under the prime operation.}
\begin{equation}
\Phi^{(1)}_{1}=0\,,\qquad \Phi^{(1)}_{2}=0,\qquad \Phi^{(1)}_{3}=0\,,\qquad
\Phi^{(2)}_{1}=0\,,\qquad \Phi^{(2)}_{2}=0,\qquad \Phi^{(3)}_{3}=0\,.
\label{eq:phase-nonbps}
\end{equation}
We note that discrete $\mathbb{Z}_2\times \mathbb{Z}_2\times\mathbb{Z}_2$ residual gauges remain with the gauge fixing condition \eqref{eq: nonbpsgauge}. One set of consistent representation (utilizing the overall $U(1)_{\text{diag}}$ to set the $U(1)$ associated with the $2^{\mathrm{nd}}$ brane to identity) for its generators is,
\begin{equation} \label{eq: dgauge}
\begin{aligned}
    &g_{1}: \quad Z^{(1k)} \longrightarrow -Z^{(1k)} \quad \mathrm{and} \quad \quad Z^{(k1)} \longrightarrow -Z^{(k1)}\, , \\
    &g_{2}: \quad Z^{(3k)} \longrightarrow -Z^{(3k)} \quad \mathrm{and} \quad \quad Z^{(k3)} \longrightarrow -Z^{(k3)}\, , \\
    &g_{3}: \quad Z^{(4k)} \longrightarrow -Z^{(4k)} \quad \mathrm{and} \quad \quad Z^{(k4)} \longrightarrow -Z^{(k4)}\, ,
\end{aligned}
\end{equation}
and the same $(-ve)$ signs for their hermitian conjugates. There is also a discrete $\mathbb{Z}_2$ parity symmetry,
\begin{equation} \label{eq: parity}
    g_{4}: \quad Z^{(ij)} \longrightarrow -Z^{(ij)} \quad \mathrm{and} \quad \Phi^{(k)}_i \longrightarrow -\Phi^{(k)}_i\, ,
\end{equation}
with the same $(-ve)$ signs for their hermitian conjugates. The potential \eqref{eq: nonbpsscalar} is invariant under this $\mathbb{Z}_2^4$ 16--fold discrete symmetry. The 8--fold gauge symmetry, $\mathbb{Z}_2^3$ acts by sign flips on bifundamentals, which corresponds to \textit{Weyl reflections} of the broken $U(1)$ factors. These transformations leave all physical observables invariant and should be quotiented out. The parity $\mathbb{Z}_2$ acts as a global (non--gauge) symmetry and leads to genuine \textit{Gribov copies} \cite{Gribov:1977wm} in the vacuum manifold. This will be particularly useful when we discuss the approximate low--energy spectra of this non--BPS system in Section \ref{sec: unique}. As we shall see, the vacuum landscape admits isolated bound state vacua and more interestingly various marginally bound state solutions as \textit{stabilizer vacuum submanifolds} which preserve subsets of the $U(1)$ gauge symmetries (stabilizer subgroups). In such cases, we have to be careful with the gauge fixing and the representative residual symmetries for proper vacua enumeration and stability analysis, as detailed in Appendix \ref{app:gauge_consistency}.

\subsubsection*{Heuristics for $E\neq0$ Extremality}

To address the first question, regarding the existence of $V=E=0$ ground state(s), it is sufficient to check whether a $V=0$ configuration exists. As the potential is a sum of squares, at $V=0\,$ the vanishing of the gradients, $\left\{ \nabla V=0 \right\}$ is automatic. Then, this is essentially the same setup as its BPS counterpart, except now the F--term equations coming from the $V_F$ are non--holomorphic. This has two consequences:
\begin{enumerate}
    \item The F--term equations are no longer invariant under the complexified $U(1)$ gauge symmetries. This forces us to explicitly look for solutions not only for $V_F=0$ but also for $V_D=0\,$ i.e. both the non--holomorphic F--term and D--term equations have to be solved simultaneously. 
    \item We have to get creative in applying the computational algebraic geometry techniques to solve for the non--holomorphic systems of polynomials. 
\end{enumerate}
As we shall see in the next section, it is best to decompose the complex $Z$ and $\Phi$ fields into their real and imaginary parts. For the abelian case, there are in total 12 complex $Z$ and 12 complex $\Phi\,$, i.e. 48 real variables. After gauge fixing and removing Goldstones, there are 21 real $Z$ and 12 real $\Phi$ leaving $21 +12=33$ real variables. However, we have 33 real F--term equations and 3 independent real D--term equations to solve. If these equations are completely generic, we should expect no solutions to this overdetermined system. Similar arguments apply for the non--abelian cases. 

In the BPS case, this apparent overdetermination is resolved because the holomorphicity of the F--terms enhances the physical gauge symmetries to their complexified counterparts, thereby generating three extra syzygies (due to complexification of the three relative $U(1)$ gauge symmetries)  that reduce the equations to a square system with 12 solutions. In the non--BPS case, by contrast, the loss of holomorphicity removes this manifest complexification, and one therefore expects the system to be genuinely overdetermined. Nevertheless, since the non--BPS configuration is microscopically assembled from BPS building blocks, one cannot \emph{a priori} exclude the possibility that some nontrivial hidden syzygies survive despite the absence of holomorphicity. Any such residual relations would render the equations dependent and could in principle allow exceptional solutions. For this reason, simple counting arguments are not sufficient, and the issue must be settled by the Gröbner basis analysis carried out in the next section. That computation shows that no additional syzygies are present in the non--BPS system and, accordingly, that there are no zero--energy solutions. Equivalently, there are no classical field configurations for which the potential vanishes, so the classical global minima occur at strictly positive energy. This conclusion is consistent with the microscopic uniqueness argument discussed in \cite{Mondal:2024qyn} and supports the absence of genuinely extremal zero--energy bound states in this non--BPS D--brane realization \footnote{The Ward identities \eqref{eq:ward} in Appendix \ref{app:gauge_consistency} encode the physical $U(1)$ gauge symmetries, but unlike in the BPS case, they do not supply enough relations among the real F--term and D--term equations to reduce the system to a square one in the absence of complexified gauge symmetry.}.

\subsection{Algorithmic Certification: The Gröbner Basis}
\label{subsec:GB}

While numerical methods as discussed here and in \cite{Chowdhury:2023wss} are indispensable for enumerating isolated vacua in zero--dimensional varieties, they are inherently probabilistic in their certificate of completeness. In contrast, the method of Gröbner bases offers an algebraic rigor capable of certifying the non--existence of solutions -- a property crucial for verifying the supersymmetry breaking of the non--BPS D--brane system. In standard supersymmetric systems, the vacuum equations are holomorphic. However, for non--BPS systems or when including D--term constraints explicitly, the stationary conditions involve both the $N$ number of fields $\Psi$ and their Hermitian conjugates $\Psi^\dagger$, rendering the system non--holomorphic/non--polynomial in the complex variables \footnote{For us $\Psi$ is the generic label we are using for the collection of bifundamentals $Z$ and the adjoints $\Phi$ fields.}. To apply standard algebraic algorithms, we must ``realify" the system. We decompose each complex field and moduli parameter into its real and imaginary components, $\Psi_{k} = x_{k} + i\, y_{k}$, and separate the constraints into real and imaginary parts.

Let $R = \mathbb{Q}[\{x_{k}\,, y_{k}\}]$ denote the polynomial ring generated by these $2N$ real components. The stationary conditions generate an ideal $I = \langle f_1, \ldots, f_s \rangle \subset R\,$. The physical vacuum moduli space is identified with the real affine variety $\mathcal{V}_{\mathbb{R}}(I)\,$. To analyze the structure of this variety, we select a monomial ordering $\prec$ (typically graded reverse lexicographic for efficiency) and compute a Gröbner basis $G = \{g_1, \ldots, g_t\}$ for the ideal $I\,$. The power of this construction lies in Hilbert's \emph{weak Nullstellensatz} \cite{cox1,AtiyahMacdonald1969}. The weak Nullstellensatz states that for an ideal in a polynomial ring over an algebraically closed field (such as $\mathbb{C}$), the affine variety is empty if and only if the ideal contains the unit element. Although our physical moduli space requires solutions over a non--algebraically closed field, we can leverage this theorem to certify inconsistency for real systems. Specifically, if the ideal generated by the polynomial constraints contains the unit element, the variety is strictly empty over the complex numbers, and a fortiori empty over the reals,
\begin{equation}
\label{eq:nullstellensatz}
    1 \in I \iff \mathcal{V}_{\mathbb{C}}(I) = \varnothing \implies \mathcal{V}_{\mathbb{R}}(I) = \varnothing \,.
\end{equation}

Algorithmically, this condition is detected if and only if the reduced Gröbner basis contains the constant unit element, i.e., $G = \{1\}$. If the computation returns $G=\{1\}$, it implies the existence of a ``certificate of inconsistency'', a set of cofactor polynomials $h_i \in R$ such that
\begin{equation}
    \sum_{i=1}^s h_i(\{x\}, \{y\}) \cdot f_i(\{x\}, \{y\}) = 1 \,.
\end{equation}

Evaluation of this identity at any putative solution $\Psi^* \in \mathbb{C}^{2N}$ (and thus any physical solution $\Psi^* \in \mathbb{R}^{2N}$) yields $0=1$, a contradiction. Thus, determining $1 \in G$ constitutes a rigorous algebraic proof that the scalar potential possesses no zero energy stationarity points anywhere in the field space.

\paragraph{Limitations for Counting and Real Geometry.}
While the detection of $G=\{1\}$ is decisive, the converse case $G \neq \{1\}$ highlights a fundamental limitation of standard algebraic geometry when applied to physical systems defined over the real numbers. Standard Gröbner basis algorithms operate implicitly over the algebraic closure (the complex numbers). Consequently, if the ideal is not the unit ideal, the basis $G$ describes the geometry of the variety in $\mathbb{C}^{2N}$. It cannot inherently distinguish between physical vacua (points in $\mathbb{R}^{2N}$) and non--physical ``parasitic'' solutions that possess non--zero imaginary components. For example, the simple real constraint $x^2 + 1 = 0$ yields a Gröbner basis $G = \{x^2+1\} \neq \{1\}$, as solutions exist in the complexification ($x = \pm i$). To rigorously enumerate solutions strictly over the reals, one must employ techniques from \emph{semialgebraic geometry}, most notably Cylindrical Algebraic Decomposition (CAD) \cite{Lee2025CylindricalAD}. CAD decomposes the ambient space into disjoint cells where the sign of each polynomial is invariant, allowing for an exact topological description of the real solution space, effectively deciding
the first--order theory of the reals (Tarski--Seidenberg theorem \cite{Tarski1951,Seidenberg1954,Basu2014AlgorithmsIR}). However, the complexity of CAD is doubly exponential in the number of variables $\left(O\left(2^{2^N}\right)\right)$, rendering it computationally intractable for quiver--like theories with dozens of scalar fields.

Fortunately, for the specific objective of certifying supersymmetry breaking, this limitation is irrelevant. Because $\mathbb{R} \subset \mathbb{C}$, the non--existence of complex solutions (guaranteed by $G=\{1\}$) logically necessitates the non--existence of real solutions. Thus, for the purpose of proving inconsistency, the efficient F4 algorithm over the rational numbers suffices, bypassing the need for prohibitive semialgebraic computations. The F4 algorithm \cite{Faugere1999F4} is a matrix--based method for Gröbner basis computation that leverages sparse linear algebra. It is particularly efficient for overdetermined systems with high rigidity and admits parallelization.

\paragraph{Application to the Non--BPS System.}
As motivated in detail in Section \ref{sec: vacua}, we apply  this Gröbner basis certification technique  to the stationary conditions of the non--BPS D--brane system. Despite the high dimensionality of the scalar manifold, the rigidity of the constraints allows the parallelizable F4 algorithm (implemented in \texttt{Macaulay2} \cite{M2}) to terminate, returning $G_{\mathrm{non-BPS}} = \{1\}$. This confirms that supersymmetry is broken not merely by the choice of vacuum expectation values, but by the inconsistency of the algebraic relations defining the critical locus. We provide a brief technical exposition of the parent algorithm, including the basis--expansion mechanism and illustrative examples, in Appendix \ref{app:GB}.

\subsubsection*{Remarks}

\paragraph{Analytic Certainty vs. Numerical Probability.}
A crucial distinction between Gröbner basis techniques and numerical methods (such as monodromy) lies in the nature of their output. Numerical methods can find isolated roots with high precision, but they generically struggle to distinguish between a system with \emph{no} solutions and a system where the numerical path--tracking simply failed to converge. In contrast, the Gröbner basis algorithm is symbolic and exact. The output $G=\{1\}$ is a definitive mathematical proof that the stationary conditions are mutually inconsistent. It leaves no room for ``missed'' vacua in remote regions of the field space.

\paragraph{Field of Definition and Exact Arithmetic.}
The computations are performed over the field of rational numbers $\mathbb{Q}\,$. This avoids the floating--point instabilities inherent in numerical approaches. Because the coefficients of the superpotential and D--terms are rational numbers (typically quantized charges, couplings and moduli), the ideal is fully defined over $\mathbb{Q}\,$. A standard result in commutative algebra ensures that if $1 \in I_{\mathbb{Q}}$ (inconsistency over rationals), then $1 \in I_{\mathbb{C}}$ (inconsistency over complex numbers) \cite{cox1}. Thus, the result holds for the full physical space of space.

\paragraph{Computational Complexity vs. Rigidity.}
It is well known that the worst--case complexity of computing a Gröbner basis is doubly exponential in the number of variables (Buchberger's criterion  \cite{1570291224270729344,cox1}). For generic superpotentials with many fields, this computation would be intractable. However, the non--BPS system under study exhibits high ``rigidity'' -- the constraints are possibly overdetermined and structurally tight. This specific property allows the F4 algorithm to converge rapidly to the unit ideal, turning what is usually a weakness of the method (complexity) into a strength (rapid detection of inconsistency).

\subsection{Low Energy Potential Landscape} \label{sec: unique}

We have established in the previous section that for the non--BPS D--brane system, there are no energy $E=0$ states. The next step would be to investigate the approximate low--energy vacuum landscape of the potential, i.e. enumerate all or at least the low--lying local minima for the potential. Just like the BPS cases, at generic values of the moduli parameters $c$, away from the degeneration limits, we expect isolated extremum points to populate the low energy vacuum landscape \footnote{We are a bit loose with the term \textit{moduli}. There are two notions of moduli in this setup. The background metric and B--fields in the compact directions, which feed into $c$ and the expectation values of the fields (like $Z$, $\Phi$, etc.) at the vacuum manifold. Both are standard terminology and should be clear from the context.}.  We can take the gauge fixed quartic potential $V$ in \eqref{eq: nonbpsscalar}, compute the zero set of its gradients (cubic equations $\left \{\nabla V =0\right \}$) and the eigenvalues of the Hessian to classify the local extremum points for a system size of 33 real variables. This expectation is partially realized with important caveats, as discussed later and explained in detail in Appendix \ref{app:gauge_consistency}. 

Before delving into the specifics of these caveats and the challenges of flat directions, it is useful to explicitly summarize the methodological pipeline used to derive the subsequent physical results. Because the non--BPS scalar potential is real, non--holomorphic, and its critical locus is populated by non--isolated continuous flat directions (stabilizer submanifolds), a single mathematical technique is insufficient to resolve the complete vacuum structure. To clarify the division of labor across our framework, the workflow proceeds as follows:
\begin{enumerate}[label=(\roman*), leftmargin=*]
    \item \textbf{Algebraic Certification:} Exact Gr\"obner bases over the field of rational numbers (Section \ref{subsec:GB}) are used to rigorously prove the absence of zero--energy global minima.
    \item \textbf{Algebraic Excision:} When continuous families of vacua are known \textit{a priori}, algebraic saturation can exactly excise these branches from the gradient ideal (Section \ref{sec: saturation}). However, this is insufficient for a global search as generic flat directions remain hidden.
    \item \textbf{Topological Regularization:} To systematically handle unknown flat directions, we apply Morse--Bott deformations (Section \ref{sec: morsebott}) and Gated Soft--Trapping (Section \ref{sec: gatedtrap}). This is a mathematically robust procedure; Morse--Bott theory guarantees that the deformation acts as a topological probe, lifting continuous flat directions into isolated critical points while preserving the structural stability of the genuine physical extrema.
    \item \textbf{Numerical Extremization:} With the flat directions topologically regularized, we fix specific numerical values for the background moduli parameters ($c^{(ij)},\,c^{(k)}$) and locate the  coordinates of the critical points using high--precision, second--order numerical extremization (the Newton--Raphson method), tracking the limits as the deformation parameter $\varepsilon \to 0\,$.
    \item \textbf{Physical Classification:} Finally, to differentiate between stable vacua, maxima, and saddle points, we explicitly evaluate the full Hessian matrix of the potential $V$ at these numerical coordinates. As detailed in Appendix \ref{app:gauge_consistency}, a critical point is classified as a stable local minimum if and only if its evaluated Hessian is positive semi--definite, with all exactly zero eigenvalues mapping purely to the Goldstone modes of appropriate gauge orbits.
\end{enumerate}
To be able to actually solve the system of equations, of course, we can also lift all the real variables to complex variables and apply the monodromy method discussed in Section \ref{sec:monodromy}, but it will be  computationally very wasteful to do so for two reasons:
\begin{enumerate}
    \item Unlike the BPS pure D--brane cases, where the equations to solve are at most quadratic in the variables, here it is cubic and due to \text{Bezout bound} \cite{Kushnirenko1976NewtonPA}, the complexity scales with the ratio $(\frac{3}{2})^{33}$. Even if we manage to solve the system, which we didn't, a measure zero subset of the solutions will be real under the homotopy continuation. So, we have to hunt for real solutions with reasonable tolerances and then compute the eigenvalues of the $33\times 33$ Hessian matrices to sort the extremum points as local minima, maxima or saddles.
    \item As discussed in \cite{Kumar:2023kpc}, in the context of BPS D--brane system, if we set all the bifundamentals $Z$ to zero, the gauge fields $X$ and the adjoint fields $\Phi$ to there commuting counterparts i.e. go to the \textit{Coulomb branch} of the superpotential, we get a non--compact submanifold $\mathcal{M}_\mathrm{critical}$ of local extrema with all the the commuting gauge and adjoint fields as flat directions in the full configuration space $\mathcal{C}$ of all fields \footnote{We are calling it non--compact as the gauge fields $X$ represent non--compact directions but the $\Phi$ are the compactification circles of the IIA theory. In the numerical setups, we have ignored the compact nature of $\Phi$, as it amounts to appropriately scaling the radius of compactification and then performing an identification to make it a circle. We don't foresee any complications due to this representation, but maybe in degeneration limits, it will matter.}. This manifold supports non--normalizable scattering states with energy (in the abelian case), 
    \begin{equation} \label{eq: critical}
E=V_{\mathrm{critical }}=\frac{1}{2}\left(\left(c^{(1)}\right)^2+\left(c^{(2)}\right)^2+\left(c^{(3)}\right)^2+\left(c^{(4)}\right)^2\right)+4 \sum_{k<l ; k, l=1}^3\left|c^{(k l)}\right|^2+4  \sum_{k=1}^3\left|c^{(4 k)}\right|^2 \,.
\end{equation}
 The same is also true for the non--BPS cases with $c^{(k\ell)}\rightarrow c'^{(k\ell)} +c''^{(k\ell)}$ (also for the $c^{(4k)}$) and poses serious challenges to the numerical enumeration of the isolated vacua. 
Moreover, $\mathcal{M}_\mathrm{critical}$ is not the only vacuum submanifold; there are many marginally bound vacuum submanifolds $\left\{\mathcal{M}\right\}$, as discussed in Section \ref{sec: landscape} and in Appendix \ref{app:gauge_consistency}.
\end{enumerate}

\subsubsection{Flat Directions}
Flat directions present both theoretical and numerical challenges in the analysis of the statics and dynamics of a potential. Typically in such scenarios involving singular perturbations, the stability and vacuum expectation values are entirely determined by subleading corrections -- such as background fluxes or non--perturbative contributions to the superpotential -- that lift the degeneracy. Dynamically, the absence of a restoring force along these flat, distinct trajectories implies that the field evolution is decoupled from the potential gradient at the tree level, leading to large field excursions that are highly sensitive to the initial data of the system.

In the context of this paper, we are interested in the classical static analysis of the potential V in \eqref{eq: nonbpsscalar}. In cases where we may possess some prior knowledge of a subset of flat directions, and we can lift them by artificially intersecting them with lines or hypersurfaces, i.e. \textit{Witness Sets} \cite{Sommese2005TheNS,Bates2023NumericalNA}. However, if we were to do a precise counting, generic methods that don't rely on an a priori understanding of the flat directions should be developed. Standard second--order methods like Newton--Raphson or Trust--Region optimization algorithms rely on the Hessian matrix $\mathcal{H}_{ij} = \partial_i \partial_j V$. In the presence of flat directions, $\mathcal{H}$ becomes singular or ill--conditioned (possessing eigenvalues $\lambda \approx 0$). This leads to two critical failures \footnote{First--order methods like Gradient descent are computationally cheap but not suitable for precise numerical computations and fail miserably near flat directions.}: 
\begin{itemize}
  \item \textbf{Matrix Singularity:} The inversion of the Hessian required for the Newton step ($\delta {\bm x} = -\mathcal{H}^{-1} \nabla V$) becomes numerically unstable, leading to divergence.  
  \item \textbf{Unbounded Drift:} In directions where the potential is effectively zero, the solver may take arbitrarily large steps, pushing the field configuration ${\bm x}$ to physically irrelevant regions (e.g., decompactification limits) or jumping between distinct basins of attraction.
\end{itemize}
Fortunately, in the monodromy method, a deformation with generic parameters $p$ suffices to lift the flat directions. It is like adding a linear deformation (a line in the space of variables with arbitrary coefficients) to the potential. Unfortunately, as we perform Parameter Homotopy continuation to $\bm{p}=0\,$, the generic $\bm{p}$ solution ends up at the extremization of the line we added, constrained to the solution manifold. If the flat direction is compact like a circle, then no issues, but for non--compact flat directions, the extremization of a line is at $\pm\infty$ and the solution will be dropped as a valid solution. There are ways to get around it using \textit{Singular Endgames} strategies \cite{Sommese2005TheNS,BatesHauensteinSommeseWampler2013}, but they are not very reliable.  We shall now describe some strategies we developed to tackle flat directions in the context of finding extremum points and apply them to the non--BPS case.

\subsubsection{Saturation of Ideals}
\label{sec: saturation}
In situations where we have prior knowledge of the flat directions like the $\mathcal{M}_\mathrm{critical}$ discussed above, we can remove them algebraically by considering the extrema variety as generated by the ideal $I=\left\{\nabla V({\bm x})\right\}$ and then enforce the condition that all (or some) hypermultiplets $Z \subset \{{\bm x}\}$ can't vanish simultaneously. This is the \textit{saturation} operation as described below through an example. When applied correctly, it solves the problem of endless iterations that plagues most analytic and numerical solvers. This can be implemented in Macaulay2 \cite{M2} or Singular \cite{Singular}.

\paragraph{Example.} To illustrate how the saturation operation isolates specific physical phases/regions, consider a standard algebraic geometry toy model that mimics mutually exclusive vacuum branches. Let the scalar fields be partitioned into two distinct sets, $\bm{x} = \{x, y\}$ and $\bm{z} = \{z, w\}$. In many effective field theories, cross--couplings in the scalar potential (such as $V \supset |\bm{x}|^2 |\bm{z}|^2$) force these two sectors to mutually annihilate in the true zero--energy vacuum. Algebraically, this cross--annihilation is exactly captured by the ideal generated by all pairwise products, $I = \langle xz, xw, yz, yw \rangle \subset \mathbb{C}[x,y,z,w]$. The complete vanishing locus $\mathcal{V}_{\mathbb{C}}(I)$ of this system consists of the union of two distinct coordinate planes; the branch $\mathcal{M}_x = \{z=w=0\}$ where the $\bm{x}$ variables are free, and the branch $\mathcal{M}_z = \{x=y=0\}$ where the $\bm{z}$ variables are free. These branches intersect non--transversally at the origin $(0,0,0,0)$. This geometry is the paradigmatic toy model for the vacuum moduli spaces of supersymmetric gauge theories, where $\mathcal{M}_x$ and $\mathcal{M}_z$ would represent intersecting Higgs and Coulomb branches \cite{Akhond:2021xio}.

Suppose we wish to strictly isolate the $\mathcal{M}_x$ branch. To do so, we enforce the physical condition that the $\bm{x}$ variables do not vanish simultaneously, i.e., $(x,y) \neq (0,0)$. Geometrically, this requires the excision of the $\mathcal{M}_z$ branch from the total space. Algebraically, the ideal describing the closure of this non--degenerate locus is obtained via the \textit{saturation} of $I$ with respect to the ideal defining the forbidden singularity at the origin, $J = \langle x, y \rangle$. The saturation operation, defined as 
\begin{equation}
    I : J^\infty = \{ f \in R \mid \forall g \in J\,, \exists\, n \in \mathbb{N} : f g^n \in I \} \,,
\end{equation}
effectively quotients the ring by the torsion submodule supported on $\mathcal{V}_{\mathbb{C}}(J)$. Evaluating this for our toy model yields
\begin{equation}
    I_{\text{sat}} = I : \langle x, y \rangle^\infty = \langle z, w \rangle\,.
\end{equation}
By systematically annihilating the primary components supported within the locus $\mathcal{V}_{\mathbb{C}}(J)$, the saturation operation collapses the original system of four quadratic constraints into a simple linear system. The viable solution manifold is strictly reduced to the irreducible smooth surface $\mathcal{V}_{\mathbb{C}}(z,w)$. In the context of our vacuum solver, this algebraic decoupling avoids the endless iterations of standard numerical solvers when exploring intersecting flat directions.

\subsubsection{Morse--Bott Regularization}
\label{sec: morsebott}

In the analysis of the vacuum structure, we seek the critical points of the potential $V: \mathbb{R}^n \to \mathbb{R}$ satisfying $\{\nabla V(\bm{x}) = 0\}$. For the non--BPS D--brane system, the solution set is rarely a simple collection of isolated roots. Instead, it typically consists of degenerate continuous components -- flat directions or stabilizer submanifolds -- denoted by $\mathcal{M} \subset \mathbb{R}^n$. On these submanifolds, the Hessian matrix contains null eigenvalues, rendering standard numerical algebraic geometry methods (which assume isolated, non--degenerate roots) inapplicable. To rigorously enumerate the vacua, we employ a \textit{symmetry--preserving Morse--Bott regularization}. We introduce a deformed potential $V_\epsilon\,$,
\begin{equation}
    V_\epsilon(\bm{x}) = V(\bm{x}) + \epsilon W(\bm{x})\,,
\end{equation}
where $\epsilon$ is a small perturbation parameter ($0 < \epsilon \ll 1$) and $W(\bm{x})$ is a generic regulating function. The function $W(\bm{x})$ is chosen to respect the exact physical symmetries (e.g., $\mathbb{Z}_2$ parity \eqref{eq: parity} or the discrete gauge symmetries \eqref{eq: dgauge}) while explicitly breaking the accidental continuous symmetries responsible for the flatness. The physical vacua are identified by solving the regularized system $\{ \nabla V_\epsilon(\bm{x}) = 0 \}$ and tracking the stable solutions in the limit $\epsilon \to 0\,$. The validity of this procedure relies on \textit{Morse--Bott theory} \cite{Bott1954NondegenerateCM,Nishikawa2025OnDM}. This framework ensures that the deformation acts as a topological probe, preserving genuine minima while resolving continuous flat directions into a discrete set of isolated points suitable for counting.

\paragraph{Structural Stability of Non--Degenerate Vacua.}
First, we consider a pre--existing isolated local minimum $\bm{x}_0$ of the unperturbed potential $V$. If $\bm{x}_0$ is non--degenerate, its Hessian matrix $\mathcal{H}_V(\bm{x}_0)$ is strictly positive definite and thus invertible. We analyze the stability of this solution by expanding the gradient condition, $\{\nabla V(\bm{x}) + \epsilon \nabla W(\bm{x}) = 0\}$ around $\bm{x}_0\,$. Let $\bm{x}(\epsilon) = \bm{x}_0 + \epsilon \bm{\delta} + \mathcal{O}(\epsilon^2)\,$. To first order, the condition becomes,
\begin{equation}
    \mathcal{H}_V(\bm{x}_0) \cdot \bm{\delta} + \nabla W(\bm{x}_0) = 0 \quad \implies \quad \bm{\delta} = - \mathcal{H}_V(\bm{x}_0)^{-1} \cdot \nabla W(\bm{x}_0)\,.
\end{equation}
Since the Hessian is invertible, the shift $\bm{\delta}$ is unique and finite. This explicit construction, guaranteed by the Implicit Function Theorem, proves that ``genuine'' minima are structurally stable; the deformation induces a predictable shift of order $\mathcal{O}(\epsilon)$ but preserves the existence and stability index of the vacuum.

\paragraph{Lifting of Flat Directions.}
Second, consider a connected submanifold of degenerate critical points $\mathcal{M} \subset \mathbb{R}^n$ (a flat valley). On $\mathcal{M}$, the unperturbed gradient vanishes identically, $\{\nabla V|_{\mathcal{M}} = 0\}\,$, and the Hessian possesses null eigenvalues along the tangent directions $T\mathcal{M}\,$. The equilibrium condition for the deformed potential requires balancing the forces, $\{\nabla V = -\epsilon \nabla W\}$. Decomposing this onto the tangent space of $\mathcal{M}$, we find
\begin{equation}
    \text{Proj}_{T\mathcal{M}} \big( \nabla V + \epsilon \nabla W \big) = 0 + \epsilon \,\text{Proj}_{T\mathcal{M}} \big( \nabla W \big) = 0\,.
\end{equation}
Thus, a necessary condition for a solution is that the regulating force must vanish along the valley. The system settles at points $\bm{x}^* \in \mathcal{M}$ where the function $W$ is extremal with respect to the flat directions. Effectively, $W$ acts as a \textit{Morse function} on $\mathcal{M}$, resolving the continuum into a discrete set of isolated vacua determined by the topology of $\mathcal{M}\,$.

\paragraph{Implementation and Validity.}
To maintain the physical sign symmetries of the original theory while lifting the moduli, we select an anisotropic quadratic form (instead of a linear form),
\begin{equation}\label{eq: deform}
    W(\bm{x}) = \sum_{i=1}^n c_i |x_i|^2\,,
\end{equation}
where the coefficients $c_i > 0$ are distinct generic real numbers. The anisotropy ($c_i \neq c_j$) is crucial to avoid introducing accidental rotational symmetries that would preserve the flatness \footnote{This choice ensures that all flat directions are lifted to isolated points while preserving the $\mathbb{Z}_2\times \mathbb{Z}_2^3$ equivariant structure of the solution set for the non--BPS D--brane counting. However, there are situations especially in the analysis of the stabilizer submanifolds where it is important to preserve the stabilizer subgroup (products of $U(1)\equiv SO(2)$). This can be easy achieved by choosing the appropriate $c_i$ to be the same.}. Physically, this deformation is equivalent to adding small soft mass terms to the Lagrangian, which lift the expectation values of massless moduli fields. The validity of this choice is ensured by two constraints, 
\begin{enumerate}
    \item \textbf{Asymptotic Stability:} 
    Although the original potential $V(\bm{x})$ is generally quartic (degree 4), it becomes constant (degree 0) along the flat directions. In these valleys, the quadratic regulator $W(\bm{x})$ (degree 2) provides the dominant confining force. This degree hierarchy ensures that the total potential remains coercive ($V_\epsilon \to +\infty$) in all directions, precluding the existence of ``runaway'' solutions or spurious critical points entering from infinity.
    
    \item \textbf{Topological Counting:} The isolated vacua obtained via this procedure allow for the computation of the Euler characteristic of the underlying vacuum (sub--)manifolds via the Poincaré--Hopf theorem, $\sum_{p \in \text{Cr}(W)} (-1)^{\lambda_p} = \chi(\mathcal{M})$, where $\lambda_p$ is the Morse index of the lifted solution \footnote{This usually works for Morse functions $W({\bm x})|_{\mathcal{M}}$ restricted to smooth compact manifold $\mathcal{M}\,$. For non--compact manifolds, it holds provided the gradient field satisfies appropriate boundary conditions at infinity. For example, if the flat direction is a circle ($\mathcal{M} \cong S^1$), the perturbation will generically yield pairs of critical points with alternating indices summing to $\chi(S^1)=0\,$. For zero--dimensional $\mathcal{M}\,$, it trivially works even in scenarios where critical points collide and have multiplicities, as checked for the BPS cases. Other kinds of degeneration are more challenging, see Section \ref{sec:conclusion}.}.
\end{enumerate}
Thus, the set of finite solutions obtained is the complete and faithful representation of the vacuum manifold.

\paragraph{Numerical Protocol.}
In practice, the limit $\epsilon \to 0$ is implemented via an annealing procedure. We perform the minimization of $V_\epsilon(\bm{x})$ for a decreasing sequence of deformation parameters (e.g., $\epsilon_k = 10^{-k}$ for $k=1, \dots, 9$). The solution $\bm{x}_{\epsilon_k}$ from the $k$--th step is used as the initial seed for the $(k+1)$--th step. A numerical vacuum $\bm{x}^*$ is accepted as a valid solution of the original theory if it satisfies two criteria,
\begin{enumerate}
    \item \textbf{Convergence:} The coordinate trajectory stabilizes, satisfying $|\bm{x}_{\epsilon_k} - \bm{x}_{\epsilon_{k+1}}| < \delta_{\text{tol}}$.
    \item \textbf{Gradient Consistency:} The gradient of the perturbed potential vanishes to the required precision, $|\nabla V_\epsilon(\bm{x}^*)| < 10^{-36}$.
\end{enumerate}
This protocol ensures that the solver does not become trapped in transient numerical artifacts and effectively tracks the Morse trajectory to the true vacuum.

\subsubsection{Gated Soft--Trapping}
\label{sec: gatedtrap}

While the polynomial deformation discussed in the context of Morse--Bott theory is generic, agnostic to prior knowledge of the flat directions and effectively lifts them locally, it introduces a global potential, $W(\bm{x}) \sim |\bm{x}|^2$ that diverges at infinity. In landscapes where one seeks to explore vacua far from the origin, or where multiple scales are attached to the vacua structure, such unbounded regulation can create artificial ``parasitic'' minima where the deformation force balances the physical gradient. With partial prior knowledge of the flat directions, we can do much better and eliminate these global artifacts. We propose a \textit{Gated Soft--Trapping} ansatz for global stability. We partition the coordinates into stiff variables $\bm{x}_S$ and flat moduli $\bm{x}_F$, and define the regulator as a localized Gaussian trap,
\begin{equation}
    W_{\mathrm{gated}}(\bm{x}_S, \bm{x}_F) = M \left( \sum_i d_i|\bm{x}_{F,i}|^2 \right) \exp\left( -\frac{\sum_j c_j|\bm{x}_{S,j}|^2}{\sigma^2} \right)\, ,
\end{equation}
where the coefficients $d_i$ and $c_j$ are generic random numbers. This deformation mimics the behavior of non--perturbative instanton corrections in string compactifications, providing two distinct advantages: 

\begin{enumerate}
    \item \textbf{Asymptotic Decoupling:} In the limit $|\bm{x}_S| \to \infty$, the regulator $W_{\mathrm{gated}} \to 0\,$. Unlike polynomial terms, this ensures that the deformation vanishes in the asymptotic bulk of the vacua moduli space, guaranteeing that any minima discovered at large field values are genuine features of the physical potential $V$.
    
    \item \textbf{Vacua Preservation:} The gradient of the regulator with respect to the stiff sector is proportional to the square of the moduli,
    \begin{equation}
        \nabla_{\bm{x}_S} W_{\mathrm{gated}} = -\frac{2\bm{x}_S}{\sigma^2} W_{\mathrm{gated}} \propto \mathcal{O}(|\bm{x}_F|^2)\,.
    \end{equation}
    As the minimization drives the flat directions to their vacuum expectation value ($\bm{x}_F \to 0$), the ``interference force'' on the stiff sector vanishes quadratically. This ensures that the soft trap does not induce a linear tadpole term, thereby preserving the exact location of the stiff vacuum $\bm{x}_S^*$ without perturbative shifts.
\end{enumerate}
This gated ansatz effectively creates a ``domain of reliability'' defined by the scale $\sigma$. Inside the gate ($|\bm{x}_S| < \sigma$), flat directions are lifted with a mass $\sim M$; outside the gate, the original potential is untouched. In the context of the non--BPS D--brane system, we do have prior knowledge of the coulomb branch of the vacua manifold $\mathcal{M}_{\mathrm{critical}}$ and other stabilizer submanifolds where a group of bifundamentals $Z$ are the stiff variables ${\bm x}_S$ and the corresponding adjoint $\Phi$ directions are the flat moduli $\bm{x}_F\,$ (see Section \ref{sec: landscape}). For example, in the case of $\mathcal{M}_{\mathrm{critical}}\,$, this deformation only gets activated when all $Z$ variables tend to zero and roll all otherwise flat $\Phi$ to zero.

\subsubsection{Vacuum Landscape of non--BPS D--brane system}
\label{sec: landscape}

With these tools, we now study the low--energy vacuum landscape of the Abelian non--BPS D--brane system. We make the following observations:
\begin{enumerate}
    \item Although the potential $V$ in \eqref{eq: nonbpsscalar} is a sum of squares, the individual squared expressions involve relative minus signs among their constituent monomials. This hints at the fact that for generic moduli parameters $c\,$, the value of the energy at the non--compact, unbounded Coulomb branch $\mathcal{M}_{\mathrm{critical}}$ Eq.~\eqref{eq: critical} where all the hypermultiplets $Z$ vanish should be parametrically higher than local isolated minima which utilizes the relative signs to lower its energy and form bound states. Note that the adjoints $\Phi$, control the separation between the different brane stacks. Specific expectation values for $\Phi$ form bound states of D--branes at these isolated local minima, and if they remain unbounded, the brane stacks are free to move out to infinity (or maximal separation on the compact tori) without any cost to energy. Even with isolated vacua, it is not clear what constitutes a ``band" of low--lying energy states that we can attribute to a ``configurational entropy" \cite{Mondal:2024qyn}.  One way to possibly characterize it is by ensuring that the ratio of energies, $\mathrm{E}_{\mathrm{isolated}}/\mathrm{E}_{\mathcal{M}_{\mathrm{critical}}}$ is small with a tight spread.   

    \item The Coulomb branch $\mathcal{M}_\mathrm{critical}$ above labels the limit in which all four D--brane stacks can move away from each other at no energy cost (unbounded). We should also expect to have other marginally bound non--compact submanifolds where any combination of the branes can move away from the bound states formed by the rest of the brane stacks at no energy cost. It can be checked that by switching off sets with a critical number of bifundamental fields $Z$, the corresponding adjoints $\Phi$ and gauge fields $X$ drop off the gradient equations, $\{\nabla V =0\}$ and remain undetermined (flat). For example, if we set all $Z^{(1k)}$ and $Z^{(k1)}$ for $k=2,3,4$ to zero, the directions $\left(\Phi_3^{(1)}-\Phi_3^{(2)}\right)\,$,
$\left(\Phi_2^{(3)}-\Phi_2^{(1)}\right)$ and 
$\left(\Phi_1^{(1)}-\Phi_1^{(4)}\right)$ are flat, so are the directions $\left(\vec{X}^{(1)}-\vec{X}^{(2)}\right)\,$, $\left(\vec{X}^{(1)}-\vec{X}^{(3)}\right)\,$ and $\left(\vec{X}^{(1)}-\vec{X}^{(4)}\right)\,$. These are \textit{accidental flat directions}, and the $1^{\mathrm{st}}$ brane effectively decouples from the rest of the three bounded D--brane stacks. This is a stabilizer vacua submanifold $\mathcal{M}_{(1)}$, invariant under the action of the $1^{\mathrm{st}}$ $U(1)$ group. Here, except for the pairs involving the $1^{\mathrm{st}}$ brane, the rest of the configuration can lower its energy efficiently, so it is expected that, energy wise it should sit above the isolated vacua but below $\mathcal{M}_\mathrm{critical}\,$. Similarly, in situations where two brane stacks decouples, the configuration energy should be higher than a single brane stack becoming marginal. 

    \item As the parent potential before gauge fixing has the $U(1)^4$ and the shift symmetries \eqref{eflatgen}, there will always be flat directions associated with these symmetries. Of these, the diagonal $U(1)_{\mathrm{diag}}$ and the shift symmetries are associated with the overall centre of mass motion of the D--brane system and are quotiented out for studying the collection of bound states. This still leaves us three flat directions due to the relative $U(1)^3$ gauge symmetries. In a sense, by making the gauge choices \eqref{eq: nonbpsgauge}, we have \textit{spontaneously broken} the global gauge symmetries, though in quantum mechanics the system can always tunnel through the barrier. Further, we have the $\mathbb{Z}_2^3$ residual gauge symmetry \eqref{eq: dgauge} and a $\mathbb{Z}_2$ parity \eqref{eq: parity}. Thus, while searching for isolated bound states and populating the low--energy vacuum landscape, we should expect at least a 16--fold degeneracy in the solution space of solutions for the local minima, with groups of 8 solutions identified due to the residual gauge symmetry. 

    \item There are instances where the above gauge fixing is not appropriate. Say, we gauge fix the $U(1)$ on $1^{\mathrm{st}}$ brane by $\Im Z^{(12)}=0\,$, but the solver lands to a potential minima where $\Re Z^{(12)}=0\,$. The \textit{gauge orbit} action on $Z^{(12)}$ has collapsed, signaling bad gauge choice. We should then move to the next gauge choice, $\Im Z^{(21)}=0\,$. If, it so happens that $Z^{(21)}=0\,$, then the directions $\left(\Phi_3^{(1)}-\Phi_3^{(2)}\right)\,$ and $\left(X_i^{(1)}-X_i^{(2)}\right)\,$ are flat and again we land on a non--compact submanifold with slightly higher energy than the isolated vacua. The gauge can be fixed by the remaining available $\Im Z^{(\ell k)}=0$ restrictions, where either $\ell$ or $k$ index is $1\,$, with the assumption that their real counterpart don't vanish, otherwise more flat directions will open up. Though these are possibilities, they don't occur for this particular D--brane system; only valid marginally bound cases are where all $Z^{(1k)}$ and $Z^{(k1)}$ for $k=2,3,4$ are set to zero and not a subset of them. Then the potential vacua genuinely preserves the $U(1)$ stabilizer subgroup and the stabilizer submanifold sits at a higher energy.

    \item From the potential, its submanifold and isolated vacua, and the Hessian analysis detailed in Appendix \ref{app:gauge_consistency}, we expect a multitude of cascading transitions when the system tries to lower its energy by rolling down from an excited metastable configuration. Suppose the system is in $\mathcal{M}_{\mathrm{critical}}\,$. The Hessian of the bifundamentals $Z$ is block diagonal in each brane stack pair $(k,\ell)\,$. Lets, focus on the pair $Z^{(12)}$ and $Z^{(21)}$. The pair is further block diagonal in the pairs $\left(\Re Z^{(12)}\,,\Re Z^{(21)}\right)$ and $\left(\Im Z^{(12)}\,,\Im Z^{(21)}\right)$ with degenerate eigenvalues,
    \begin{equation}
\lambda_{ \pm}=\frac{1}{2}\left(\left|\vec{X}^{(1)}-\vec{X}^{(2)}\right|^2 +2\left |\Phi_3^{(1)}-\Phi_3^{(2)}\right|^2 \pm \sqrt{\left(c^{(1)}-c^{(2)}\right)^2+\left|c'^{(1 2)}+c''^{(12)}\right|^2}\right) \,,
\end{equation}
where the associated $X$ and $\Phi$ fields only appear in this block and are  themselves accidental flat directions. Given a set of values for the moduli parameters $c$, we notice that at large values of the free fields, $\mathcal{M}_{\mathrm{critical}}$ is marginally stable, but as we bring the brane stacks closer, i.e. lower these values, at first two new accidental flat direction opens up in the $\left(\Re Z^{(12)}\,,\Re Z^{(21)}\right)$ and $\left(\Im Z^{(12)}\,,\Im Z^{(21)}\right)$ planes as the eigenvalue $\lambda_{-}$ goes to zero followed by the emergence of two unstable directions along the same planes \cite{Walcher:2004tx} \footnote{\label{foot: flatdir} Formally, the simultaneous vanishing of two eigenvalues identifies the vacuum as a non--Morse critical point of corank 2 (see Point 6 and Appendix \ref{app:Stability}). In the Thom--Arnol'd classification, the generic unfolding for two state variables is the \textit{Elliptic} or \textit{Hyperbolic Umbilic} ($D_4^{\pm}$), which necessitates cubic terms in the potential \textit{germ} (of the form $x^3 - 3xy^2$ or $x^3 + y^3$). However, restricting our analysis to the sub--sector spanned by $\left(\Re Z^{(12)}\,,\Re Z^{(21)}\right)$ and $\left(\Im Z^{(12)}\,,\Im Z^{(21)}\right)$, and identifying the flat directions $\left|\vec{X}^{(1)}-\vec{X}^{(2)}\right|^2 +2\left |\Phi_3^{(1)}-\Phi_3^{(2)}\right|^2$ as dynamical control parameters, the bifurcation is governed by a single $U(1)$--charged complex eigenvector $\rho\,$, associated with the degenerate $\lambda_{-}$ eigenvalue. The inherent $O(2)$ equivariance of this sector forces the effective potential to be a function of the invariant $|\rho|^2$, eliminating the cubic terms requisite for $D_4\,$. Consequently, the singularity is non--generic; the dynamics reduce to the codimension--1 unfolding of the \textit{symmetric Cusp catastrophe} ($A_{+3}$), with the normal form $V_{\text{eff}}(\rho; \lambda) = |\rho|^4 - \lambda_{-} |\rho|^2$. This describes a \textit{supercritical pitchfork bifurcation} of revolution.}. 

At this point we note that for a pair of branes stacks $(k\, ,\ell)$ unbounded from the other bounded pair, the Hessian is again block diagonal in accordance with the stabilizer subgroups. If we focus on the stability along the directions, $\left(Z^{(12)}\,,Z^{(21)}\right)$, we notice that for the unbounded brane pair $(3\,,4)\,$, the  fields, $\left(\vec{X}^{(1)}-\vec{X}^{(2)}\right) =0$ and $\left(\Phi_3^{(1)}-\Phi_3^{(2)}\right)$ get fixed by the stable bounded pair $(1\, ,2)\,$. Similarly, for degeneration pairs $\left\{(1\,,2)\,,(1\,,3)\,,(1\,,4)\,,(2\,,3)\,,(2\,,4)\right\}$, the fields $\left(\vec{X}^{(1)}-\vec{X}^{(2)}\right)$ and $\left(\Phi_3^{(1)}-\Phi_3^{(2)}\right)$ are flat and can be tuned to stabilize directions $\left(Z^{(12)}\,, Z^{(21)}\right)$ except for $(1\,,2)$ where it will remain unstable for the same reason as in the case of  $\mathcal{M}_{\mathrm{critical}}\,$. So, as we lower the values of fields $\left(\vec{X}^{(1)}-\vec{X}^{(2)}\right)$ and $\left(\Phi_3^{(1)}-\Phi_3^{(2)}\right)\,$ the system is expected to roll to either of the lower energy marginally stable stabilzer vacua submanifolds, $\left\{\mathcal{M}_{(1,3)}\,,\mathcal{M}_{(1,4)}\,,\mathcal{M}_{(2,3)}\,,\mathcal{M}_{(2,4)}\right\}$ or $\mathcal{M}_{(3,4)}\,$. If it is the former,  lowering the values further will push the system towards even lower energy stabilizer submanifolds $\left\{\mathcal{M}_{(1)}\,,\mathcal{M}_{(2)}\right\}$ where the fields are flat and tunable or $\left\{\mathcal{M}_{(3)}\,,\mathcal{M}_{(4)}\right\}$ where the fields are fixed by stability along the directions $\left(Z^{(12)}\,, Z^{(21)}\right)\,$. If the system happens to land on either of $\left\{(1)\,,(2)\right\}$ degeneration,  further lowering the values will roll the system to any of the isolated stable vacua at even lower energies. These are the options for the quantum system/particle, as almost all such transitions would require crossing an energy barrier by tunneling between vacua. For a classical particle, however, a local \textit{bifurcation} analysis using the Splitting Lemma (see Footnote \ref{foot: flatdir} and Appendix \ref{app:Stability}) indicates that there is only one path along the flat direction valley, leading directly to $\mathcal{M}_{(3,4)}\,$ where $\left(\vec{X}^{(1)}-\vec{X}^{(2)}\right)=0$ and $\left (\Phi_3^{(1)}-\Phi_3^{(2)}\right)=0\,$. Nevertheless, this is merely one instance of a rather large number of possible cascade chains, where the selection of the decay channel depends entirely on the values of the moduli parameters $c$ and the relative energy differences.

\item So far, we have laid down our expectations for a particular set of moduli parameter values. Except for the hope that the low--lying stable isolated vacua count for the non--BPS case should match the corresponding BPS count, we don't have any other convincing argument for the counts to remain the same under moduli deformations. On the contrary, from studies in generic complex dynamical systems \cite{Strogatz:2018,Sen1999NonBPS,Bergman1999Stable}, we should expect stability jumps (phase transitions) as we move through the \textit{control parameter} space. This possibility is explored in some detail in Appendix \ref{app:Stability}. 
    
    \item  In setting up the numerics to enumerate the isolated vacua and characterize these marginally stable stabilizer submanifolds, the overall scale of the moduli parameters $c$ is not important. Except for the flat directions, scaling all $c\rightarrow \lambda c\,$, scales all fields $Z\rightarrow \sqrt{\lambda} Z$ and $\Phi \rightarrow \sqrt{\lambda} \Phi$ for a valid vacuum solution. For numerical stability it is best to pick $\mathcal{O}(1)$ values for $c$ which should result in $\mathcal{O}(1)$ solutions in $Z$ and $\Phi\,$. It also means that for a given set of values for the moduli parameters $c$, the characteristics of the vacuum manifold remain the same for the values $\lambda c\,$. However, the numerics for the stabilizer submanifolds should be handled with great care as detailed in Appendix \ref{app:gauge_consistency}. Conjugacy classes of moduli parameter values that are not related by scaling can exhibit bifurcations in the vacuum manifold structure. An obvious degeneration limit is when all $c\rightarrow0$ and the full vacuum landscape collapses to zero energy.
      
\end{enumerate}
The last point in particular makes it clear that the optimal way forward is to use a combination of the physics inspired  \textit{Morse Deformations} and the \textit{Gated Soft--Trapping} ideas applied to the potential V as in \eqref{eq: nonbpsscalar}, with combinations of bifundamentals $Z$ identified with the stiff variables $\bm{x}_S$ and the appropriate combinations of the adjoints with the flat moduli $\bm{x}_F\,$. We fix some generic numerical values for the moduli parameters $c$ and perform an \textit{adaptive Monte--Carlo} to seed second order Newton--Raphson minimizers (like available in Mathematica \cite{Mathematica}) for local extremization around the seeds in a finite dimensional space spanned by appropriate members of the  $Z$ and $\Phi$ variables as dictated by the cases mentioned in Appendix \ref{app:gauge_consistency}. We have been reasonably careful with error analysis and adjusted various numerical tolerances at each step based on the loss of significant precision digits as numbers flow through the code. For example, we made sure to differentiate between genuine local (marginal--)minima and spurious local minima arising from our lifting of the flat directions. Switching on the Morse deformations and running the annealing process with a cutoff at $\mathcal{O}(10^{-9})$ lets the solver converge faster to a marginally stable vacuum in the stabilizer submanifold cases maintaining high precision. Once, the flat directions are identified, the validity of a possible (marginally--)stable vacuum is checked by vanishing of the gradients of the unperturbed potential $\left\{\nabla V\approx 10^{-36}\right\}$ (which makes the $Z$ and $\Phi$ values correct up to 12 digits of precision), followed by appropriately precise computation of Hessian eigenvalues to certify stability. Finally, we observe that the symmetry constraints defining the stabilizer submanifolds $\mathcal{M}_{(i,j)}$ and $\mathcal{M}_{\mathrm{critical}}$ drastically reduce the effective dimensionality of the configuration space. This simplification renders the system amenable to exact analytical solutions, providing a rigorous benchmark for the numerical results.

\subsubsection*{Results}
We have performed the numerical computations for several sets of moduli parameters, including real and complex $c^{\prime(k\ell)}$ and $c^{\prime\prime(k\ell)}$, as well as real $c^{(k)}$, and the conclusions largely agree. There are 3 doubly degenerate, isolated vacuum solutions that are reasonably deep in the potential landscape and very easy to find, even with low precision, across all sets of moduli parameters we tested. From here onward, the results change a bit. We find that the number of isolated stable local minima (vacua) is 6 (doubly degenerate) for almost all sets of moduli, except for some, where the number fluctuates a bit. This is also true for the stabilizer submanifolds. As discussed earlier in the current non--BPS case, we believe the system should behave like any other general complex system, and typically they exhibit stability jumps across \textit{bifurcation walls} partitioning the \textit{control parameter} (moduli) space. It is also entirely possible that even after maintaining at least 12 digits of precision, we are missing or adding ``spurious" stable local minima. Except for doing some very preliminary study for stabilizer submanifolds where most bifundamentals are put to zero, we have not ventured into studying these jumps systematically in this paper. Even though, for a large system size  like ours, it may seem impossible, in Appendix \ref{app:Stability} we presented a summary of the framework and argued that we often don't need to look at the full system to perform a stability analysis against moduli deformations. We leave the analysis for the future, if and when adequate computational resources become available to us.   

For now, we present the results for a choice of moduli parameters that yield 6 doubly degenerate, isolated, stable minima. We show only 4 decimal places, but the solutions have at least 12 digits of precision. The moduli parameters are,
\begin{align}
c^{(1)} &= 2, \quad
c^{(2)} = 4, \quad
c^{(3)} = 6, \quad
c^{(4)} = -12\,,
\end{align}
satisfying the constraint $\displaystyle \sum_{k=1}^{4} c^{(k)} = 0$, together with,
\begin{align}
c^{\prime(12)} &= \frac{2}{3}, \quad
c^{\prime(13)} = \frac{3}{5}, \quad
c^{\prime(14)} = \frac{5}{7}, \quad
c^{\prime(23)} = \frac{7}{11}, \quad
c^{\prime(24)} = \frac{11}{13}, \quad
c^{\prime(34)} = \frac{13}{17}, \\
c^{\prime\prime(12)} &= \frac{2}{3}, \quad
c^{\prime\prime(13)} = \frac{3}{5}, \quad
c^{\prime\prime(14)} = \frac{5}{7}, \quad
c^{\prime\prime(23)} = \frac{7}{11}, \quad
c^{\prime\prime(24)} = \frac{11}{13}, \quad
c^{\prime\prime(34)} = \frac{13}{17}\,,
\end{align}
as the $c^{(k\ell)}$ always enter the potential as sums, $c'^{(k\ell)}+c''^{(k\ell)}\,$. The doubly degenerate global minimum of the potential is at $2.5880\,$ and the upper bound on the energy landscape is the value of the potential for the Coulomb branch  $\mathcal{M}_\mathrm{critical}$ at $150.5033\,$, which is parametrically above the global minima value. 

The vacuum structure is summarized in Table \ref{tab:vacuum-summary}. Under the Morse regularization, the critical points are neither lost nor mislabeled; only the flat direction, if any, is localized. We have kept track of the flat directions, and the first 6 entries are the isolated local minima. Note that for all entries, the solver was able to find 16 degenerate solutions, but even if we have just found one solution, the 16 different degenerate solutions can be constructed from the $\mathbb{Z}_2^4$ discrete transformations \eqref{eq: dgauge} and \eqref{eq: parity}. We have also checked that after modding out by the 8--fold discrete gauge identifications, two genuine $\mathbb{Z}_2$ solutions remain for each entry and hence, we have 6 doubly degenerate stable vacua \footnote{In the corresponding abelian BPS system, one may also specialize to real values of the moduli parameters $c^{(k\ell)}$, so that the holomorphic F--term equations have real coefficients and are therefore also invariant under complex conjugation. The same observation applies more generally to the BPS systems with matrix-valued fields; on a real slice of the parameter moduli space, the holomorphic equations have real coefficients and are invariant under complex conjugation, so the resulting SUSY vacua are compatible with the discrete $\mathbb{Z}_2 \times \mathbb{Z}_2$ symmetry generated by parity and complex conjugation.}. As locally each pair is like a double--well potential, it is conceivable that instanton corrections will introduce a small split and the ground state will become unique. There are solutions at higher energies above the highest stable isolated solution $20.4035$ representing the stabilizer submanifolds of the marginal bound states,  however, the action of the discrete symmetries is truncated and meaningful only on the non--flat directions. In the stabilizer submanifolds $\mathcal{M}_{(i)}\,$, the stabilizer subgroup is $U(1)$ and the discrete gauge symmetries are 4--fold $\mathbb{Z}_2^2\,$. Similarly, in $\mathcal{M}_{(i,j)}$ as most $Z$ and $\Phi$ are zero, the $\mathbb{Z}_2$ parity can be identified with the $\mathbb{Z}_2$ discrete gauge symmetry and results in a unique marginal minima.    

We also observe that the low--lying energy ``band'' exhibits a relative spread of $\left(\frac{\mathrm{E}_{\mathrm{isolated}}}{\mathrm{E}_{\mathcal{M}_{\mathrm{critical}}}}\right) \sim [10^{-2},\,10^{-1}]$. Given this significant width, we remain cautious in interpreting these states as the direct microstate duals of the black hole entropy. As expected, while scanning through multiple sets of moduli parameters, we observed an energy separation between the group of isolated stable solutions and the stabilizer submanifolds. But, within the stabilizer submanifold group, contrary to expectation, sometimes the energy ordering of the marginally bound states shows level crossing ($\mathcal{M}_{(4)}$ in Table \ref{tab:vacuum-summary}), though, always bounded from above by the columb branch  $\mathcal{M}_{\mathrm{critical}}\,$.   It would be useful to determine the theoretical energy barrier separating the isolated and submanifold vacua, and to conduct a detailed, systematic analysis of the moduli dependence. It seems that the entropy of the abelian non--BPS black hole can be approximately attributed to isolated vacua; we wonder whether there are robust arguments guaranteeing this numerical evidence.

\begin{table}[h] 
\centering
\begin{tabular}{|c|c|c|c|c|} 
\hline
\textbf{Potential $(V)$} &  \textbf{Count} & \textbf{Unique} & \textbf{Type}  \\
\hline
2.5880   & 16 & 2 (DD)& I \\
4.7724    & 16 & 2 (DD)& I \\
9.0624    & 16 & 2 (DD)& I  \\
11.2015     & 16 & 2 (DD)& I  \\
13.3367    & 16 & 2 (DD)& I  \\
20.4035      & 16  &2 (DD) & I  \\
33.0195    & 8  & 2 (DD)& $\mathcal{M}_{(1)}$   \\
45.3006    & 8 &  2 (DD) & $\mathcal{M}_{(2)}$  \\
45.6198    & 8  & 2 (DD) & $\mathcal{M}_{(3)}$  \\
60.1469    & 2 & 1 (ND)& $\mathcal{M}_{(1,2)}$  \\
72.8701    & 2 & 1 (ND)& $\mathcal{M}_{(1,3)}$  \\
93.3400    & 2  & 1 (ND)& $\mathcal{M}_{(2,3)}$  \\
127.1530    & 8 & 2 (DD)& $\mathcal{M}_{(4)}$  \\
140.7430    & 2  & 1 (ND)& $\mathcal{M}_{(2,4)}$   \\
142.3921    & 2  & 1 (ND) &$\mathcal{M}_{(3,4)}$   \\
143.0242    & 2  & 1 (ND) & $\mathcal{M}_{(1,4)}$   \\
150.5033    & 1  & 1 (ND) & $\mathcal{M}_{\mathrm{critical}}$  \\
\hline
\end{tabular}
\caption{Vacuum manifold summary showing  the energy of the non--BPS potential $V$ \eqref{eq: nonbpsscalar}, degeneracies picked up by the solver and the type of vacuum; isolated (bound states) or stabilizer submanifolds (marginally bound states). The ``Unique" column shows the number of unique local minima, either Doubly Degenerate (DD) or Non--degenerate (ND) in energy. The explicit potential values listed here are obtained via the numerical extremization step (Newton--Raphson) at fixed background moduli parameters, subsequent to the algebraic and topological regularizations of the flat directions.}
\label{tab:vacuum-summary}
\end{table}

\subsubsection*{Instanton Connectivity in High Dimensions}

In the standard semiclassical approximation, the low--energy spectrum of the theory is described by a collection of perturbative quantum bound states, each strictly localized within the harmonic basin of an isolated local minimum. However, the global dynamics of the landscape is governed by the non--perturbative mixing of these states via instanton transitions, with tunneling rates $\Gamma_{ab} \propto e^{-S_E[\bm{x}_{cl}]/\hbar}$. In high--dimensional configuration spaces ($\dim(\mathcal{C}) \gg 1$), standard ``shooting'' methods for solving the Euclidean Euler--Lagrange equations are often rendered intractable by the exponential Lyapunov instability of the inverted potential dynamics. A robust computational alternative is to recast the instanton search as a geometric optimization problem within the loop space of the manifold \cite{E_Ren_VandenEijnden_2002}. By discretizing the trajectory into a chain of $N$ images $\displaystyle \{\bm{y}_k\}_{k=0}^N$ connecting vacua $\bm{x}_a$ and $\bm{x}_b$, one minimizes the discretized geometric action functional $ S_{geo} \approx \sum_{k} |\bm{y}_{k+1} - \bm{y}_k| \sqrt{2(V(\bar{\bm{y}}_k) - E)}$. This approach, conceptually analogous to the String Method or Nudged Elastic Band (NEB) algorithms \cite{Jonsson_Mills_Jacobsen_1998,Henkelman2000NEB}, relaxes an initial trial path directly onto the geodesic of the underlying Jacobi metric. Crucially, this circumvents the necessity of explicitly locating saddle points, allowing the algorithm to automatically capture ``corner--cutting'' effects where the tunneling trajectory bypasses high--curvature regions of the potential to optimize the interplay between barrier height and path length. It will be a worthwhile challenge to implement this for the non--BPS systems, as it should give a refined understanding of the spectrum and whether it can fully account for the proposed microstate counting of the non--BPS black holes entropy \cite{Sen:2007qy,Goldstein:2005hq}.

\section{Conclusion and Future Directions}\label{sec:conclusion}

This paper studies the vacuum structure of 4--charge extremal black holes in Type IIA string theory compactified on $T^{6}$ using the pure D--brane description. The system consists of three stacks of D2--branes wrapping mutually orthogonal two--cycles of $T^{6}$ and a stack of D6--branes (BPS), or anti--D6--branes (non--BPS), wrapping the entire $T^6$. The low--energy dynamics is described by a matrix quantum mechanics obtained by dimensional reduction of a four--dimensional $\mathcal{N}=1$ gauge theory, with adjoint and bifundamental matter fields arising from open strings stretched between the different brane stacks. For the $\tfrac{1}{8}$--BPS configurations, the supersymmetric vacua are obtained by solving the F--term constraints modulo the complexified gauge symmetry and the flat directions (Goldstone modes) associated with broken supersymmetries. For generic values of the background moduli, the resulting SUSY vacuum manifold consists of a finite set of isolated points and can be identified with a zero--dimensional algebraic variety. The degree of this variety computes the $14^{\text{th}}$ helicity trace index $B_{14}$, which captures the microscopic degeneracy and hence the entropy of the corresponding black holes. An important technical tool used in this work is the monodromy method to determine the complete set of isolated solutions to the F--term equations. By embedding the physical system into a parameterized family and exploiting the monodromy action on the solution space, all branches of solutions can be generated starting from a single seed configuration. Particular care is taken in fixing the complexified gauge symmetry and the shift symmetries of the adjoint fields, a subtlety that becomes essential for higher--rank non--abelian charge configurations. Using this approach, the supersymmetric vacua are computed explicitly for the $(1,1,1,5)$ and $(1,1,1,6)$ 4-charge pure D--brane systems. In both cases, the number of solutions matches precisely the degeneracies predicted by the U--dual D1--D5--P--KK monopole description, extending earlier results to higher charges.

The paper also analyzes abelian 4--charge non--BPS configurations where the bosonic field content and the gauge symmetries are identical to those of the BPS case, but supersymmetry is completely broken. The non--BPS theory assigns different $\mathcal{N}=1$ subalgebras to different brane triplets via appropriate R--symmetry rotations, leading to modified superpotential couplings. The vacuum manifold of the non--BPS system is analyzed by first examining the possibility of zero--energy configurations. Unlike the BPS case, the gauge symmetries do not admit a complexified extension, and the F--terms are not holomorphic, both of which pose technical challenges for applying algebraic geometric techniques suited to analytic domains. Efficient use of analytical Gr\"obner basis techniques reveals the absence of zero energy microstate for the abelian non--BPS extremal black holes. The numerical study of the vacuum landscape reveals an interesting structure. For generic values of the moduli parameters, the potential admits 12 isolated vacua, organized into 6 pairs of degenerate local minima corresponding to bound states in the non--BPS D--brane system. Assuming that instanton corrections lift the degeneracy, the system admits a unique ground state at non--zero energy, and the resulting twelve stable local minima are expected to account for the microscopic degeneracy underlying the black hole entropy. But, at higher energies, there are non--compact stabilizer submanifolds representing various marginally stable bound state configurations of the D-brane system. 

 In the absence of supersymmetric protection against moduli deformations, the stability of these non--BPS configurations is expected to change discontinuously across bifurcation walls in moduli space. The macroscopic moduli--independent attractor entropy should be understood as a \textit{thermodynamic envelope} valid only in the large--charge limit where the supergravity solution is stable. Our analysis of the abelian non--BPS system is far away from this limit, and the vacuum landscape we find is a collection of semiclassical microstates. The attractor mechanism ``coarse--grains'' over these details, predicting the area of the horizon only if it forms. However, the \textit{microscopic realizability} of this horizon depends critically on the stability of the underlying D--brane constituents. Furthermore, even if a stable horizon forms, its entropy receives significant quantum and higher--derivative corrections. This is consistent with the ``Fake Supergravity'' formalism \cite{Goldstein:2005hq}, where non--BPS attractors are described as critical points of a fake superpotential; such critical points are dynamically fragile, and their stability is naturally conditional on the moduli parameters of the potential.

Overall, our results demonstrate that both BPS and non--BPS 4--charge black holes can be analyzed within a unified algebraic framework based on a pure D--brane description. In the BPS case, supersymmetry ensures a finite set of isolated vacua whose cardinality is captured by an index. By contrast, the absence of supersymmetry leads to a qualitatively different vacuum structure; the non--BPS system admits no zero--energy configurations, but does possess a doubly degenerate global minimum and hence a well--defined ground state after instanton corrections. The methods developed here provide a concrete and systematically improvable approach to studying such systems directly at the level of microscopic D--brane dynamics, and may be applied more broadly to related problems, including the ongoing search for de Sitter vacua in string compactifications \cite{McAllister:2024lnt,Banlaki:2018ayh,McAllister:2025qwq,Wrase:2010ew,Kachru:2003aw}.

\subsection*{Future Directions}
The analysis presented in this paper opens up several directions for further investigation.

\paragraph{Extension to higher--charge and non--abelian configurations.}While we have explicitly analyzed the $(1,1,1,5)$ and $(1,1,1,6)$ systems in the BPS sector and focused on the abelian version of the non--BPS theory, it would be interesting to extend the present methods to higher charge configurations like $(1,1,N_3,N_4)$, $(1,N_2,N_3,N_4)$ or $(N_1,N_2,N_3,N_4)$ for BPS and non--BPS systems. It will be good to have a generating function in the pure D--brane picture.

\paragraph{Resolution of the $(1,1,2,3)$ system.}
An important open problem concerns the microscopic counting of the BPS $(1,1,2,3)$ charge configuration, which is expected -- by virtue of S--duality -- to have the same helicity trace index as the $(1,1,1,6)$ charge configuration. We have attempted to analyze this system using the monodromy method within the pure D--brane framework. Interestingly, the resulting count appears to organize itself as $
3584 \;=\; 5616 - 2032 \, $,
where $5616$ and $2032$ are precisely the numbers of supersymmetric vacua obtained for the $(1,1,1,6)$ and $(1,1,1,5)$ systems, respectively. While this structure suggests a nontrivial relation between these charge sectors, it does not reproduce the expected U--dual result of $5616$ for the $(1,1,2,3)$ configuration. Understanding the origin of this discrepancy, and whether it reflects a subtlety in the pure D--brane description, the gauge fixing, or the Monodromy method, remains an important open question. Resolving this issue would provide a stringent test of the present microscopic framework and its consistency with U--duality. Maybe the resolution lies in ways the Monodromy can fail as discussed in Appendix \ref{app:master_model}.

\paragraph{Degenerate limits.}
Our analysis has been restricted to generic values of the background moduli, for which the SUSY vacua are isolated. Although SUSY protects against stability jumps under generic moduli deformations, special degeneration limits -- where some background parameters vanish -- can lead to qualitatively different behavior; the vacuum space may acquire higher--dimensional components, isolated solutions can migrate to infinity, or collide and develop singular loci. A sensible way forward is to  \textit{projectivize} the affine variety and compute the Topological Euler Characteristics ($\chi$) of the projective variety and the variety at infinity separately and subtract to recover the Euler Characteristic of the original affine variety \cite{cox1}. Unfortunately, it doesn't work for a simple reason. Say, the variety is two isolated points for $x^2-p=0$ and $\chi=2\, $. As we deform $ p\rightarrow 0$ at $p=0$, we get to the variety $x^2=0$, and the roots collide. In both cases, Witten Index should be counted as 2, but the Euler Characteristic doesn't register the ``thickness" of a point, i.e. it sees the \textit{radical} ideal, which is $x$ for $x^2$ and gives $\chi=1\,$. So for singular varieties, $\text{Witten Index} \neq \chi$ unless we systematically keep track of the \textit{Milnor numbers} etc., which is hard for systems of the sizes as ours.

\paragraph{Twined index computation.}
An important possible extension is the computation of the $\mathbb{Z}_2$--twined index $B_6^{\,g}$ within the pure D--brane framework \cite{Sen:2009md}. Implementing the $\mathbb{Z}_2$ symmetry requires restricting to a special locus in moduli space, which in turn necessitates setting four parameters $c^{(k\ell)}$ to zero. As a preliminary step, we analyzed the abelian system with $c^{(13)}=0$ and the gauge choice $Z^{(12)}=Z^{(23)}=Z^{(14)}=1$. A naive analysis yields only four solutions. However, tracking the $c^{(13)}\to 0$ limit or embedding in a projective space to compactify infinity, shows that in eight of the twelve generic solutions the separation between branes $1$ and $3$ diverges such that the combination $c^{(13)}\left(\Phi^{(3)}_2-\Phi^{(1)}_2\right)$ remains finite. After an appropriate field redefinition, the expected twelve solutions are recovered, indicating that the $c^{(13)}\to 0$ limit corresponds to a degeneration rather than a genuine reduction of vacua. An even richer structure emerges when two of the $c$ parameters are taken to vanish. Then the isolated roots escaping to infinity collide at infinity and develop multiplicities. We then run into problems discussed in the context of degeneration limits. When all four $c^{(k\ell)}$ are set to zero, a gauge--invariant Hilbert series computation instead signals the appearance of a four--dimensional vacuum manifold. Developing a systematic treatment of these degeneration limits and extracting the protected twined index $B_6^{\,g}$ therefore, remains an important direction for future work.

\paragraph{Dynamics, stability, and the landscape of non--BPS microstates.}
The rich structure of isolated stationary points and vacuum submanifolds identified in this analysis suggests that the localized quantum states supported by these classical vacua constitute the semiclassical microstates dual to the non--BPS black hole entropy. Validating this conjecture requires mapping the global dynamical flow and stability of the potential landscape across the full moduli space. At the classical level, a systematic study of particle trajectories would reveal the basins of attraction and separatrices that organize the phase space, distinguishing true trapping regions from transient directions. This classical analysis provides the skeleton for a full quantum mechanical treatment. As moduli parameters vary, the ensemble of stable vacua evolves via creation/annihilation bifurcations and stability exchanges. To rigorously compute the thermodynamic entropy $\sim \log \Omega$, one must move beyond the perturbative spectrum to include non--perturbative effects; tunneling between shallow minima will mix the semiclassical basis states, requiring the use of exact WKB, resurgent analysis, and Lefschetz--thimble decompositions to resolve the spectral properties \cite{Witten:2010cx, Behtash2015Toward}. The ultimate goal is to determine if the moduli--dependent count of these possibly dynamically robust quantum states matches the macroscopic entropy expected from the dual supergravity solution, particularly across the walls of stability where the topological structure of the vacuum manifold changes.

\paragraph{Broader applicability of algebraic geometric methods.}
The combination of monodromy techniques, Gröbner basis methods, and Morse--Bott regularization employed in this work is not specific to the pure D--brane system considered here. These tools may provide a systematic framework for analyzing other D--brane configurations, quiver quantum mechanics, and more general questions concerning vacuum structure in string compactifications \cite{Mehta:2012wk}.  Beyond their numerous applications in physics -- where one is often confronted with large polynomial or transcendental systems, or with the problem of extremizing potentials possessing flat directions -- these techniques are universal and have broad applicability across several analytical disciplines in science and engineering \cite{en10.1115/1.1649965,en7367874,chemfaulstich2023algebraic,Gross2015NumericalAG}. One particularly timely direction involves potential applications to loss landscapes in machine learning, which display structural similarities to the complicated potentials with flat directions discussed here \cite{He:2017set}.

\subsection*{Acknowledgments}
We thank Ashoke Sen and Dileep Jatkar for useful discussions regarding some of the open problems discussed in this paper. A.C. acknowledges the Strings 2024 held at CERN Geneva, String--Math 2024 at ICTP, Quantum Information Quantum Field Theory \& Gravity Conference 2024 at ICTS Bengaluru,  Monsoon meet on Gauge Gravity 2025, IISc and School \& Workshop on Number Theory and Physics (co--organized by Abhiram Kidambi) held at ICTP,  Trieste in 2024 \& 2025 for their hospitality and creating stimulating environments where part of the work was completed. A.C. also acknowledges the NSM 2024 held at IIT Ropar and ISM 2025 held at IIT Bhubaneswar for supporting the string community in India and for useful discussions. The work of A.C. is supported by the IIT Bhubaneswar Seed Grant SP--103.  

\subsection*{Note}
While preparing this manuscript, we became aware of related work by Swapnamay Mondal \emph{et al.}, which was submitted to the arXiv around the same time as the first version of this paper \cite{Kumar:2026aca}. There is partial overlap in the analysis of the non--BPS system. The results are overall consistent, though certain aspects of the non--BPS analysis differ.

\appendix

\section{Monodromy: Illustrative Examples}
\label{app: monoexamples}

This appendix is the result of our humble attempt to demystify the inner workings of the Monodromy Method used in Section \ref{sec:monodromy} through a series of illustrative examples. For an example, where it might fail, see Appendix \ref{app:master_model}. 

\subsubsection*{Example 1: The Univariate Quadratic}

\begin{figure}[t]
\centering
\begin{tikzpicture}[font=\small, scale=0.9]

\begin{scope}[xshift=-4cm]
  \draw[->] (-3.5,0) -- (2.5,0) node[right] {$\Re(p)$};
  \draw[->] (0,-2.2) -- (0,2.2) node[above] {$\Im(p)$};
  \filldraw[red] (-1,0) circle (2pt) node[below=4pt, text=red] {$\Delta: p=-1$};
  \coordinate (pstar) at (1.0,0.0);
  \filldraw[black] (pstar) circle (1.5pt) node[] at (1.25,-0.2) {$p^*$};
  \draw[thick, blue, ->, >=stealth] 
    plot [domain=0:360,samples=200] ({-1.0 + 2.0*cos(\x)}, {2.0*sin(\x)});
  \node[blue] at (-0.6,1.5) {Loop $\gamma$};
\end{scope}

\begin{scope}[xshift=4cm]
  \draw[->] (-2.2,0) -- (2.2,0) node[right] {$\Re(x)$};
  \draw[->] (0,-2.2) -- (0,2.2) node[above] {$\Im(x)$};

  \filldraw[black] (1.414,0) circle (1.5pt) node[below right] {$+\sqrt{2}$};
  \filldraw[black] (-1.414,0) circle (1.5pt) node[below left] {$-\sqrt{2}$};

  \draw[thick, red, ->, >=stealth]   plot [domain=0:180]   ({sqrt(2)*cos(\x)}, {sqrt(2)*sin(\x)});
  \node[red]   at (0, 1.7) {$+\sqrt{2} \to -\sqrt{2}$};

  \draw[thick, orange, ->, >=stealth] plot [domain=180:360] ({sqrt(2)*cos(\x)}, {sqrt(2)*sin(\x)});
  \node[orange] at (0, -1.7) {$-\sqrt{2} \to +\sqrt{2}$};
\end{scope}

\end{tikzpicture}
\caption{
Monodromy for the family $x^{2}-1-p=0\,$. 
\textbf{Left:} A loop $\gamma$ in the parameter space based at $p^{*}=1$ encircling the discriminant singularity $p=-1\,$. 
\textbf{Right:} The lifted paths in the solution space. As $p$ completes a full rotation, the two solution sheets exchange, visualizing the action of the fundamental group $\pi_1(U)$ on the fiber.}
\label{fig:monodromy-illustration}
\end{figure}
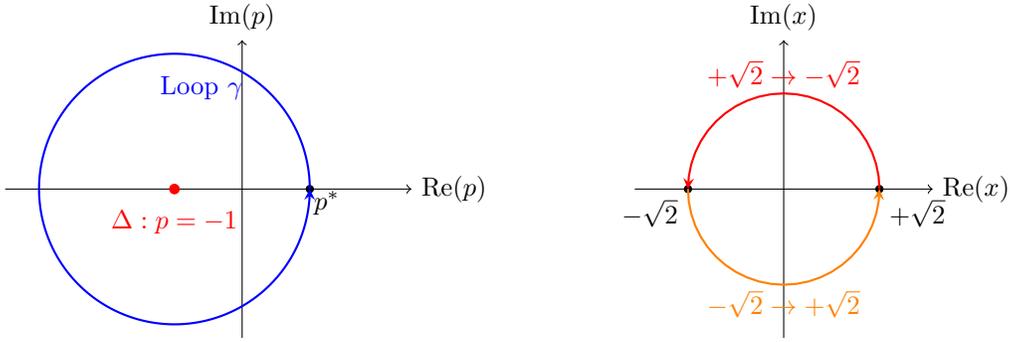

Consider the simplest polynomial equation, $x^{2}-1=0\,$. To analyze this within the monodromy framework, we embed it into the one--parameter deformation family,
\begin{equation}
    F(x;p) = x^{2} - 1 - p = 0\,.
\end{equation}
The target system is recovered at $p=0\,$. For a generic parameter $p$, the fiber $\pi^{-1}(p)$ consists of the two roots $x(p) = \pm\sqrt{1+p}\,$. The discriminant locus $\Delta$ is defined by the vanishing of the derivative $\partial_x F = 2x\,$, which implies $x=0$ and consequently $p=-1\,$, i.e. $\Delta = \{-1\}\,$. We select a generic base point $p^{*}=1\,$, where the fiber consists of the distinct seeds $\{ +\sqrt{2}, -\sqrt{2} \}$ and  construct a closed loop $\gamma$ in the parameter space starting at $p^*$  encircling the singularity at $p=-1\,$,
\begin{equation}
    p(\tau) = -1 + 2\, e^{2\pi i \tau}, \qquad \tau\in[0,1]\, .
\end{equation}
As $p(\tau)$ traverses this loop, we track the solutions by analytic continuation. The function $\sqrt{1+p}$ is multi--valued; a full rotation of $2\pi$ around the branch point adds a phase of $e^{i\pi} = -1\,$. Consequently, the path beginning at $+\sqrt{2}$ evolves continuously into $-\sqrt{2}$, and vice versa (see Figure \ref{fig:monodromy-illustration}). This loop induces the transposition $(1\,2)$ in the symmetric group $S_2\,$, generating the full monodromy group. Note that the sum of the solutions (the trace) remains $x_1(\tau) + x_2(\tau) = 0$ for all $\tau$, which is trivially linear (constant), satisfying the trace test. Finally, a parameter homotopy from $p^{*}$ back to $p=0$ deforms the seeds to the target solutions $\pm 1\,$.

\subsubsection*{Example 2: A Multivariate System}
\label{sec:example2}

To demonstrate the monodromy method in a multivariable setting, we study the system,
\begin{equation}
\label{eq:toy-system-detailed}
\begin{aligned}
x^{2}+y^{2}-1 &= p_{1}\,,\\
x^{2}-y &= p_{2}\,,
\end{aligned}
\end{equation}
viewed as a family $F(x,y;p)=0$ with parameters $p=(p_{1},p_{2}) \in \C^2$. This example is sufficiently simple to allow exact algebraic manipulations while still exhibiting the features relevant to large systems; a non--trivial discriminant locus, multiple isolated solutions, branch exchange under monodromy, and the need for numerical path--tracking.

\paragraph{Exact solutions at the physical point.}
At the physical parameters $p=(0,0)$ the system reduces to
\begin{equation}
\begin{cases}
x^{2}+y^{2}=1\,,\\
y=x^{2}\,,
\end{cases}
\end{equation}
which implies $x^{2}+x^{4}=1\,$. Setting $t=x^{2}$ yields the quadratic $t^{2}+t-1=0$ with roots $\displaystyle t_{\pm}=\frac{-1\pm\sqrt{5}}{2}\,$.
The four isolated solutions forming the fiber $\pi^{-1}(0,0)$ are,
\begin{equation}
\label{eq:explicit-solutions}
\begin{aligned}
(x_{1},y_{1}) &= \bigl(+\sqrt{t_{+}}\;,\ t_{+}\bigr)\approx (+0.786,\ 0.618)\,,\\
(x_{2},y_{2}) &= \bigl(-\sqrt{t_{+}}\;,\ t_{+}\bigr)\approx (-0.786,\ 0.618)\,,\\
(x_{3},y_{3}) &= \bigl(+ i\sqrt{|t_{-}|}\;,\ t_{-}\bigr)\approx (+1.272\,i,\ -1.618)\,,\\
(x_{4},y_{4}) &= \bigl(- i\sqrt{|t_{-}|}\;,\ t_{-}\bigr)\approx (-1.272\,i,\ -1.618)\,.
\end{aligned}
\end{equation}

\paragraph{Discriminant locus.}
The discriminant locus $\Delta\subset\mathbb{C}^{2}$ consists of parameter values where the system becomes singular. The Jacobian matrix with respect to $(x,y)$ is
\begin{equation}
J(x,y)=\begin{pmatrix}2x & 2y\\[2pt] 2x & -1\end{pmatrix},
\qquad
\det J = -2x(1+2y)\,.
\end{equation}
The condition $\det J = 0$ yields two irreducible components of the discriminant:
\begin{enumerate}
  \item \textbf{Case $x=0$ :} Substituting into \eqref{eq:toy-system-detailed} implies $y=-p_{2}$ and $p_{1}=y^{2}-1 = p_{2}^{2}-1\,$. Thus,
  \begin{equation}
  \Delta_{1}=\{(p_{1},p_{2})\mid p_{1}=p_{2}^{2}-1\},
  \end{equation}
  which defines a parabola.
  \item \textbf{Case $1+2y=0$ :} Here $y=-\tfrac12$, which implies $x^{2}=p_{2}-\tfrac12\,$. Substituting into the first equation yields $p_{1}=p_{2}-\tfrac54$. Thus,
  \begin{equation}
  \Delta_{2}=\{(p_{1},p_{2})\mid p_{1}=p_{2}-\tfrac54\}\,,
  \end{equation}
  which defines a line.
\end{enumerate}
Any loop in parameter space that winds around these components (without intersecting them) may produce non--trivial monodromy.









\begin{figure}[t]
\centering
\begin{tikzpicture}[font=\small,scale=1.1]

\draw[->] (-3.0,0) -- (3.2,0) node[right] {$\Re(p_{1})$};
\draw[->] (0,-2.4) -- (0,3.2) node[above] {$\Re(p_{2})$};
\draw[->, dotted] (0.0,0) -- (3,1) node[right] {$\Im(p_1)$};

\draw[thick, orange, domain=-2:2, samples=240]
      plot ({\x*\x - 1}, {\x});
\node[orange] at (2.5,2.6) {$\Delta_{1}$};

\draw[thick, red, domain=-2:3.0]
      plot ({\x - 1.25}, {\x});
\node[red] at (-2.2,-1.6) {$\Delta_{2}$};

\coordinate (pstar) at (1,0);
\filldraw (pstar) circle (2pt) node[below right] {$p^{*}=(1,0)$};


\coordinate (pstar) at (1,0);
\filldraw (pstar) circle (2pt) node[below right] {$p^{*}=(1,0)$};

\def\R{1.6}
\coordinate (C) at ({1-\R},0); 

\draw[blue, thick, densely dotted]
  (C) circle (\R);

\node[blue] at (2.4,-1.1) {\small Loop $\gamma$ (projection)};





\end{tikzpicture}

\caption{\small
Real--slice projection of the discriminant locus 
$\Delta=\Delta_{1}\cup\Delta_{2}$ for the two–equation system.  
The parabola $\Delta_{1}$ (orange) and line $\Delta_{2}$ (red) mark 
parameters where the Jacobian becomes singular.  
The dotted blue curve shows a monodromy loop $\gamma$ anchored at 
$p^{*}=(1,0)$, which winds around the discriminant locus in the full 
complex plane while avoiding it in this projection.  
Such loops generate nontrivial monodromy among the four solutions 
above $p^{*}$.}
\label{fig:discriminant-twoeq}
\end{figure}
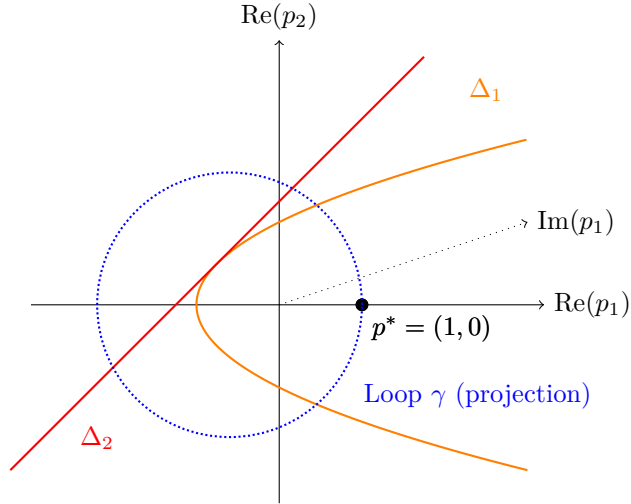

\paragraph{Seed selection and numerical setup.}
We choose a generic base point $p^{*}=(1,0)$ far from $\Delta$. At this point, the system is non--singular. In practice, we require only a single seed solution $s^{(1)} \in \pi^{-1}(p^*)\,$. This can be found by standard Newton iteration on a random initial guess, as the polynomial system is small.

\paragraph{Lifting a loop (Predictor--Corrector).}
Let \(\gamma:[0,1]\to\mathbb{C}^{2}\) be a closed loop with \(\gamma(0)=\gamma(1)=p^{*}\) lying in the complement of \(\Delta\,\). We track the solution along $\gamma$ by solving the homotopy condition $H(x,y,\tau)=F(x,y;\gamma(\tau))=0\,$. This is performed numerically via adaptive steps \footnote{The loop is a closed path in the full complex parameter space $\mathbb{C}^2$. Figure~\ref{fig:discriminant-twoeq} shows its projection onto the real slice; the actual path may detour into $(\Im p_1, \Im p_2)$ to remain in the complement of $\Delta\,$.}:
\begin{enumerate}
  \item \textbf{Predictor:} Use a tangent predictor (e.g., Runge--Kutta) to estimate $(x,y)$ at $\tau+\delta\,$.
  \item \textbf{Corrector:} Apply Newton's method to converge back to the curve $H=0\,$.
  \item \textbf{Adaptation:} Adjust step--size $\delta$ based on convergence speed.
\end{enumerate}
The endpoint of the path provides a solution in the fiber $\pi^{-1}(p^*)\,$.

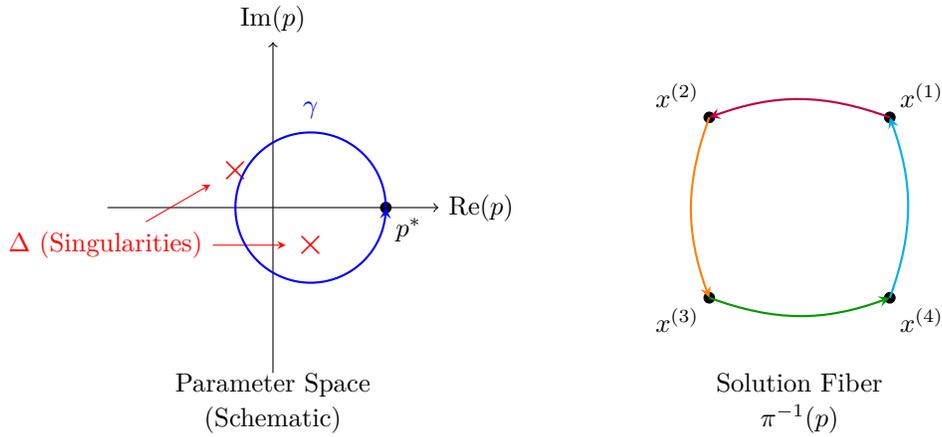
\begin{figure}[t]
\centering
\begin{tikzpicture}[font=\small, scale=1.0]

\begin{scope}[xshift=-3.5cm]
    \draw[->] (-2.2,0) -- (2.2,0) node[right] {$\Re(p)$};
    \draw[->] (0,-2.2) -- (0,2.2) node[above] {$\Im(p)$};
    \node[align=center] at (0,-2.6) {Parameter Space \\ (Schematic)};
    
    \node[red] (sing1) at (-0.5, 0.5) {\Large $\times$};
    \node[red] (sing2) at (0.5, -0.5) {\Large $\times$};
    
    \node[red, anchor=east] (DeltaLabel) at (-0.8, -0.5) {$\Delta$ (Singularities)};
    \draw[->, red, thin, >=stealth] (DeltaLabel) -- (sing1);
    \draw[->, red, thin, >=stealth] (DeltaLabel) -- (sing2);
    
    \coordinate (pstar) at (1.5, 0);
    \filldraw[black] (pstar) circle (2pt) node[below right] {$p^*$};
    
    \draw[thick, blue, ->, >=stealth] 
        plot [domain=0:360, samples=100] ({0.5 + 1.0*cos(\x)}, {1.0*sin(\x)});
    \node[blue] at (0.5, 1.3) {$\gamma$};
\end{scope}

\begin{scope}[xshift=3.5cm]
    \node[align=center] at (0,-2.6) {Solution Fiber \\ $\pi^{-1}(p)$};
    
    \coordinate (S1) at (1.2, 1.2);
    \coordinate (S2) at (-1.2, 1.2);
    \coordinate (S3) at (-1.2, -1.2);
    \coordinate (S4) at (1.2, -1.2);
    
    \filldraw (S1) circle (2pt) node[above right] {$x^{(1)}$};
    \filldraw (S2) circle (2pt) node[above left] {$x^{(2)}$};
    \filldraw (S3) circle (2pt) node[below left] {$x^{(3)}$};
    \filldraw (S4) circle (2pt) node[below right] {$x^{(4)}$};
    
    \draw[thick, purple, ->, >=stealth, bend right=20] (S1) to (S2);
    \draw[thick, orange, ->, >=stealth, bend right=20] (S2) to (S3);
    \draw[thick, green!60!black, ->, >=stealth, bend right=20] (S3) to (S4);
    \draw[thick, cyan, ->, >=stealth, bend right=20] (S4) to (S1);
\end{scope}

\end{tikzpicture}
\caption{Schematic representation of the monodromy action. \textbf{Left:} A loop $\gamma$ in parameter space encircles the discriminant locus $\Delta$ (marked by red crosses). These crosses represent the singularities where solutions collide. \textbf{Right:} Lifting this loop to the solution space induces a permutation of the four solutions (shown as $1 \to 2 \to 3 \to 4 \to 1$).}
\label{fig:monodromy-lift-twoeq}
\end{figure}

\paragraph{Monodromy sweep and stopping criterion.}
Starting with the initial seed set \(S_{1}=\{s^{(1)}\}\), we iteratively
\begin{itemize}
  \item generate a random loop \(\gamma\) in $\C^2$ that avoids $\Delta\,$,
  \item lift \(\gamma\) starting from each known solution $s \in S_k\,$.
  \item collect the endpoints; any new solutions are added to form $S_{k+1}\,$.
\end{itemize}
For irreducible algebraic curves, the monodromy group acts transitively. We terminate when the number of solutions stabilizes or matches the expected Bezout bound \cite{Kushnirenko1976NewtonPA} (in this case 4, see Figure \ref{fig:monodromy-lift-twoeq}).

\paragraph{Homotopy to the physical system.}
Once the full fiber \(S\) at \(p^{*}\) is recovered, we perform a final straight--line homotopy
\begin{equation}
\gamma_{\text{phys}}(t)=(1-t)p^{*}+t(0,0)\,, \quad t\in[0,1]\,.
\end{equation}
Tracking the four solutions along this path yields the target solutions at $p=(0,0)$, matching the explicit values in \eqref{eq:explicit-solutions}.

\subsection*{Example 3: A Transcendental System}

To demonstrate the method’s applicability beyond polynomial systems, we consider the transcendental equation
\begin{equation}
F(x;p) \equiv x + e^{-x} - p = 0 \,,
\end{equation}
which provides a minimal prototype for the competition between perturbative and non--perturbative contributions in semiclassical quantum field theory. Physically, one may interpret $x$ as a vacuum expectation value determined by the extremization of an effective potential, while $p$ plays the role of an external source or coupling. In particular, this equation arises as the saddle--point condition associated with an effective potential of the form,
\begin{equation}
V_{\mathrm{eff}}(x) = \frac{1}{2}x^{2} - e^{-x}\,,
\end{equation}
where the quadratic term represents the leading perturbative contribution, and the exponential term models a non--perturbative instanton effect with unit fugacity. Coupling the system linearly to a source $p$ yields the stationarity condition $\partial V_{\mathrm{eff}}/\partial x = p\,$, which reproduces the transcendental relation above. Equivalently, the same structure emerges from the semiclassical evaluation of a zero--dimensional path integral
\begin{equation}
Z(p) = \int_{\mathcal C} \mathrm{d}y \,
\exp\!\left[-\frac{1}{\hbar}\left(\frac{1}{2}y^{2} - e^{-y}\right)
+ \frac{p}{\hbar}y \right],
\end{equation}
whose dominant saddle points in the limit $\hbar \to 0$ satisfy,
\begin{equation}
\frac{\mathrm{d}}{\mathrm{d}y}\!\left(\frac{1}{2}y^{2} - e^{-y} - py\right)=0\,.
\end{equation}
Identifying the saddle $y=x$ leads precisely to $F(x;p)=0$, making this equation a concrete toy model for saddle competition between perturbative vacua and instanton--induced corrections, and a natural testing ground for \textit{resurgence} and \textit{transseries techniques} \cite{Dunne:2012ae,Aniceto:2018bis}.

\paragraph{Lambert $W$ Function.}
The exact solutions are given by the branches of the Lambert $W$ function \cite{CorlessLambertW}. Rewriting the equation as $(p-x)e^{x} = 1$, the solutions are
\begin{equation}
    x_k(p) = p + W_k(-e^{-p}), \quad k \in \mathbb{Z}\,.
\end{equation}
As $|p| \to \infty$ (weak coupling limit), the asymptotic behavior reveals distinct physical sectors,
\begin{itemize}
    \item \textbf{The Perturbative Vacuum ($k=0$):} For large positive $p$, the principal branch behaves as $x_0 \approx p\,$. This is the classical solution where the exponential term $e^{-x}$ is exponentially suppressed.
    \item \textbf{The Non--Perturbative Sectors ($k \neq 0$):} The other branches are dominated by the exponential term. Asymptotically, $x_k \approx -\ln(p) + 2\pi i k\,$. These correspond to ``instanton" vacua that do not exist in the strict $e^{-x} \to 0$ limit.
\end{itemize}

\paragraph{The Singular Ladder.}
The discriminant locus is defined by the vanishing derivative $\partial_x F = 1 - e^{-x} = 0\,$, implying $x \in 2\pi i \mathbb{Z}\,$. Mapping these critical points to the parameter space yields an infinite vertical ladder of singularities
\begin{equation}
    \Delta = \{ p_n = 1 + 2\pi i n \mid n \in \mathbb{Z} \}\,.
\end{equation}
The most physically significant singularity is at $p_0=1\,$. On the real line $p \in \mathbb{R}\,$, the function $p(x) = x+e^{-x}$ is convex with a global minimum at $(0,1)\,$:
\begin{itemize}
    \item For $p > 1$, the horizontal line $y=p$ intersects the curve $y=p(x)$ twice, yielding two real solutions (the ``perturbative" vacuum $x_{\text{pert}} \approx p$ and the ``instanton" vacuum $x_{\text{np}}  \approx -\ln p$).
    \item At $p=1$, these two real branches collide and merge. It is important to note that the labels $x_{\text{pert}} \approx p$ and $x_{\text{np}} \approx -\ln p$ describe the asymptotic behavior in the weak coupling limit ($p \gg 1$). In the vicinity of the critical point $p=1\,$, these approximations break down. A local expansion reveals that the branches merge not logarithmically, but according to the universal square--root scaling $x \sim \pm \sqrt{2(p-1)}$ characteristic of a \textit{saddle--node bifurcation}.
    \item For $p < 1$, the intersection vanishes on the real line; the solutions move off--axis becoming complex conjugates.
\end{itemize}

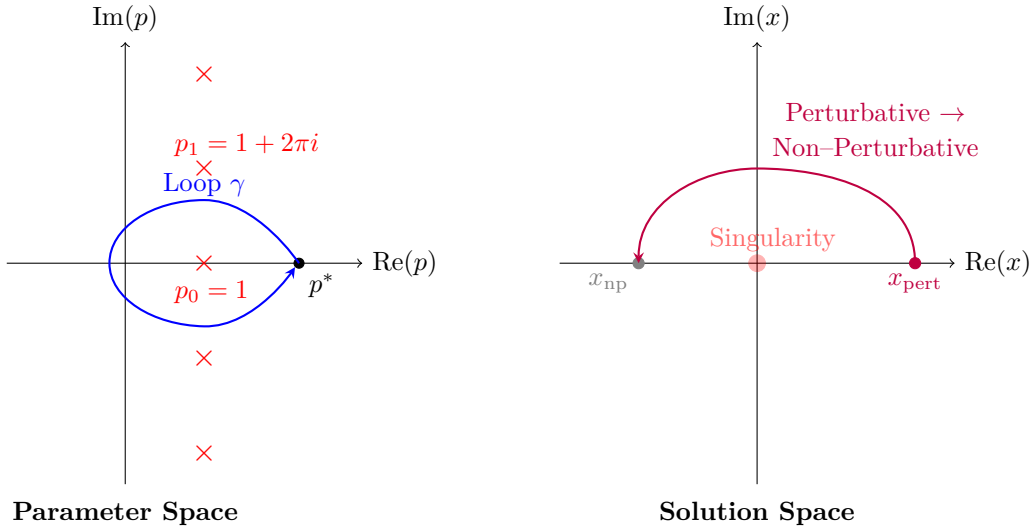
\begin{figure}[t]
\centering
\begin{tikzpicture}[font=\small, scale=1.05]

\begin{scope}[xshift=-4.0cm]
    \draw[->] (-1.5,0) -- (3.0,0) node[right] {$\Re(p)$};
    \draw[->] (0,-2.8) -- (0,2.8) node[above] {$\Im(p)$};
    \node[align=center, anchor=north] at (0,-2.9) {\textbf{Parameter Space}};
    
    \foreach \n in {-2,-1,0,1,2} {
        \node[red] at (1, \n*1.2) {\large $\times$};
    }
    \node[red, anchor=north west] at (0.5, -0.1) {$p_0=1$};
    \node[red, anchor=west] at (0.5, 1.5) {$p_1=1+2\pi i$};
    
    \coordinate (pstar) at (2.2, 0); 
    \filldraw[black] (pstar) circle (1.8pt) node[below right] {$p^*$};
    
    \draw[thick, blue, ->, >=stealth] 
        plot [smooth, tension=1] coordinates {(2.2,0) (1.0, 0.8) (-0.2, 0) (1.0, -0.8) (2.15, -0.05)};
    
    \node[blue] at (1.0, 1.0) {Loop $\gamma$};
\end{scope}

\begin{scope}[xshift=4.0cm]
    \draw[->] (-2.5,0) -- (2.5,0) node[right] {$\Re(x)$};
    \draw[->] (0,-2.8) -- (0,2.8) node[above] {$\Im(x)$};
    \node[align=center, anchor=north] at (0,-2.9) {\textbf{Solution Space}};
    
    \coordinate (Xpert) at (2.0, 0);   
    \coordinate (Xnp) at (-1.5, 0);   
    
    \filldraw[purple] (Xpert) circle (2pt) node[below] {$x_{\text{pert}}$};
    \filldraw[gray] (Xnp) circle (2pt) node[below left] {$x_{\text{np}}$};
    
    \filldraw[red, opacity=0.3] (0,0) circle (3pt);
    \node[red, opacity=0.6] at (0.2, 0.3) {Singularity};

    \draw[thick, purple, ->, >=stealth] 
        (Xpert) to[out=90, in=0] (0, 1.2) to[out=180, in=90] (Xnp);
        
    \node[purple, align=center] at (1.5, 1.7) {Perturbative $\to$ \\ Non--Perturbative};
\end{scope}

\end{tikzpicture}
\caption{Monodromy and Resurgence in the system $x + e^{-x} = p\,$.
\textbf{Left:} The discriminant locus forms an infinite vertical ladder at $\Re(p)=1\,$. The loop $\gamma$ encircles the critical point $p_0=1\,$.
\textbf{Right:} At the real base point $p^* > 1$, two real solutions exist: the perturbative vacuum $x_{\text{pert}} \approx p$ and the non--perturbative vacuum $x_{\text{np}} \approx -\ln p\,$. Following the loop $\gamma$ causes these two distinct physical sectors to exchange, demonstrating that they are branches of a single analytic function.}
\label{fig:transcendental-monodromy}
\end{figure}

\paragraph{Monodromy as Stokes Wall--Crossing.}
We select a base point in the ``physical" region $p^* > 1$ (weak coupling).
If we track the perturbative solution $x_{\text{pert}} \approx p^*$ along a loop $\gamma$ that encircles the primary singularity at $p=1$, the analytic continuation forces the solution to switch branches.
Upon returning to $p^*$, the system does not return to the perturbative vacuum $x_0\,$, but arrives at the first non--perturbative branch $x_{-1}\,$.
This geometric mechanism illustrates \emph{Stokes phenomena} \cite{Stokes1864,Berry1989} -- moving the parameters in the complex plane allows the ``perturbative" physics to mix smoothly with ``non--perturbative" sectors, a phenomenon inaccessible to standard Taylor expansions (see Figure \ref{fig:transcendental-monodromy}).

\section{Monodromy Edge Case: Composite Vacua}
\label{app:master_model}

We highlight possible failure modes of the monodromy method using a single model based on a composite order parameter. This structure captures the behavior of theories with discrete gauge symmetries or orbifold constraints, where the defining equations depend on a composite invariant $u = \Psi^k$ rather than the fundamental field $\Psi$. Consider a theory where the vacuum structure is determined by a quadratic constraint on $u\,$,
\begin{equation}
\label{eq:master_model}
    P(u; p) = u^2 + \alpha(p) u + \beta(p) = 0 \,, \quad \text{with} \quad u = \Psi^k \,.
\end{equation}
Here $p$ represents the coordinates of the moduli space, and $\alpha(p), \beta(p)$ are parameter--dependent coefficients \footnote{We can also think of the $p$ as parameters which capture the deformation of the original quadratic constraint under the monodromy method of solving a system of equations. See Section \ref{sec:monodromy}.}. The total number of vacua is $N=2k\,$. Depending on the specific choice of coefficients, this system exhibits two distinct topological obstructions.

\subsection*{Case I: Reducibility (Disjoint Sectors)}
Consider the specific parameterization where the effective potential factorizes globally over the moduli space. For example, let $\alpha(p) = -3p$ and $\beta(p) = 2p^2$. The master equation becomes,
\begin{equation}
    (u - p)(u - 2p) = 0 \implies (\Psi^k - p)(\Psi^k - 2p) = 0 \,.
\end{equation}
The discriminant of the quadratic in $u$ is a perfect square, $\Delta_u = p^2$. Consequently, the roots $u_1(p) = p$ and $u_2(p) = 2p$ are single--valued functions that never exchange.
The solution variety splits into two disjoint invariant sectors (orbits) corresponding to the two factors.
Because the roots $u_1$ and $u_2$ define distinct scales, no continuous deformation can bridge the gap between them. The monodromy group is \textit{intransitive} ($G \subsetneq S_{2k}$), modeling physical scenarios with \textit{Superselection Sectors}.

\subsection*{Case II: Imprimitivity (Phase Locking)}
Now consider the generic case where $\alpha(p)$ and $\beta(p)$ are chosen such that the quadratic discriminant $\Delta_u$ is not a perfect square (e.g., $\alpha=p, \beta=1$). The roots $u_\pm(p)$ can now be swapped by encircling the locus $\Delta_u = 0\,$. While transitivity is theoretically possible, the monodromy group is \textit{imprimitive}.
Generating the full group requires loops around two topologically distinct loci,
\begin{enumerate}
    \item \textbf{The Cluster Locus ($\Delta_u=0$):} A loop here swaps the values $u_+ \leftrightarrow u_-\,$. In the fiber, this exchanges the \emph{set} of solutions $\{\Psi^k = u_+\}$ with the set $\{\Psi^k = u_-\}$, acting on the clusters as whole blocks.
    \item \textbf{The Origin Locus ($\beta=0$):} This corresponds to a symmetry enhancement point $u=0\,$. A loop here forces $u \to e^{2\pi i} u\,$, inducing the action $\Psi \to e^{2\pi i/k} \Psi\,$.
\end{enumerate}
Standard algorithms often sample loops that encircle $\Delta_u$ but miss $\beta=0\,$. In this failure mode, the solver correctly finds one solution from each cluster but fails to generate the internal $\mathbb{Z}_k$ phase copies.

\begin{figure}[h]
\centering
\begin{tikzpicture}[font=\small, scale=1.0]
\begin{scope}[xshift=-4.5cm]
    \node at (0, 2.5) {\textbf{Case I: Reducible}};
    \node[gray, scale=0.8] at (0, 2.1) {Factorized Coefficients};
    \draw[thick, blue] (0,0) circle (0.8);
    \draw[thick, red] (0,0) circle (1.4);
    \node[blue] at (0, 0.2) {$|\Psi|^k \sim |p|$};
    \node[red] at (0, -1.7) {$|\Psi|^k \sim 2|p|$};
    \node[align=center, scale=0.8] at (0, -2.5) {Orbits are permanently\\disjoint};
\end{scope}
\begin{scope}[xshift=4.5cm]
    \node at (0, 2.5) {\textbf{Case II: Imprimitive}};
    \node[gray, scale=0.8] at (0, 2.1) {Generic Coefficients};
    \draw[gray, dashed] (-1, 0) circle (0.6);
    \foreach \a in {90, 210, 330} \filldraw[black] [shift={(-1,0)}] (\a:0.6) circle (1.5pt);
    \draw[gray, dashed] (1, 0) circle (0.6);
    \foreach \a in {90, 210, 330} \filldraw[black] [shift={(1,0)}] (\a:0.6) circle (1.5pt);
    \draw[thick, blue, <->, >=stealth] (-0.8, 0.8) to[bend left] (0.8, 0.8);
    \node[blue, above, scale=0.8] at (0, 1.1) {Cluster Swap ($\Delta_u$)};
    \draw[thick, green!50!black, ->, >=stealth] (-1.8, 0) arc (180:360:0.8);
    \node[green!50!black, below, scale=0.8, align=center] at (-1, -0.8) {Phase Rotation\\(Requires $\beta=0$)};
\end{scope}
\end{tikzpicture}
\caption{The two topologies of the Master Model \eqref{eq:master_model}. \textbf{Left:} If coefficients factorize, the solution space splits into disjoint sectors (rings) that never mix. \textbf{Right:} In the generic case, solutions group into clusters. Generic loops swap clusters (blue) but fail to rotate phases (green) unless the origin locus is explicitly targeted. Both issues are resolved by the $\epsilon$--deformation.}
\label{fig:master_topology}
\end{figure}
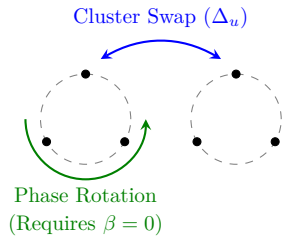
\subsection*{Unified Resolution: Symmetry--Breaking Homotopy}
\label{sub:gamma_trick}

To strictly overcome both obstructions without prior knowledge of the symmetry structure, we employ a \textit{Symmetry--Breaking Deformation} (the ``Gamma Trick''). We embed the physical system $F(x; p) = 0$ into a deformed family,
\begin{equation}
    \mathcal{H}(x; p, \epsilon) = F(x; p) + \epsilon \cdot L(x) = 0 \,,
\end{equation}
where $L(x)$ is a generic linear function and $\epsilon \in \mathbb{C}$ is a small parameter.
For $\epsilon \neq 0\,$, the deformation destroys the composite structure ($u=\Psi^k$) and any factorization. The monodromy group becomes the full symmetric group $S_{2k}\,$, ensuring that a single seed generates the complete fiber. We then perform a final parameter homotopy $\epsilon \to 0$ to recover the physical solutions.

\section{Gröbner Bases}
\label{app:GB}

In this appendix, we provide a short technical exposition of Gröbner basis theory. Our goal is to demystify the ``black box'' nature of the algorithms used in Section \ref{subsec:GB} and to provide the reader with the tools to interpret the output of computer algebra systems like \texttt{Macaulay2} \cite{M2} or \texttt{Singular} \cite{Singular}. While the physical discussion in the main text focused on the application to realified non--BPS systems, the underlying algebraic machinery is universal to polynomial rings over fields. We focus particularly on the behavior of polynomial ideals where the number of generators in the basis exceeds the number of physical constraints -- a phenomenon often confused with redundancy, but which is in fact crucial for the algorithmic solvability of the system.

\subsection{Ideals, Orderings, and the Division Algorithm}

Let $R = \mathbb{C}[x_1, \dots, x_n]$ be the polynomial ring in $n$ variables over the field of complex numbers. The physical constraints of our theory (F--terms and D--terms) define a finite set of polynomials $F = \{f_1, \dots, f_s\} \subset R\,$. The algebraic object of study is the ideal generated by these constraints,
\begin{equation}
    I = \langle f_1, \dots, f_s \rangle = \left\{ \sum_{i=1}^s h_i f_i \; \Big| \; h_i \in R \right\} \,.
\end{equation}
Physically, $I$ contains all algebraic consequences of the static equations of motion. By the Hilbert Basis Theorem \cite{cox1,AtiyahMacdonald1969}, $R$ is a \emph{Noetherian ring}, guaranteeing that every ideal is finitely generated. However, the specific generating set $F$ derived from the potentials like \eqref{eq: scalar} and \eqref{eq: nonbpsscalar} is rarely the most useful one for extracting solutions. To transform $F$ into a canonical form, we must first impose a strict hierarchy on the terms in the polynomial ring.

\subsubsection*{Monomial Orderings}

A \textit{monomial ordering} $\prec$ is a total ordering on the set of monomials $\mathbb{Z}_{\ge 0}^n$ that is \emph{admissible}, meaning it satisfies two axioms:
\begin{enumerate}
    \item \textbf{Well--ordering:} Every non--empty set of monomials has a minimal element. This ensures that any polynomial division algorithm eventually terminates.
    \item \textbf{Multiplicative Compatibility:} If $m_1 \prec m_2$, then $m \cdot m_1 \prec m \cdot m_2$ for any monomial $m$. This ensures the ordering is preserved under polynomial multiplication.
\end{enumerate}
The choice of ordering dictates the algebraic structure of the final basis. We highlight two orderings relevant to our computations:
\begin{itemize}
    \item \textbf{Lexicographic (Lex):} Analogous to dictionary order ($x_1 \succ x_2 \succ \dots \succ x_n$). This ordering is elimination--based; a Gröbner basis in Lex order often takes a triangular form (e.g., the last element depends only on $x_n$), facilitating explicit solution finding. However, the coefficients in a Lex basis can grow doubly exponentially with the number of variables, making it computationally prohibitive for large systems.
    
    \item \textbf{Graded Reverse Lexicographic (Grevlex):} Monomials are compared first by total degree $|\alpha| = \sum \alpha_i\,$. Ties are broken by the \emph{reverse} lexicographic order: $x^\alpha \succ x^\beta$ if the \emph{last} non--zero entry of $\alpha - \beta$ is \textit{negative}. This ordering tends to produce bases with the smallest coefficients and lowest degrees. It is the standard choice for calculating Hilbert series or checking ideal membership (e.g., verifying vacuum inconsistency via $1 \in I$).
\end{itemize}

\subsubsection*{The Initial Ideal}

Given a fixed ordering $\prec\,$, every non--zero polynomial $f \in R$ has a unique \textit{Leading Monomial}, denoted $\mathrm{LM}(f)$. The initial Ideal of $I$, denoted $\mathrm{in}_\prec(I)$, is the monomial ideal generated by the leading terms of \emph{all} polynomials in $I\,$,
\begin{equation}
    \mathrm{in}_\prec(I) = \langle \mathrm{LM}(f) \mid f \in I \setminus \{0\} \rangle \,.
\end{equation}
Crucially, this is generally \emph{much larger} than the ideal generated merely by the leading terms of the initial generators $f_i\,$. A finite set $G = \{g_1, \dots, g_t\} \subset I$ is defined to be a \textit{Gröbner basis} if its leading terms suffice to generate the full initial ideal,
\begin{equation}
    \langle \mathrm{LM}(g_1), \dots, \mathrm{LM}(g_t) \rangle = \mathrm{in}_\prec(I) \,.
\end{equation}
When this condition holds, the multivariate division algorithm becomes well--defined, i.e. division of any polynomial $p \in R$ by $G$ yields a \emph{unique} remainder $r$, independent of the order in which the divisors $g_i$ are applied.

\subsection{Buchberger's Algorithm and Basis Expansion}

A common misconception in the application of algebraic geometry to physics is that the Gröbner basis $G$ is merely a subset or a linear rearrangement of the original equations of motion $F$. In reality, $G$ is often significantly larger than $F$ ($|G| \gg |F|$). This phenomenon, known as \textit{basis expansion}, is not a redundancy but a necessary revelation of the variety's geometry. The original generators $f_i$ may conceal algebraic dependencies (\textit{syzygies}) that prevent the division algorithm from being unique. Buchberger's algorithm makes these dependencies manifest through the construction of \textit{S--polynomials} (Syzygy polynomials), which are designed to cancel the leading terms of pairs of generators \cite{cox1},
\begin{equation}
    S(f_i, f_j) = \frac{L}{\mathrm{LT}(f_i)} \cdot f_i - \frac{L}{\mathrm{LT}(f_j)} \cdot f_j \,, \quad \text{where } L = \mathrm{lcm}(\mathrm{LM}(f_i), \mathrm{LM}(f_j)) \,.
\end{equation}
If the remainder of $S(f_i, f_j)$ upon division by the current basis is non--zero, it represents a new, independent algebraic constraint hidden within the ideal. This remainder is added to the basis, and the process iterates until all S--polynomials reduce to zero. We now demonstrate this with two explicit examples.

\subsubsection*{Example I: Basis Expansion (The Twisted Cubic)}

Consider a physical model with three fields, ${x, y, z}$ subject to two vacuum constraints. We have the ideal $I = \langle f_1, f_2 \rangle \subset \mathbb{C}[x,y,z]$ with Lex ordering $x \succ y \succ z\,$,
\begin{equation}
    f_1 = x^2 - y \,, \quad f_2 = xy - z \,.
\end{equation}
\paragraph{Iteration 1.}
$\mathrm{LM}(f_1) = x^2$, $\mathrm{LM}(f_2) = xy\,$. The $\mathrm{lcm}$ is $x^2y\,$.
\begin{equation}
    S(f_1, f_2) = y(x^2 - y) - x(xy - z) = -y^2 + xz \,.
\end{equation}
In Lex order, $x \succ y\,$, so the leading term is $xz\,$. This cannot be divided by $x^2$ or $xy\,$. We add the new generator,
\begin{equation}
    f_3 = xz - y^2 \,.
\end{equation}

\paragraph{Iteration 2.}
We compute $S(f_2, f_3)$ with $\mathrm{lcm}(xy, xz) = xyz\,$,
\begin{equation}
    S(f_2, f_3) = z(xy - z) - y(xz - y^2) = -z^2 + y^3 \,.
\end{equation}
Rearranging in Lex order ($y \succ z$) we get $y^3 - z^2$. This depends only on $y,z$ and is not divisible by the current leading terms $\{x^2, xy, xz\}$. So, we add to the set of generators,
\begin{equation}
    f_4 = y^3 - z^2 \,.
\end{equation}
All subsequent S--polynomials reduce to zero (e.g., $S(f_1, f_3) = x(y^2)-z(x^2) = xy^2 - x^2z\,$, which reduces via $f_2$ and $f_3$ to zero). The final basis has expanded from 2 to 4 elements,
\begin{equation}
    G = \{ x^2 - y, \; xy - z, \; xz - y^2, \; y^3 - z^2 \} \,.
\end{equation}
Physically, this expansion reveals
that while the inputs constrained only $x^2$ and $xy$, the vacuum structure forces a specific
relationship between $y$ and $z\,$ $ (y^3 = z^2)$ that was not explicit in the potential.

\subsubsection*{Example II: Inconsistency over $\mathbb{C}$}

We address the exact situation of the non--BPS system, i.e. proving $V(I) = \varnothing$ over $\mathbb{C}\,$. Consider $I = \langle f_1, f_2, f_3 \rangle\,$,
\begin{equation}
    f_1 = x^2 + y^2 - 1 \,, \quad f_2 = x^2 - y^2 - 4 \,, \quad f_3 = xy - 1 \,.
\end{equation}
In the actual problem, we wanted to check that no real solutions to these kinds of equations exist, but we managed to show that no complex solutions exist, which, of course, includes the reals.
\paragraph{Step 1: Elimination.}
S--polynomials between $f_1, f_2$ effectively perform Gaussian elimination on the quadratic terms (reducing leading terms $x^2$ against each other):
\begin{align}
    S(f_1, f_2) &= 1 \cdot f_1 - 1 \cdot f_2 = 2y^2 + 3 \implies g_1 = y^2 + \tfrac{3}{2} \,. \\
    S(f_1, g_1) &\xrightarrow{\text{reduce}} x^2 - \tfrac{5}{2} \implies g_2 = x^2 - \tfrac{5}{2} \,.
\end{align}
The ideal is now $\langle x^2 - \tfrac{5}{2}, \, y^2 + \tfrac{3}{2}, \, xy - 1 \rangle$.

\paragraph{Step 2: The Contradiction.}
The algorithm tests consistency between the variable values fixed by $g_1, g_2$ and the constraint $f_3\,$. Consider the S--polynomial relations involving $f_3\,$. A standard reduction sequence effectively computes $(xy)^2 - x^2 y^2\,$,
\begin{equation}
    1^2 - (\tfrac{5}{2})(-\tfrac{3}{2}) = 1 - (-\tfrac{15}{4}) = 1 + \tfrac{15}{4} = \tfrac{19}{4} \neq 0 \,.
\end{equation}
Since the ideal contains the non--zero constant $\tfrac{19}{4}\,$ and $I$ is an ideal, the computation for the non--BPS D--brane system proceeds analogously, involving thousands of polynomial reductions to ultimately derive the unit element. 

\section{Geometric Consistency of Gauge--Restricted Minimization}
\label{app:gauge_consistency}

The numerical analysis presented in Section~\ref{sec:nonbps} relies on a gauge--fixed minimization strategy; the scalar potential $V(\mathbf{\Psi})$ is minimized along a submanifold (slice) $\mathcal{S} \subset \mathcal{C}$ defined by the vanishing of specific field components. This procedure reduces the computational complexity of the problem but requires geometric justification. Specifically, one must ensure that a critical point of the restricted functional $V_{\mathrm{red}} \equiv V|_{\mathcal{S}}$ corresponds to a stationary point of the full, gauge--invariant action, and that the stability of the vacuum is correctly diagnosed by the restricted Hessian. The validity of this reduction depends on the transversality of the gauge slice with respect to the gauge orbits. In this appendix, we stress the local geometric conditions under which the restricted minimization is exact, identifying the explicit failure modes associated with stabilizer subgroups and \textit{orbit collapse} \cite{Henneaux:1992ig}. This is the analysis we do for each choice of the moduli parameters $c\,$; in Appendix \ref{app:Stability}, we sketch the possible analysis of stability jumps across bifurcation walls as we move around in the moduli space. 

\subsection{Gradient Orthogonality and Constrained Stationarity}

Let $\mathcal{C}$ denote the configuration space of scalar fields (with components $\Psi_I$), endowed with a $G$--invariant physical inner product $\langle \cdot, \cdot \rangle$. The potential $V: \mathcal{C} \to \mathbb{R}$ is assumed to be $C^2$ and invariant under a compact Lie group $G$ of dimension $D$. The infinitesimal gauge variations are generated by anti--Hermitian operators $T^a$ ($a=1,\dots,D$), and the invariance of the potential implies the point--wise \textit{Ward identity},
\begin{equation}
\label{eq:ward}
\langle \nabla V(\mathbf{\Psi}), \, T^a \mathbf{\Psi} \rangle = 0 \qquad \forall a\,.
\end{equation}
Geometrically, this identity states that the gradient $\nabla V$ is everywhere orthogonal to the tangent space of the gauge orbit, $T_{\mathbf{\Psi}}(\mathcal{O}_{\mathbf{\Psi}}) = \mathrm{span}\{ T^a \mathbf{\Psi} \}$. We define a local gauge slice $\mathcal{S}$ in the neighborhood of a candidate vacuum via $D$ smooth, holonomic constraints $F^\alpha(\mathbf{\Psi}) = 0$, for $\alpha = 1, \dots, D$. Consider a configuration $\mathbf{\Psi}^* \in \mathcal{S}$ which is a stationary point of the restricted potential, $V_{\mathrm{red}} \equiv V|_{\mathcal{S}}$. By the method of Lagrange multipliers, the gradient of the full potential at $\mathbf{\Psi}^*$ must be normal to the slice. Thus, there exist multipliers $\lambda_\alpha$ such that
\begin{equation}
\label{eq:LM_expansion}
\nabla V(\mathbf{\Psi}^*) = \sum_{\alpha=1}^D \lambda_\alpha \, \nabla F^\alpha(\mathbf{\Psi}^*)\,.
\end{equation}
The configuration $\mathbf{\Psi}^*$ is a genuine physical solution of the equations of motion ($\left \{\nabla V = 0\right \}$) if and only if the constraint forces vanish, i.e., $\lambda_\alpha = 0\,$. To determine these multipliers, we project Eq.~\eqref{eq:LM_expansion} onto the gauge tangent directions. Taking the inner product with the orbit generators $T^a \mathbf{\Psi}^*$ and utilizing the Ward identity \eqref{eq:ward}, we obtain a homogeneous linear system,
\begin{equation}
0 = \langle \nabla V(\mathbf{\Psi}^*), \, T^a \mathbf{\Psi}^* \rangle = \sum_{\alpha=1}^D \lambda_\alpha \underbrace{\langle \nabla F^\alpha(\mathbf{\Psi}^*), \, T^a \mathbf{\Psi}^* \rangle}_{\mathcal{J}_{a\alpha}}\,.
\end{equation}
The matrix $\mathcal{J}_{a\alpha}(\mathbf{\Psi}^*)$ is the \textit{Faddeev--Popov operator} associated with the gauge fixing condition. The solvability of the system depends strictly on the rank of this matrix.

\paragraph{Transversality and the Gradient Norm Test.}
If the slice is \emph{transverse} to the orbit at $\mathbf{\Psi}^*$ (implying $\det \mathcal{J} \neq 0$), the kernel of $\mathcal{J}^\top$ is trivial, necessitating the unique solution $\lambda_\alpha = 0\,$. In this generic case, constrained stationarity implies full stationarity, i.e. $\left \{\nabla V(\mathbf{\Psi}^*) = 0\right \}\,$. However, if $\det \mathcal{J} = 0$, the slice fails to intersect the gauge orbit transversally. This frequently occurs in numerical optimization if the gauge--fixed fields vanish ($\Psi_{b_\alpha} \to 0$). Physically, we distinguish two scenarios:
\begin{itemize}
    \item \textbf{Stabilizer Vacuum:} Consider a vacuum $\mathbf{\Psi}^*$ invariant under a non--trivial continuous stabilizer subgroup $H \subseteq G$ i.e. the generators $\{T^h\}$ of $H$ annihilate the vacuum $\left (T^h \mathbf{\Psi}^* = 0\right)$ and the symmetry generated by $H$ is unbroken.  Consequently, the corresponding rows of the Faddeev--Popov matrix $\mathcal{J}$ vanish identically. However, by the \textit{equivariance} of the gradient $\left \{\nabla V(g\mathbf{\Psi}) = g\nabla V(\mathbf{\Psi})\right \}$, $\nabla V(\mathbf{\Psi}^*)$ is also invariant under $H$. For the $U(1)$ charged sectors of the abelian non--BPS case considered in this paper, the only invariant vector is the zero vector, implying $\nabla V$ vanishes in these directions. Thus, the equation system is consistent with $\lambda_\alpha = 0\,$.

    \item \textbf{Improper Gauge Choice:} The symmetry is spontaneously broken by a field or composite $\Phi$, but the gauge is inadvertently fixed on a vanishing field $\Psi_{\mathrm{fixed}}$.
    \begin{itemize}
            \item \textbf{With Mixing:} If the potential contains linear mixing terms (e.g., $\Psi_{\mathrm{fixed}}^\dagger \Psi_{\mathrm{not-fixed}}$), the VEV of $\Psi_{\mathrm{not-fixed}}$ typically sources a non--zero gradient for $\Psi_{\mathrm{fixed}}$ (\textit{tadpole} problem). If constrained to vanish, $\Psi_{\mathrm{fixed}}$ requires a non--zero constraint force ($\lambda_\alpha \neq 0$) to maintain stationarity. The solver converges to a spurious point where $\left \{\nabla V_{\mathrm{full}} \neq 0\right \}$. To detect this, we implement a \textit{Gradient Norm Test} where upon convergence of the restricted solver, we compute the norm of the full gradient $||\nabla V(\mathbf{\Psi}^*)||$ in the unconstrained space. We accept the solution if and only if,
        \begin{equation}
        \frac{\|\nabla V(\mathbf{\Psi}^*)\|}{1+|V(\mathbf{\Psi}^*)|} < \varepsilon_{\rm rel},\qquad
        \varepsilon_{\rm rel}\in[10^{-36},\,10^{-39}]\,.
        \end{equation}

        \item \textbf{Without Mixing:} If the sectors are decoupled at linear order (e.g., $|\Psi_{\mathrm{not-fixed}}|^2 |\Psi_{\mathrm{fixed}}|^2$), $\Psi_{\mathrm{fixed}}=0$ is a valid stationary point $\left \{\nabla V = 0\right \}$. The Gradient Norm Test accepts this point, but the stability in the gauge--fixed direction remains unverified (see Case III below).
    \end{itemize}
\end{itemize}
Consequently, a vanishing gradient norm is a necessary but insufficient condition for vacuum validation; the classification requires spectral analysis.

\subsection{Hessian Structure and Spectral Sufficiency}

The physical classification of the vacuum relies on the spectral properties of the Hessian operator. In generic constrained optimization, stability is governed by the Hessian of the Lagrangian, $D^2 \mathcal{L} = D^2 V - \sum \lambda_\alpha D^2 F^\alpha$, restricted to the constraint surface.
However, the unitary gauge employed in this work imposes \emph{linear} constraints ($F^\alpha = \mathrm{Im}\,\Psi_{b_\alpha}$). Linearity implies vanishing constraint curvature ($D^2 F^\alpha = 0$). Consequently, the Hessian of the Lagrangian reduces strictly to the Hessian of the potential,
\begin{equation}
D^2 \mathcal{L} \equiv \mathcal{H}(\mathbf{\Psi}^*)\,.
\end{equation}
This identity holds regardless of the values of the Lagrange multipliers. Thus, stability is determined by the projection of the potential's Hessian onto the tangent space of the slice $\mathcal{S}$. We define the reduced Hessian as,
\begin{equation}
\label{eq:Hred_def}
\mathcal H_{\mathrm{red}} \equiv P_{\mathcal S}\,\mathcal H(\mathbf{\Psi}^*)\,P_{\mathcal{S}}^\top\,,\qquad
P_{\mathcal S}=I - N (N^T N)^{-1} N^T,
\end{equation}
where \(N=[\nabla F^1,\dots,\nabla F^D]\) is the matrix of constraint normals and $P_{\mathcal{S}}$ is the orthogonal projector onto $T_{\mathbf{\Psi}^*}\mathcal{S}$. We analyze the validity of the reduced Hessian in three regimes.

\paragraph{Case I: Regular Vacua (Broken Symmetry).}
Consider a vacuum $\mathbf{\Psi}^*$ where the gauge action is free (trivial stabilizer), and the gauge fixing is regular ($\det \mathcal{J} \neq 0$). This is the standard Higgs phase. The full Hessian $\mathcal{H}$ possesses exact zero modes (Goldstone bosons) corresponding to the broken generators. Since the slice is transverse to the orbits, $P_{\mathcal{S}}$ induces an isomorphism between the physical tangent space and the slice, effectively filtering out the unphysical Goldstone modes. In this regime, the positive definiteness of $\mathcal{H}_{\mathrm{red}}$ is the necessary and sufficient condition for stability \footnote{The gauge-fixed potential retains a residual discrete symmetry corresponding to the \textit{Weyl group} of the broken $U(1)$, specifically $\Psi_b \to (-1)^{q_b}\Psi_b\,$. This guarantees that critical points with non--zero components in odd--charged sectors appear as $\mathbb{Z}_2$ \textit{Gribov copies} \cite{Gribov:1977wm} and should be identified.}.

\paragraph{Case II: Stabilizer Vacua (Unbroken Symmetry).}
This regime is accessed when the charged fields $\{\Psi_k\}$ associated with the gauge fixing acquire vanishing VEVs, preserving a stabilizer subgroup $H \subseteq G$. Geometrically, this corresponds to a change in the orbit stratum -- the gauge orbit $\mathcal{O}_{\mathbf{\Psi}^*}$ becomes \textit{degenerate}, and its dimension drops from $\mathrm{dim}(G)$ to $\mathrm{dim}(G) - \mathrm{dim}(H)$. Consequently, the Faddeev--Popov operator becomes singular ($\det \mathcal{J} = 0$), and the standard slice construction ceases to be transverse. Despite this geometric singularity, the Hessian identity $D^2 \mathcal{L} \equiv \mathcal{H}$ remains valid. To analyze the spectrum, we decompose the tangent space into sectors labeled by the irreducible representations of the stabilizer $H \cong U(1)$. The non--trivial structure appears in the charged sector \footnote{We can consider all fields not charged under the stabilzer $H$ as ``neutral". The full Hessian is block--diagonal in the \{neutral, charged sectors\} basis, and their stability can be analyzed separately. Also, the stabilzer subgroup can be a product of more than one $U(1)\,$, with further block--diagonal structure. To optimize the spectral analysis, we exploit the inherent sparsity of the Hessian, which reflects the physical decoupling of sectors in the scalar potential. We map the non--zero structure of the Hessian to the boolean adjacency matrix $\mathcal{A}$ of an undirected graph $G=(V, E)$, defining $\mathcal{A}_{ij} = 1$ if $|\mathcal{H}_{ij}| > 0$ and $0$ otherwise. The vertices $V$ represent field components, while the edges $E$ correspond to non--vanishing interactions. We identify the connected components of this graph using a standard traversal algorithm (e.g., Breadth--First Search \cite{CLRS2009}). These components define a permutation matrix $P$ that transforms the Hessian into a block--diagonal form, $P^T \mathcal{H} P = \mathrm{diag}(\mathbf{h}_1, \dots, \mathbf{h}_k)$. Since the characteristic polynomial factorizes, $\det(\mathcal{H} - \lambda \mathbb{I}) = \prod_i \det(\mathbf{h}_i - \lambda \mathbb{I})$, the full spectrum is obtained by independently diagonalizing these smaller blocks. This procedure reduces the computational complexity from $\mathcal{O}(N_{\mathrm{total}}^3)$ to $\sum \mathcal{O}(n_i^3)$, where $n_i$ is the dimension of the $i$--th block.}. In the real basis of field components $\Psi = (\mathrm{Re}\,\Psi_1, \mathrm{Im}\,\Psi_1, \dots)^T$, the generator of the stabilizer acts as a block--diagonal symplectic matrix,
\begin{equation}
\mathcal{T} = \bigoplus_{k} q_k J, \quad \text{with} \quad J = \begin{pmatrix} 0 & -1 \\ 1 & 0 \end{pmatrix}\,.
\end{equation}
Gauge invariance requires the Hessian to commute with this generator, i.e. $[\mathcal{H}_{\mathrm{charged}}, \mathcal{T}] = 0\,$. We can determine the structure of the allowed blocks using Schur's Lemma for real representations:
\begin{enumerate}
    \item \textbf{Same--Charge Mixing ($q_i = q_j$):} The block $\mathbf{H}_{ij}$ connects two identical representations. The commutation condition $[\mathbf{H}_{ij}, J] = 0$ requires $\mathbf{H}_{ij}$ to be in the commutant algebra of the symplectic group. Restricting to symmetric matrices (as Hessians must be), the only solution is \textit{isotropic},
    \begin{equation}
    \mathbf{H}_{ij} = m_{ij}^2 \,\mathbb{I}_2 = \begin{pmatrix} m_{ij}^2 & 0 \\ 0 & m_{ij}^2 \end{pmatrix}\,.
    \end{equation}

    \item \textbf{Opposite--Charge Mixing ($q_i = -q_j$):} The block $\mathbf{H}_{ij}$ connects a representation to its conjugate. The condition involves the generator with opposite signs, $\mathbf{H}_{ij}(-J) = J\mathbf{H}_{ij}\,$, implying the block must \textit{anticommute} with $J$ (i.e., $\{ \mathbf{H}_{ij}, J \} = 0$). The most general real symmetric matrix satisfying this is \textit{anti--isotropic}, spanned by the Pauli matrices $\sigma_1$ and $\sigma_3\,$,
    \begin{equation}
    \mathbf{H}_{ij} = \begin{pmatrix} u_{ij} & v_{ij} \\ v_{ij} & -u_{ij} \end{pmatrix}\,.
    \end{equation}
\end{enumerate}
This structure -- isotropic diagonal blocks and anti--isotropic off--diagonal blocks -- ensures that while the Hessian is not block--diagonal in the real/imaginary basis (due to the mixing terms $u, v$), it respects the complex structure induced by the stabilizer. Specifically, the matrix commutes with the operator $\mathcal{T}$, which satisfies $\mathcal{T}^2 = -\mathbb{I}$ (normalized to unit charge). This implies that the tangent space possesses a linear complex structure preserved by the Hessian. Consequently, the spectrum is \textit{doubly degenerate} --  the eigenspaces are complex vector spaces, and every real eigenvalue has an even multiplicity. 

For vacuum validation, we define the \textit{Complex Hessian}, $\mathcal{H}_\mathrm{complex}$ by mapping the system to the holomorphic basis $\{\Psi_k, \Psi_k^\dagger\}$ in the charged sector. This Hermitian matrix encapsulates the full stability information, including the mixing terms, while naturally factoring out the spectral degeneracy. The Hessian for the neutral section can be analyzed separately. As no zero modes are associated with the the unbroken stabilizer group $H$, the full Hessian has $\mathrm{dim}(G/H)$ zero modes and the reduced Complex Hessian with all the other $U(1)$ groups gauge fixed has none. However, \textit{accidental flat directions} are always a possibility and in the abelian non--BPS case some of the adjoint fields in the neutral sector are flat at stabilizer extrema points (submanifolds). 

\paragraph{Case III: Singular Gauge Choice (Spectral Blindness).}
This pathological regime arises when the minimization converges to a point where the gauge--fixed field vanishes ($\Psi_{\mathrm{fixed}} = 0$), but the gauge symmetry is spontaneously broken by a distinct field or sector ($\langle \Psi\rangle_{\mathrm{not-fixed}} \neq 0$). 
Geometrically, this represents a complete failure of the gauge slice. The constraint $\mathrm{Im}\,\Psi_{\mathrm{fixed}}=0$ fails to align with the broken generator i.e. the slice is parallel to the gauge orbit rather than transverse to it. This misalignment has two critical spectral consequences. First, it fails to remove the unphysical Goldstone mode associated with the broken generator (which resides in the $\Psi_{\mathrm{not-fixed}}$ sector). This mode manifests as a zero eigenvalue in the reduced Hessian $\mathcal{H}_{\mathrm{red}}$ (there maybe be other accidental flat directions).
Second, and more dangerously, the vacuum suffers from \textit{spectral blindness}. Unlike Case II, where the unbroken symmetry enforced mass degeneracy between real and imaginary components, the symmetry breaking VEV $\langle \Psi\rangle_{\mathrm{not-fixed}}$ can generate interaction terms (e.g., $V \supset \kappa \langle \Psi\rangle_{\mathrm{not-fixed}} \Psi_{\mathrm{fixed}}^2 + \mathrm{h.c.}$) that explicitly split these masses. The Hessian block for the fixed field takes the form,
\begin{equation}
\mathcal{H}_{\mathrm{fixed}} \approx \begin{pmatrix} M^2 + \delta & 0 \\ 0 & M^2 - \delta \end{pmatrix}\,,
\end{equation}
where $\delta \propto \kappa \langle \Psi\rangle_{\mathrm{not-fixed}}\,$. The reduced Hessian detects only the real component ($M^2+\delta$). It is \textit{blind} to the curvature along the imaginary  $(\mathrm{Im}\,\Psi_{\mathrm{fixed}})$ axis ($M^2-\delta$). If the splitting is large enough such that $M^2 - \delta < 0$, the vacuum is physically unstable (a saddle point), yet $\mathcal{H}_{\mathrm{red}}$ may report complete stability (positive definiteness).

To prevent false validations, we implement a \textit{Complement Curvature Check}. Whenever a solution exhibits vanishing gauge--fixed fields coexisting with non--zero VEVs in other sectors, we explicitly compute the Hessian eigenvalues in the orthogonal complement of the slice (the directions removed by the constraints). If a negative eigenvalue is detected, the solution is flagged as unstable. In such cases, the solver is restarted with a valid gauge choice aligned with the symmetry--breaking field $\Psi_{\mathrm{not-fixed}}\,$. This realignment not only restores transversality but also resolves potential \textit{tadpole} instabilities (where linear mixing terms like $\Psi_{\mathrm{fixed}}^\dagger \Psi_{\mathrm{not-fixed}}$ may drive $\Psi_{\mathrm{fixed}}$ away from zero, i.e. source a non--zero gradient for $\Psi_{\mathrm{fixed}}$), ensuring the stationary point is genuine.


\section{Stability, Bifurcations, and Moduli Dependence}
\label{app:Stability}

In Section \ref{sec: landscape}, our numerical analysis alluded to a non--trivial dependence vacuum landscape of the non--BPS D--brane system on the background moduli parameters. Specifically, we noticed that the number of expected isolated stable local minima sometimes jumped as we roamed in the moduli space (constant metric and B--fields in the compactified directions, FI parameters, etc.). This behavior stands in sharp contrast to the BPS systems, where the number of vacua is typically governed by topological invariants -- such as the Witten index or intersection numbers -- which are protected against continuous deformations of the background. In this appendix, we collect some ideas from bifurcation theory that may be useful for studying such parameter dependencies. It is likely that the discrepancy is not a numerical artifact but a fundamental feature of \textit{real algebraic geometry}. While the counting of complex solutions to holomorphic BPS equations is protected by algebraic bounds (such as the Bernstein--Khovanskii--Kushnirenko theorem \cite{Sturmfels1998PolynomialEA,Kaveh2008AlgebraicEA}), the stability of non--BPS vacua is governed by the \textit{spectral flow} of the Hessian matrix. As the moduli parameters vary, the system undergoes structural bifurcations crossing ``walls of marginal stability'' -- where vacua can be created, annihilated, or destabilized.

\subsection{Hessian Structure and Stability Constraints}

We consider an effective scalar potential $V(\bm{\Psi})$ constructed as a sum of squares of constituent terms (representing F--terms and D--terms),
\begin{equation}
    V(\bm{\Psi}) = \frac{1}{2} \sum_{a=1}^N f_a(\bm{\Psi}; p)^2\,,
\end{equation}
where $\bm{\Psi} \in \mathbb{R}^n$ denotes the dynamical fields and $p \in \mathcal{P}$ represents the set of \textit{control parameters} (moduli). The E.O.M. defining the vacuum structure is $\{\nabla_i V(\bm{\Psi}) = \sum_a f_a(\bm{\Psi}) (\partial_i f_a) = 0\}$. To analyze the local stability of a critical point $\bm{\Psi}^*$, we examine the Hessian matrix,
\begin{equation} \label{eq:HessianDecomp}
    \mathcal{H}_{ij}(\bm{\Psi}) = \underbrace{\sum_a (\partial_i f_a)(\partial_j f_a)}_{\mathcal{G}_{ij}} + \underbrace{\sum_a f_a(\bm{\Psi}) (\partial_i \partial_j f_a)}_{\mathcal{Q}_{ij}}\,,
\end{equation}
where, $\mathcal{G}_{ij}$ is the Gram matrix of the Jacobian vectors $\bm{J}_i = \partial_i \vec{f}$ and hence universally positive semi--definite. The second term $\mathcal{Q}$ is the curvature term. This decomposition highlights the structural dichotomy between the BPS and non--BPS sectors:
\paragraph{The BPS Limit ($V=0$).}
A SUSY vacuum satisfies the stronger condition $f_a(\bm{\Psi}^*) = 0$ for all $a\,$. Consequently, the second term in Eq.~\eqref{eq:HessianDecomp} vanishes identically ($\mathcal{Q}_{ij} = 0$) and the Hessian $
    \mathcal{H}_{ij}^{\text{BPS}} = \mathcal{G}_{ij} \succeq 0\, $, ensures that a BPS solution can never be a maximum or an unstable saddle point; it is strictly stable (or possesses flat directions). Furthermore, if the functions $f_a$ are polynomials, the system $f_a(z)=0$ defines an affine algebraic variety and the Bernstein--Khovanskii--Kushnirenko (BKK) theorem guarantees that the number of isolated complex roots is bounded by the mixed volume of the associated Newton polytopes and is invariant for generic choices of coefficients $p\,$. Thus, BPS vacua are algebraically protected;  they cannot be created or destroyed via local bifurcations in the moduli space, ensuring the robustness of the BPS index.

\paragraph{The Non--BPS Cases ($V>0$).}
For a non--BPS vacuum, supersymmetry is explicitly or spontaneously broken such that $V(\bm{\Psi}^*) > 0\,$. This implies $f_a(\bm{\Psi}^*) \neq 0$ for some active constraints. Consequently, $\mathcal{Q}_{ij}$ is non--zero and generally indefinite. Positivity is not structurally guaranteed; the local stability of the vacuum is determined by the competition between the stabilizing metric term $\mathcal{G}_{ij}$ and the potentially destabilizing curvature term $\mathcal{Q}_{ij}\,$. If the vacuum energy (magnitude of $f_a$) is sufficiently large, a negative eigenvalue of $\mathcal{Q}$ can dominate $\mathcal{G}$, rendering the Hessian indefinite. This makes the stability of non--BPS vacua a dynamical property sensitive to the moduli $p$, necessitating a bifurcation analysis.

\subsection{Bifurcation Classification  and Dimensional Reduction}
We now study the dependence of the vacuum structure on the control parameters $p \in \mathcal{P}\,$. The space of physical vacua is defined as the \textit{Critical Locus} $\mathcal{L} \subset \mathbb{R}^n \times \mathcal{P}\,$,
\begin{equation}
    \mathcal{L} = \left\{ (\bm{\Psi}, p) \;\big|\; \nabla_{\bm{\Psi}} V(\bm{\Psi}, p) = 0 \right\}\,.
\end{equation}
For a generic parameter configuration $p\,$, the potential $V_{p}(\bm{\Psi}) \equiv V(\bm{\Psi}, p)$ is a \textit{Morse Function} if all its critical points are non--degenerate ($\det \mathcal{H} \neq 0$). In such regions, the Implicit Function Theorem guarantees that the vacua $\bm{\Psi}^*(p)$ evolve as smooth, unique trajectories in the field space as $p$ is varied locally.

\paragraph{Hessian Singularity and Root Collision.}
Structural stability breaks down at the \textit{Bifurcation Set} (or Discriminant Locus) $\Delta \subset \mathcal{P}\,$ where the Hessian becomes singular, 
\begin{equation}
    \Delta = \left\{ p \in \mathcal{P} \;\big|\; \exists\, \bm{\Psi} \in \mathcal{L} \text{ s.t. } \det \mathcal{H}(\bm{\Psi}) = 0 \right\}\,.
\end{equation}
To explicitly show that crossing $\Delta$ corresponds to a collision of solutions, consider a first--order perturbation of the vacuum solutions $\{\nabla V(\bm{\Psi}^*(p), p) = 0\}$ along a trajectory parameterized by $p\,$,
\begin{equation}
    \frac{d}{dp} \left( \nabla V \right) = \mathcal{H}(\bm{\Psi}^*) \cdot \frac{d\bm{\Psi}^*}{d p} + \frac{\partial (\nabla V)}{\partial p} = 0\, \implies \frac{d\bm{\Psi}^*}{d p} = - \mathcal{H}^{-1} \cdot \frac{\partial (\nabla V)}{\partial p}\,.
\end{equation}
As $p$ approaches the bifurcation set $\Delta\,$, the $\det \mathcal{H} \to 0$ and if the rank of the Hessian drops by 1 (the generic codimension--1 case), the inverse matrix $\mathcal{H}^{-1}$ develops a pole along the direction of the null eigenvector. Consequently, the rate of change $|d\bm{\Psi}^*/d p|$ diverges. Geometrically, this implies that the tangent to the trajectory becomes vertical in the $\mathbb{R}^n \times \mathcal{P}$ fibration, signaling a fold where two or more branches of solutions merge and potentially annihilate.

\paragraph{Dimensional Reduction via the Splitting Lemma.} 
D--brane EFTs typically involve a large number of moduli ($n \gg 1$). However, the local dynamics of the bifurcation are determined solely by the degenerate sector. We appeal to the \textit{Splitting Lemma} \cite{Montaldi:2012}, which states that near a degenerate critical point with a Hessian of rank $n-k\,$, there exists a local coordinate transformation separating the system into ``active'' variables (the null space) and ``passive'' variables (the non--degenerate subspace). The potential locally splits as (\textit{germ}),
\begin{equation}
    V(\bm{\Psi}) \cong V_{\text{eff}}(y_1, \dots, y_k; p) + \sum_{i=k+1}^n \pm x_i^2\,,
\end{equation}
where $y_i$ span the kernel of $\mathcal{H}$ (the massless modes) and $x_i$ represent the massive modes.
Note that this reduction effectively integrates out the massive modes by solving their equations of motion $\partial_{x_i} V = 0$ as a function of the light modes, $x_i = x_i(y)$.
This reduction implies that the stability analysis of the full high--dimensional system is mathematically isomorphic to the analysis of the low--dimensional effective potential $V_{\text{eff}}\,$. This justifies the possible classification of instabilities using elementary \textit{catastrophe theory}.

\subsection*{Classification of Instabilities and Universality}

It is our well--founded hope that the effective potential $V_{\text{eff}}$ of the massless modes submits to a Catastrophe Theory classification \cite{Montaldi:2012}, specifically the relationship between the number of control parameters (codimension) and the polynomial degree required to capture the topology of the singularity (determinacy).

\paragraph{Coordinate Definition and Determinacy.}
We define the local coordinates $\bm{\Psi}^*$ as fluctuations around the critical point $\bm{\Psi^*}$, such that the bifurcation occurs at the origin $\bm{\Psi}=0$\,. A critical point is defined as \textit{$k$--determined} if the first non--vanishing term in its Taylor expansion is of order $k\,$. A central theorem of \textit{singularity theory} (Thom's Finite Determinacy Theorem \cite{Thom1956LesSD}) states that for a $k$--determined potential, there exists a diffemorphism  $y \to \tilde{y}(y)$ (preserving the origin) that eliminates all higher--order terms $\mathcal{O}(y^{k+1})$.
For example, a perturbed potential $V(y) = c_4 y^4 + c_6 y^6 + \dots$ (with $c_4 \neq 0$) can be mapped exactly to the canonical form $V(\tilde{y}) = \tilde{y}^4$. These higher--order terms are ``structurally irrelevant'' because they do not alter the local topology established by the leading term. Here, we touch upon two such classifications which might show up in the general non--BPS D--brane analysis. 

\paragraph{Generic Case: The Fold Catastrophe ($A_2$).}
Consider a general potential with no $\mathbb{Z}_2$ discrete symmetries. To find a bifurcation point, one must tune a single control parameter (codimension 1) to satisfy the condition that the Hessian eigenvalue vanishes ($V''=0$).
In this generic scenario, the third derivative $V'''$ is non--zero (3--determined). A general unfolding would take the form $V(y) \sim y^3 + p_2 y^2 + p_1 y$. However, in the absence of symmetry, the quadratic term can always be eliminated by a linear coordinate shift $y \to y - p_2/3\,$. Thus, the effective potential reduces to the canonical \textit{Fold Catastrophe}, $
    V_{\text{fold}}(y; p) = y^3 + p y\,$. As $p$ crosses zero, a stable vacuum and an unstable saddle collide and annihilate, changing the real root count to $\Delta N = \pm 2\,$.

\paragraph{Symmetric Case: The Pitchfork Bifurcation ($A_3$).}
 The potential respects a discrete parity symmetry, and we expand around the symmetric locus ($\bm{\Psi}=0$), implying the fluctuation potential must be an even function, identically forbidding all odd--order derivatives ($V''' \equiv 0$). When the parameter $p$ is tuned to make the mass term vanish ($V''=0$), the leading non--zero term is automatically the quartic term $V^{(4)}$ (4--determined). Unlike the generic case, we cannot shift away the quadratic term without violating the parity symmetry. The system falls into the \textit{Cusp (or Pitchfork) Catastrophe} universality class $V_{\text{pitch}}(y; p) = y^4 + p y^2\,$.  This describes the \textit{Supercritical Pitchfork Bifurcation} \footnote{Higher codimension singularities, such as the Swallowtail ($A_4$, codimension 3) or the Umbilic series ($D_k, E_k$), appear only when multiple control parameters are tuned simultaneously to eliminate lower--order terms and sometimes fit Arnold's ADE classification \cite{Arnold1993DynamicalVIII}. In a generic one--parameter scan, the probability of intersecting such a locus is zero.}:
\begin{itemize}
    \item \textbf{Symmetry Restoration ($2 \to 1$):} For $p < 0$, two symmetric stable vacua exist at $y \approx \pm \sqrt{-p/2}$. As $p \to 0^+$, these vacua collide with the central saddle at $y=0\,$.
    \item \textbf{Complexification:} The roots do not vanish but migrate to the complex branch ($y \in i\mathbb{R}$), preserving the algebraic root count. This confirms that the variation in $N_{\text{vac}}$ is a spectral transition where real stable solutions are exchanged for complex saddle points.
\end{itemize}
An analogous realization of the pitchfork bifurcation, governing the stability of the non--BPS stabilizer submanifolds, is examined in Footnote \ref{foot: flatdir} of the main text.

\subsection*{Global Constraints and Allowed Transitions}
\label{sub:PoincareHopf}

While Catastrophe Theory governs the local geometry of bifurcations, the global consistency is enforced by topological constraints on the field configuration space $\mathcal{C}$ primarily by the Poincaré--Hopf theorem, which relates the local stability indices of critical points to the global Euler characteristic $\chi(\mathcal{C})$. For a non--degenerate critical point $\bm{\Psi}^*$, the topological index $I(\bm{\Psi}^*)$ is defined by the sign of the Hessian determinant,
\begin{equation}
    I(\bm{\Psi}^*) = \text{sgn} \det \mathcal{H}(\bm{\Psi}^*) = (-1)^{n_-}\,,
\end{equation}
where $n_-$ is the Morse index. A stable vacuum ($n_-=0$) has index $I = +1$ and a saddle point ($n_-=1$) has an index $I = -1$ (assuming codimension--1 instability). For a compact $\mathcal{C}$ and a coercive potential, the sum of indices is a topological invariant, $\sum_{p \in \text{Cr}(V)} I(\bm{\Psi}^*_p) = \chi(\mathcal{C})\,$ and dictates that it must be conserved across any bifurcation. This conservation law strictly limits the ``allowed" transitions in the vacuum landscape. For example, it forbids a $\Delta N_{vac} = +1$ (e.g., $0 \to 1$) transition. A stable vacuum cannot appear alone; it must either split from an existing vacuum (Pitchfork) or emerge from the complex plane with a saddle partner (Fold). This also provides some shielding against numerical noise. Any bifurcation that fails to conserve $\sum (-1)^{n_-}$ indicates a pathology in the EFT (e.g., a singularity moving to infinity or a breakdown of the coercivity condition).

\subsection{Illustrative Example: Geometric Frustration}
\label{sub:ExplicitFrustration}

To illustrate the universality of the bifurcation mechanism in settings where SUSY is explicitly broken, we analyze a minimal model driven by ``geometric frustration'' \cite{Moessner2006Geometrical}; a D--term preference for a nonzero radius is pitted against an F--term attraction to the origin together with a flux deformation that aligns the VEV along a real or imaginary axis. With unit dimensionless couplings, the total potential
\begin{equation}
\label{eq: gfpotential}
V(\Psi)=\tfrac{1}{2}(|\Psi|^2-\xi)^2 + \tfrac{1}{2}|\Psi|^2 + \tfrac{1}{2}(\Psi^2+\bar\Psi^2-\gamma)^2
\end{equation}
depends on the Fayet--Iliopoulos scale \(\xi\) and the real flux parameter \(\gamma\). Writing \(\Psi=x+iy\), the potential becomes
\begin{equation}
V(x,y)=\tfrac{1}{2}(x^2+y^2-\xi)^2 + \tfrac{1}{2}(x^2+y^2) + \tfrac{1}{2}(2x^2-2y^2-\gamma)^2 .
\end{equation}
 A SUSY vacuum would require simultaneously \(V_D=0\) (\(|\Psi|^2=\xi\)) and \(V_F=0\) (\(|\Psi|^2=0\)), which are incompatible conditions for any \(\xi>0\). Consequently, the model is explicitly non--SUSY with a strictly positive global minimum \(V_{\min}>0\) throughout the allowed parameter space.

\paragraph{Phase structure.} Solving the stationarity conditions reveals four distinct regimes depending on the interplay between \(\xi\) and \(\gamma\,\):

\begin{itemize}
\item \textbf{Phase I (Wigner / Origin).} The trivial solution \(x=y=0\) is the unique physical critical point. It is locally stable when
\begin{equation}
1-2\xi\pm4\gamma >0 \quad\Longleftrightarrow\quad \xi < \frac{1-4|\gamma|}{2}.
\end{equation}
In this regime, all other critical points have exited the physical spectrum, leaving the origin as the sole vacuum.

\item \textbf{Phase II (Aligned, real axis).} For \(y=0\), a nontrivial real solution exists with $
x_0^2 = \frac{2\xi + 4\gamma -1}{10}\,$, and is real (physical) when \(2\xi +4\gamma -1>0\,\). Its transverse stability is governed by the mass of the orthogonal fluctuations, \(M_y^2 \equiv \partial_y^2 V|_{(x_0,0)}\,\), which evaluates to $
M_y^2=\frac{8}{5}\big(\gamma - 2\xi +1\big)\,$.
Thus, the aligned vacuum is stable against imaginary perturbations provided $
\gamma > 2\xi -1\,$.

\item \textbf{Phase III (Anti--aligned, imaginary axis).} For \(x=0\,\), an imaginary solution exists with $
y_0^2 = \frac{2\xi - 4\gamma -1}{10}\,$, 
which is physical when \(2\xi -4\gamma -1>0\,\). Its transverse stability condition is $
\gamma < -(2\xi -1)\,$.

\item \textbf{Phase IV (Broken / Mixed vacuum).} Solving the two coupled equations simultaneously yields a mixed vacuum,
\begin{equation}
x^2 = \frac{2\xi + \gamma -1}{4},\qquad
y^2 = \frac{2\xi - \gamma -1}{4}.
\end{equation}
A valid mixed solution with both \(x^2>0\) and \(y^2>0\) exists in the wedge $
\xi > \frac{1+|\gamma|}{2}\,$.
This phase represents the global compromise between the circular D--term preference and the axis--aligning flux deformation.
\end{itemize}

\paragraph{Bifurcations and topology of transitions.} The boundaries between these phases correspond to distinct topological bifurcations in the \((\xi,\gamma)\) plane:

\begin{enumerate}
\item \textbf{Vacuum creation (Saddle--Node).} The axis solutions appear at the boundaries where their VEVs vanish,
\begin{equation}
2\xi + 4\gamma -1 = 0 \quad(\text{real axis}),\qquad
2\xi - 4\gamma -1 = 0 \quad(\text{imag.\ axis}).
\end{equation}
Crossing these lines from Phase I into Phases II/III, the origin becomes unstable (saddle) and ejects a pair of stable vacua along the axis. 

\item \textbf{Stability exchange (Supercritical Pitchfork).} The axis solutions lose transverse stability exactly when the mixed vacuum appears. This occurs when the transverse mass vanishes,
\begin{equation}
M_y^2=0 \;\Longleftrightarrow\; \gamma = 2\xi -1\,,
\end{equation}
and similarly for the anti--aligned solution at \(\gamma = -(2\xi -1)\,\). Crossing these lines, the axis--aligned minimum transforms into a saddle point, bifurcating into two stable mixed vacua off the axis.
\end{enumerate}
It is crucial to note that the massless mode appearing at the critical lines \(\gamma_c=\pm(2\xi -1)\) signals a classical topological transition (change of stability), not a restoration of supersymmetry. The underlying physics is geometric; as parameters cross the critical line, the slice of field space that minimizes the potential must reorient from an axis-aligned locus to an off--axis mixed configuration.

\paragraph{Organizing Center.}
All bifurcation lines intersect at the unique codimension--2 point $
(\xi_c, \gamma_c) = (1/2, 0)\,$. 
At this ``organizing center'', the quadratic terms in the potential vanish identically ($1-2\xi \pm 4\gamma = 0$), rendering the origin a highly degenerate critical point dominated by quartic couplings. This singularity organizes the entire phase diagram; it is the vertex where the Wigner phase ends and the three symmetry--breaking phases (Aligned, Anti--aligned, and Mixed) emerge simultaneously.

\paragraph{The BPS to non--BPS Phase Transition (Massless Limit).}
It is instructive to examine the limit where the soft--breaking mass term $\tfrac{1}{2}|\Psi|^2$ is removed from the potential \eqref{eq: gfpotential}. In this limit, the vacuum structure is determined entirely by the geometric intersection of the D--term condition ($|\Psi|^2 = \xi$, a circle) and the F--term flux condition ($x^2 - y^2 = \gamma/2$, a hyperbola). A supersymmetric ground state ($V_{\min}=0$) exists if and only if these two loci intersect in the real plane. Solving the system $x^2+y^2=\xi$ and $2x^2-2y^2=\gamma$ yields
\begin{equation}
    x^2 = \frac{2\xi + \gamma}{4}\,, \qquad y^2 = \frac{2\xi - \gamma}{4}\,.
\end{equation}
These squared VEVs remain non--negative if and only if $
    |\gamma| \le 2\xi \,$, and defines a \emph{BPS Phase} in parameter space where supersymmetry is preserved. The boundaries $\gamma = \pm 2\xi$ mark a topological phase transition:
\begin{itemize}
    \item \textbf{BPS Phase ($|\gamma| \le 2\xi$):} The circle and hyperbola intersect at four points (or two tangent points at the boundary). The vacuum energy is strictly zero.
    \item \textbf{Spontaneously Broken Phase ($|\gamma| > 2\xi$):} The geometric constraints become incompatible (the hyperbola pulls completely outside the circle). The system is forced into a frustrated state with $V_{\min} > 0$, breaking supersymmetry spontaneously.
\end{itemize}
In our model Eq.~\eqref{eq: gfpotential}, the reintroduction of the mass term $\frac{1}{2}|\Psi|^2$ acts as a universal deformation that lifts the entire BPS region, converting the BPS vacua into the stable non--BPS states analyzed above.

\bibliographystyle{JHEP}
\bibliography{references}

\end{document}